\RequirePackage{fix-cm}
\documentclass[smallextended]{svjour3}       % onecolumn (second format)
\smartqed  % flush right qed marks, e.g. at end of proof
\usepackage{cite}
\usepackage{graphicx,epstopdf,epsfig}
\graphicspath{{figures/}}
\DeclareGraphicsExtensions{.eps}
\usepackage{amsmath}
\usepackage{amssymb}
\newcommand{\R}{\mathbb{R}}
\usepackage{array}
\usepackage{fixltx2e}
\usepackage{url}
\usepackage{booktabs}
\usepackage{float}
\usepackage{hyperref}
\hypersetup{colorlinks,linkcolor={blue},citecolor={blue},urlcolor={blue}}
\usepackage[caption=false,font=scriptsize,labelfont=sf,textfont=sf]{subfig}
\usepackage[linesnumbered,ruled,vlined]{algorithm2e}
 \usepackage{mathptmx}      % use Times fonts if available on your TeX system
\usepackage{latexsym}
 \journalname{Biological Cybernetics}
\begin{document}

\title{
	Modelling \textit{Drosophila} Motion Vision Pathways for Decoding the Direction of Translating Objects Against Cluttered Moving Backgrounds
	\thanks{
		This research has received funding from the European Union’s Horizon 2020 research and innovation programme under the Marie Sklodowska-Curie grant agreement No 691154 STEP2DYNA and No 778602 ULTRACEPT. 
		Corresponding authors: Shigang Yue (syue@lincoln.ac.uk) and Qinbing Fu (qifu@lincoln.ac.uk).
	}
}

\titlerunning{
	Decoding the Direction of Translating Objects Against Cluttered Moving Backgrounds
}        % if too long for running head

\author{Qinbing Fu	\and
        Shigang Yue
}


\institute{Q. Fu and S. Yue \at
              Machine Life and Intelligence Research Centre\\
              Guangzhou University\\
              Guangzhou, China\\
           \and
           Q. Fu and S. Yue \at
              Computational Intelligence Lab/Lincoln Centre for Autonomous Systems\\
              University of Lincoln\\
              Lincoln, United Kingdom\\
}


\maketitle

\begin{abstract}
	
	Decoding the direction of translating objects in front of cluttered moving backgrounds, accurately and efficiently, is still a challenging problem. 
	In nature, lightweight and low-powered flying insects apply motion vision to detect a moving target in highly variable environments during flight, which are excellent paradigms to learn motion perception strategies. 
	This paper investigates the fruit fly \textit{Drosophila} motion vision pathways and presents computational modelling based on cutting-edge physiological researches. 
	The proposed visual system model features bio-plausible ON and OFF pathways, wide-field horizontal-sensitive (HS) and vertical-sensitive (VS) systems. 
	The main contributions of this research are on two aspects: 
	1) the proposed model articulates the forming of both direction-selective (DS) and direction-opponent (DO) responses, revealed as principal features of motion perception neural circuits, in a feed-forward manner; 
	2) it also shows robust direction selectivity to translating objects in front of cluttered moving backgrounds, via the modelling of spatiotemporal dynamics including combination of motion pre-filtering mechanisms and ensembles of local correlators inside both the ON and OFF pathways, which works effectively to suppress irrelevant background motion or distractors, and to improve the dynamic response. 
	Accordingly, the direction of translating objects is decoded as global responses of both the HS and VS systems with positive or negative output indicating preferred-direction (PD) or null-direction (ND) translation. 
	The experiments have verified the effectiveness of the proposed neural system model, and demonstrated its responsive preference to faster-moving, higher-contrast and larger-size targets embedded in cluttered moving backgrounds.
	

\end{abstract}

\section{Introduction}
\label{Sec: introduction}

Intelligence is one of the amazing products through millions of years of evolutionary development, with which the features of biological visual systems have been gradually learnt and acknowledged as powerful model systems towards building robust artificial visual systems. 
In nature, for the vast majority of animal species, a critically important feature of visual systems is the perception and analysis of motion that serves a wealth of daily tasks for animals \cite{Borst2011(review-motion),Borst2015(common-circuit-motion)}. 
Seeing the motion and direction in which a chased prey, a striking predator, or a mating partner is moving, is of particular importance for their survival.

Direction selective neurons, with responsive preference to specific directional visual motion, have been identified in flying insects, like locusts \cite{DSNs-1990(Rind-locust-DSNs)}, and flies \cite{Borst2011(review-motion)}. 
Each group of the direction selective neurons responds selectively to a specific optic flow (OF)-field representing the spatial distribution of motion vectors on the field of vision. 
Accordingly, the visual motion cues as feedback signals provided by such neurons are applied for ego-motion control of flying insects.

Recent decades have witnessed much progress on unravelling the underlying neurons, pathways and mechanisms of insects' motion vision systems \cite{Fu-ALife-review}. 
Notably, the fruit fly \textit{Drosophila} has been disseminated as a prominent paradigm to study motion perception strategies \cite{Riehle-fly-ON-OFF-1984,Nicolas-1989(DSN-Insect-Neurons),Borst2011(review-motion),Borst2015(common-circuit-motion),Borst-2010(review-fly-vision),Borst-2014(review-fly)}. 
More specifically, direction-selective (DS) and direction-opponent (DO) responses represented by the \textit{Drosophila} motion vision pathways have been identified as two essential features in the neural circuits \cite{Mauss-Opposing-Motions-Fly,Fly2016(direction-selectivity-fly),Badwan-Dynamic-nonlinearities-EMD}. 
The former indicates that neurons respond differently to stimuli moving in different directions, that is, the directional motion yielding the largest response is termed the preferred direction (PD); the latter denotes that neurons are also inhibited by stimuli in the opposite direction, i.e., the null (or non-preferred) direction (ND). 
How to realise such diverse direction selectivity in motion sensitive visual systems is thus attractive to not only biologists but also computational modellers for addressing real-world motion detection problems.

Although some biological and computational models have demonstrated the DS and DO responses resembling the neural circuits to decode the direction of translational OF, those models are faced with the following challenges: 
\begin{enumerate}
	\item The biological models focus on explaining the forming of DS and DO responses on neuronal or behavioural level, which have been tested by merely simple synthetic stimuli, e.g. sinusoidal gratings and the like \cite{Joesch-2010(ON-OFF-fly),Joesch_2013(functional-ON-OFF),Maisak_2013(T4-T5-fly),Eichner2011(2Q-motion),Clark_2011(6Q-model-fly),Fly2016(direction-selectivity-fly),Circuit-genetic(genetic-push-motion)}. 
	Flying insects nevertheless can detect and track a moving target in front of more cluttered backgrounds mixed with irrelevant motion or distractors. 
	Are these models able to reproduce the similar DS and DO responses when dealing with the highly variable statistics of natural environments? 
	This is yet lack of investigation.
	\item It is still a challenging problem for artificial visual systems to accurately decode the direction of foreground translating objects, by extracting merely meaningful motion cues embedded in a cluttered moving background. 
	The vast majority of bio-inspired models are efficient for motion perception, but deficient in effective mechanisms to deal with highly variable backgrounds. 
	In addition, the requirements of both energy-efficient and real-time visual processing exclude many segmentation or learning based methods.
	\item Most bio-inspired motion detection models derive from a classic theory of Hassenstein-Reichardt Correlation (HRC, or referred as `Reichardt detectors') \cite{HR-1956(EMD-1956),EMD-1989(principles-review)}. 
	The HRC-based models are sensitive to the temporal frequency of visual stimuli across the view rather than the true velocity. 
	Accordingly, a pronounced shortcoming of such methods is the dynamic response in speed tuning of translational OF perception \cite{Frye2015(EMDs-basic),Zanker-1999(EMD-speed-tuning)}.
\end{enumerate}

In this article, according to the latest physiological researches and our preliminary studies on the \textit{Drosophila} motion vision systems \cite{Fu2017(fly-DSNs-IJCNN),Fu-2017(ROBIO-fixation),Fu-ROBIO-2018}, we present a thorough modelling study to mimic the visual processing in \textit{Drosophila} ON and OFF motion vision pathways through multiple layers, from initial photoreceptors to internal lobula plate tangential cells (LPTCs), in a computational manner. 
Differently to previous related methods, the emphasis herein is laid behind the OF level. 
More specifically, we highlight the modelling of spatiotemporal dynamics in the proposed neural system model including 
1) the combination of spatial and temporal motion pre-filtering mechanisms prior to generating the DS and DO responses, 
2) the ensembles (or multi-connected) of local ON-ON and OFF-OFF motion correlators inside the ON and OFF pathways in horizontal and vertical directions. 
The former works effectively to suppress irrelevant background motion flows or distractors to a large extent, and to achieve edge selectivity revealed in motion detection neural circuits. 
The latter can enhance the dynamic response to translating objects in front of a cluttered moving background, and alleviate the impact by temporal frequency of visual stimuli. 
Accordingly, two wide-field systems, i.e., the horizontal-sensitive (HS) and the vertical-sensitive (VS) systems, integrate the LPTCs' responses to decode the principal direction of foreground translating objects against cluttered moving backgrounds.

The rest of this paper is structured as follows. 
Section \ref{Sec: literature} reviews the related works. 
Section \ref{Sec: formulation} presents the formulation of the proposed visual system model. 
Section \ref{Sec: setting} describes the experimental setting. 
Section \ref{Sec: result} illustrates the results. 
Section \ref{Sec: discussion} presents further discussions. 
Section \ref{Sec: conclusion} concludes this paper.

\section{Literature Review}
\label{Sec: literature}

Within this section, we concisely review the related works in the areas of 1) a few categories of motion sensitive neural models inspired by flying insects, 2) physiological research on the \textit{Drosophila} motion vision pathways, 3) different combinations of the EMD in the ON and OFF channels. 
The nomenclature is given in Table \ref{Tab: Tab1}.

\begin{table}[t!]
	\begin{center}
		\begin{tabular}{l|l||l|l}
			\toprule
			\multicolumn{4}{c}{acronym and full-name}\\
			\cmidrule{1-4}
			EMD&elementary motion detector&HRC&Hassenstein-Reichardt Correlation\\
			DS&direction-selective&DO&direction-opponent\\
			HS&horizontal-sensitive&VS&vertical-sensitive\\
			PD&preferred direction&ND&null (or non-preferred) direction\\
			OF&optic flow&LMC&lamina monopolar cell\\
			vDoG&variant of Difference of Gaussians&FDSR&fast-depolarising-slow-repolarising\\
			LPTC&lobula plate tangential cell&LPi&lobula plate-intrinsic\\
			LGMD&lobula giant movement detector&STMD&small target movement detector\\
			\bottomrule
		\end{tabular}
	\end{center}
	\caption{Nomenclature in this paper}
	\label{Tab: Tab1}
\end{table}

\subsection{Motion Sensitive Neural Models}

Flying insects have tiny brains, but compact visual systems for decoding diverse motion features varying in directions and sizes. 
Some identified neurons and corresponding circuits have been investigated as robust motion sensitive neural models, as reviewed in \cite{Fu-ALife-review}.

In the locust's visual brains, two lobula giant movement detectors (LGMDs), i.e., the LGMD-1 and the LGMD-2 have been modelled as quick and robust looming detectors specialising in collision perception \cite{Fu-2018(LGMD1-NN),Fu-LGMD2-TCYB,Fu2017a(LGMDs-IROS),Fu-2016(LGMD2-BMVC)}. 
The LGMD models respond selectively to movements in depth, with the most powerful response to objects that signal frontal collision threats. 
A good number of models have been applied for collision detection against various scenarios including ground vehicles, mobile robots and unmanned aerial vehicles \cite{Fu-ALife-review,Fu-TAROS(review)}.

Inspired by the flies and bees, a considerable number of OF-based collision sensing visual systems mimics the functions of bilateral compound eyes, at ommatidium level. 
More specifically, there are several categories of methods to realise such signal processing. 
The HRC theory originates the elementary motion detector (EMD)-based models correlating two signals in space, by multiplication with one delayed \cite{EMD-1989(principles-review)}; such a method is effective to enhance the PD motion. 
Another famous mechanism is called the ``Barlow-Levick" model by non-linearly suppressing the ND motion \cite{Barlow-1965(rabbit-DSNs)}, which is recently collaborated with the HRC mechanism in constructing fly motion detectors \cite{Fly-DS-2017(emergence-direction-selectivity),Fly2016(direction-selectivity-fly)}. 
In addition, Franceschini proposed a velocity-tuned method depending on the ratio between the photoreceptor angles in space and the time delay for each pairwise contrast detection photoreceptor \cite{Nicolas-1989(DSN-Insect-Neurons),Nicolas-insects-robot-vision}; subsequently, it has been called the ``time-of-travel" scheme \cite{Moeckel-Liu-time-to-travel-model,Vanhoutte-time-of-travel-OF-micro-flying-robot}. 
As a variation of the EMD, a few methods were proposed to decode or estimate the angular velocity accounting for various flight behaviours of bees \cite{OF-angular-velocity-PLoS-2009,Cope2016(model-angular-bee),huatian-IJCNN-2019,huatian-AIAI-2019}. 
Benefiting from the computational efficiency and robustness, many OF-based methods have been applied for near-range navigation of flying robots and micro aerial vehicles, as reviewed in \cite{Nicolas-2014(Review-Fly-Robot),Serres2017(review-optic-flow)}.

With distinct size selectivity, the small target movement detectors (STMDs) in flying insects, like the dragonflies, respond selectively to moving objects of very small size (subtended an angle of less than 10 degrees) \cite{Fu-ALife-review}. 
Wiederman proposed seminal works to detect small dark object motions embedded in natural scenes, via correlating ON and OFF channels in motion detection circuits \cite{Wiederman2008(STMD-clutter),Wiederman2013(ON-OFF-correlation)}. 
They also combined its functionality with the EMD structure to implement the direction selectivity \cite{wiederman2013biologically}. 
Recently, the STMD models have been successfully implemented in a ground robot to track small targets in natural backgrounds \cite{bagheri2017autonomous}, and in on-line system of an airborne vehicle for small-field object detection and avoidance \cite{Alvarez-Small-Object-Avoidance-2019}.

\subsection{Physiological Research on the Fly Motion Vision Pathways}

\begin{figure}[t!]
	\begin{center}
		\includegraphics[width=0.45\textwidth]{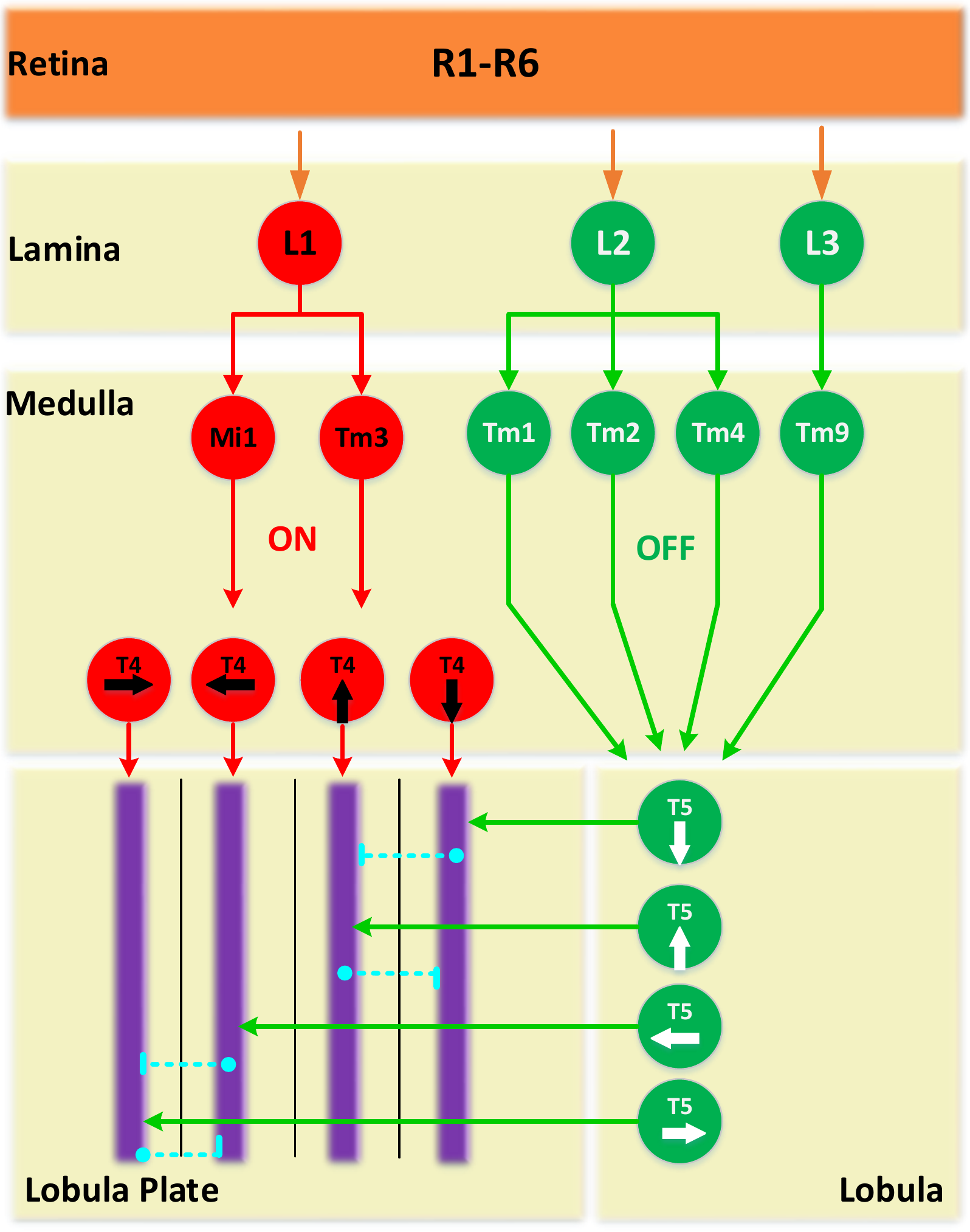}
	\end{center}
	\caption{
		Schematic diagram of the \textit{Drosophila} ON and OFF motion vision pathways with five neuropile layers: 
		the first retina layer R1--R6 neurons convey motion information to lamina monopolar cells (LMCs, i.e., L1, L2, L3); 
		the signals are then split into parallel ON and OFF channels denoted by different coloured neurons and pathways; 
		the directionally selective signals are carried via T4 and T5 cells to four sub-layers of the lobula plate, where T4 and T5 cells with the same PD signals converge on the same dendrites of the tangential cells; 
		the inhibition is conveyed via lobula plate-intrinsic (LPi) interneurons (dashed lines) between stratified neighbouring layers in the lobula plate.
	}
	\label{Fig: neuro-layers}
\end{figure}

Our proposed model is based on an important physiological theory that motion information is processed in parallel ON and OFF pathways \cite{Borst-Review-2019-Fly}. 
As illustrated in Fig. \ref{Fig: neuro-layers}, we can summarise the following steps of the \textit{Drosophila}'s preliminary visual processing: 
\begin{enumerate}
	\item The motion perception starts from the retina layer with photoreceptors (R1--R6) which conveys received brightness to LMCs in the lamina layer. 
	\item The LMCs encode motion by luminance increments (ON) and decrements (OFF). 
	The motion information is separated into parallel channels: the L1 with its downstream Mi1 and Tm3 interneurons in the medulla layer convey onset or light-on response to succeeding T4 neurons in the medulla layer; whilst the L2, L3 with their downstream Tm1, Tm2, Tm4 and Tm9 interneurons relay offset or light-off response to subsequent T5 neurons in the lobula layer \cite{Rister-2007(dissection-motion-channel),Fly2016(direction-selectivity-fly),Strother2014(direct-observation-on-off),Fisher-L3(wide-field-local)}. 
	\item The DS responses of ON and OFF contrasts are produced by the T4 and T5 cells in a feed-forward manner, respectively. 
	The selectivity to four cardinal directions is well separated in different groups of T4 and T5 neurons \cite{Maisak_2013(T4-T5-fly)}.
	\item The LPTCs in four stratified sub-layers of the lobula plate integrate the T4 and T5 signals, where the same DS responses converge on the same sub-layer. 
	Meanwhile, the LPi interneurons convey inhibition to adjacent sub-layers through sign-inverting interactions, thus forming the DO responses \cite{Mauss-Opposing-Motions-Fly}. 
	Finally, two directionally selective systems, i.e., the HS and VS systems pool the responses from LPTCs towards sensorimotor control \cite{Joesch-2010(ON-OFF-fly)}.
\end{enumerate}

\subsection{EMD in the ON and OFF Channels}

\begin{figure}[t!]
	\begin{center}
		\subfloat[EMDs]{\includegraphics[width=0.18\textwidth]{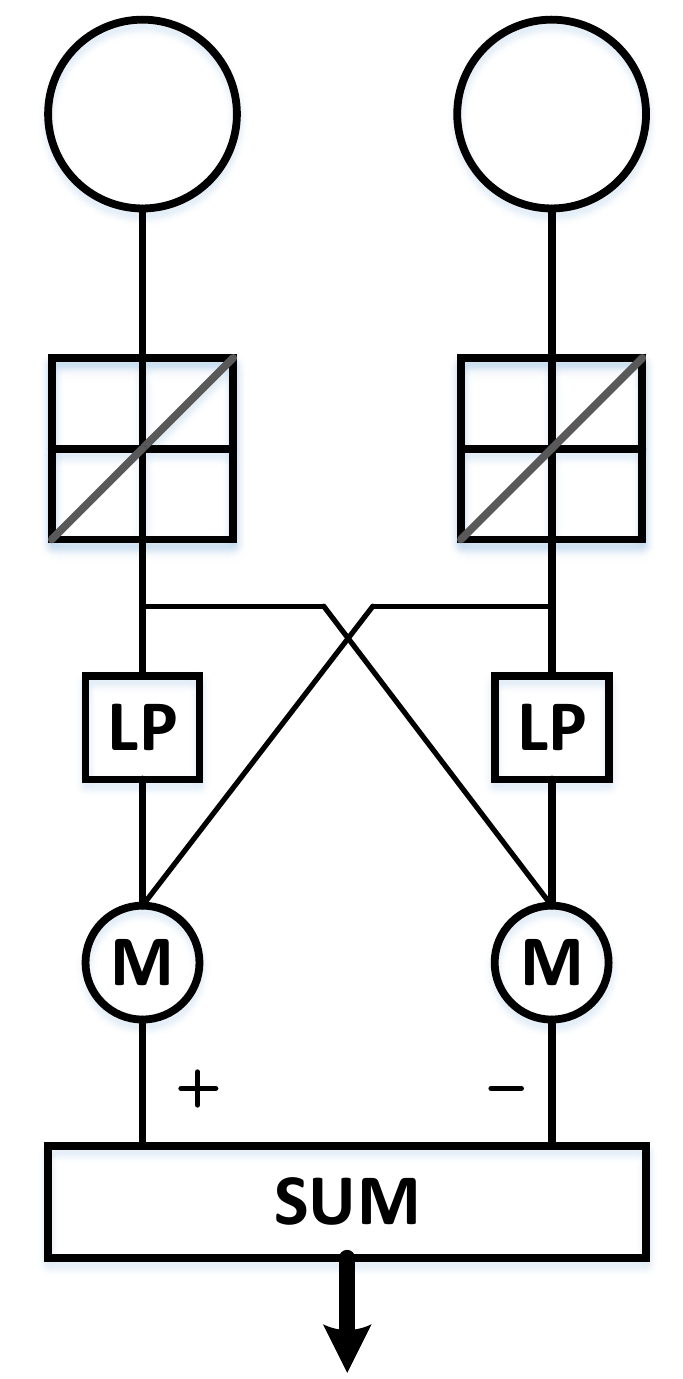}
			\label{Fig: EMDs}}
		\hfil
		\subfloat[4-Q EMDs]{\includegraphics[width=0.42\textwidth]{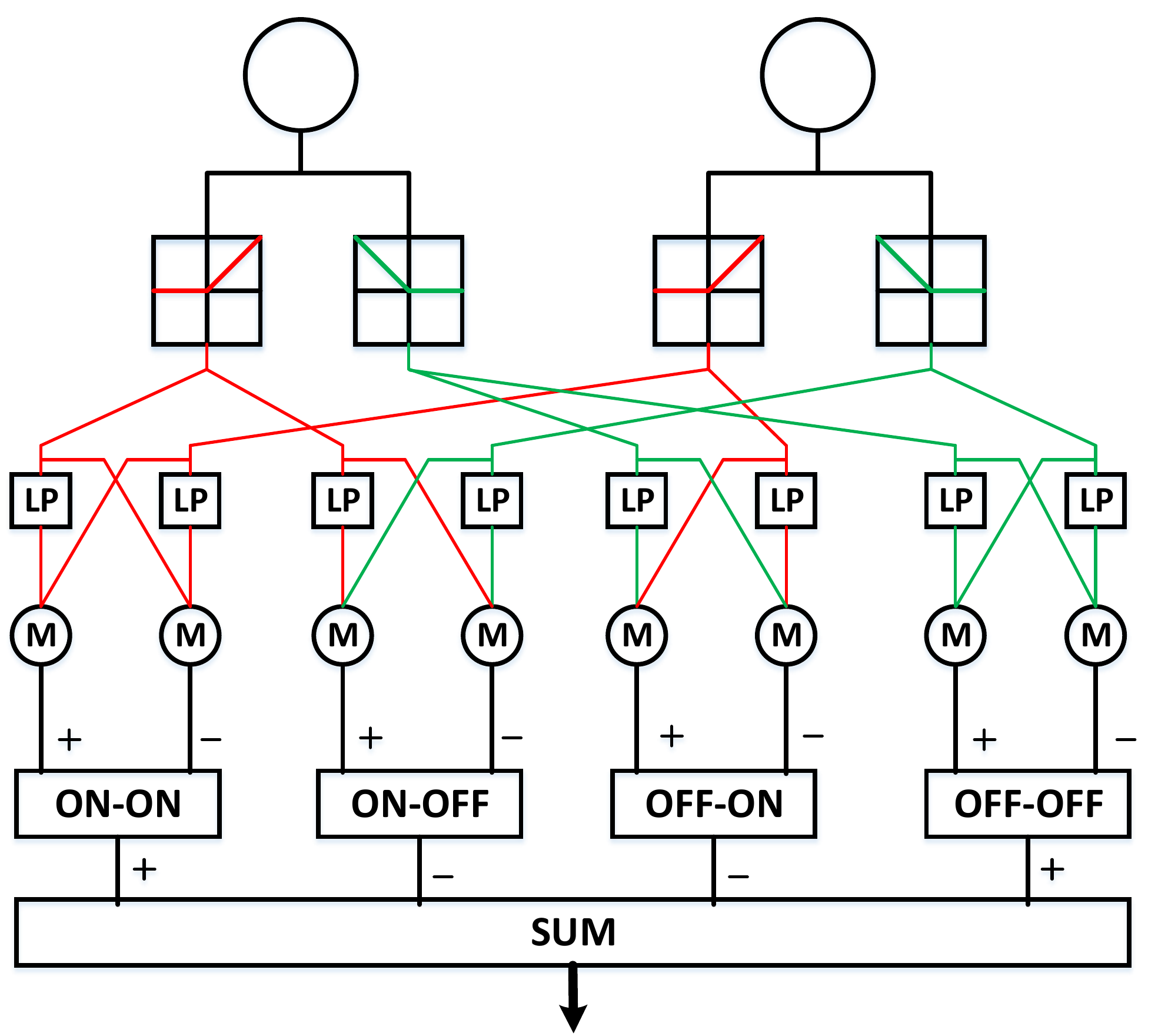}
			\label{Fig: 4Q-EMDs}}
		\hfil
		\subfloat[2-Q EMDs]{\includegraphics[width=0.33\textwidth]{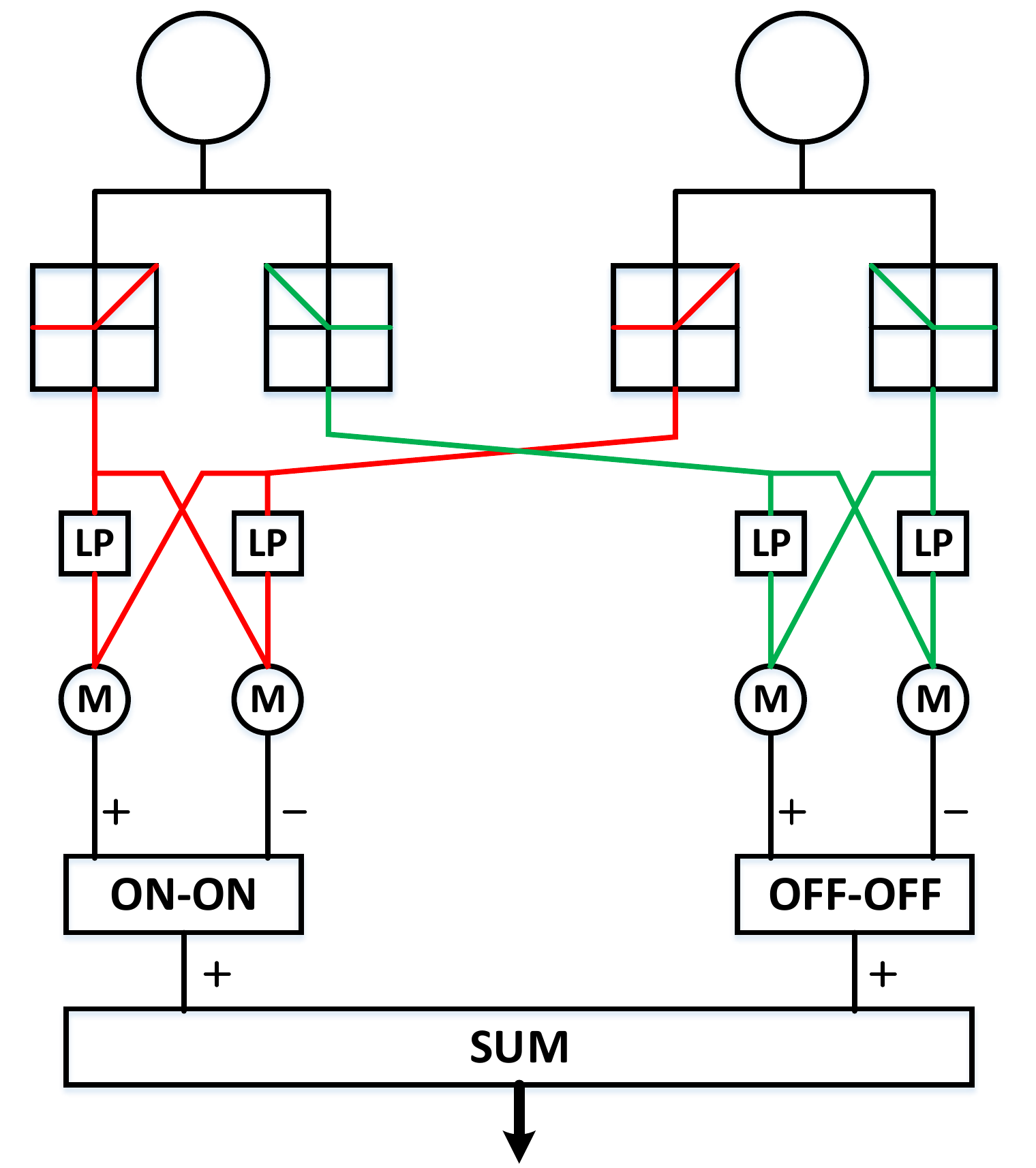}
			\label{Fig: 2Q-EMDs}}
		\vfil
		\subfloat[6-Q EMDs]{\includegraphics[width=0.7\textwidth]{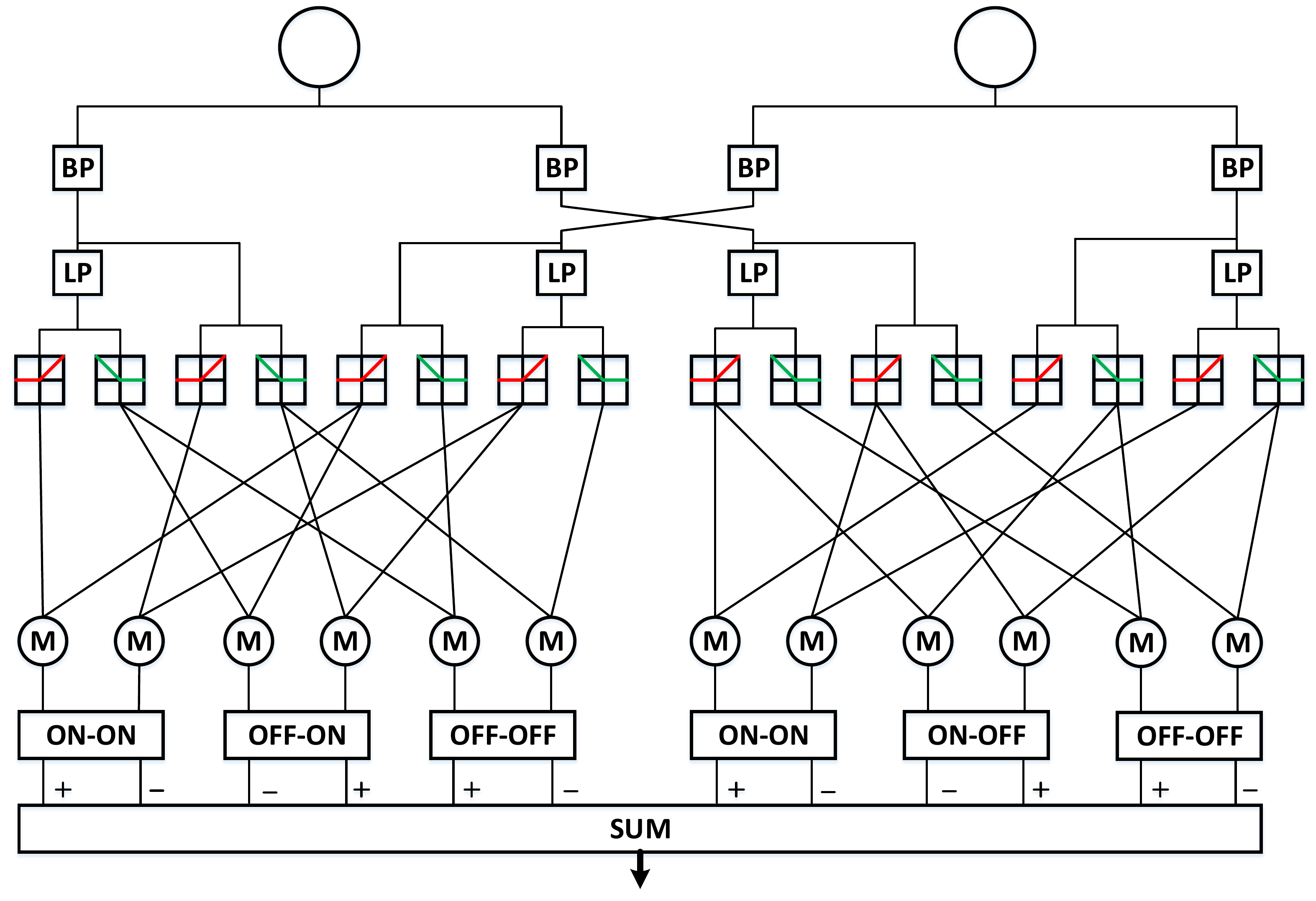}
			\label{Fig: 6Q-EMDs}}
	\end{center}
	\caption{
		Different combinations of the EMD in the ON and OFF channels: 
		LP, BP and M components indicate the low-pass filtering, the band-pass filtering and the multiplication processes.
	}
	\label{Fig: ON-OFF-models}
\end{figure}

Based on the EMD, there are several theories representing different combination forms to encode the spatiotemporal signal flows in the ON and OFF channels \cite{Joesch_2013(functional-ON-OFF)}, as shown in Fig. \ref{Fig: ON-OFF-models}. 
Importantly, since the signals are already directionally selective before collectively arriving at the stratified LPTCs \cite{Maisak_2013(T4-T5-fly),Badwan-Dynamic-nonlinearities-EMD}, all these models could well explain the neural computation inside the ON and OFF channels.

More concretely, in the former EMD models, e.g., \cite{Iida_2000(fly-visual-odometer),Zanker_2005(motion-signal-outdoor),Zanker-1999(EMD-speed-tuning)}, visual information is processed in a single pathway with the basic format between every pairwise photoreceptors (Fig. \ref{Fig: EMDs}). 
After the identification of ON and OFF channels, the motion information is split into different places, however can further interact even with the opposite polarity signal flows. 
Different combinations have been investigated through either the electro-physiological recordings from the LPTCs \cite{Eichner2011(2Q-motion)} or the behavioural experiments \cite{Clark_2011(6Q-model-fly)}. 
The 4-quadrant (4-Q) detectors model with communications between both the same and the opposite polarity signals (see Fig. \ref{Fig: 4Q-EMDs}) is fully equivalent to the full-HRC, namely the EMD model in Fig. \ref{Fig: EMDs}. 
The second important model is the 2-Q structure in Fig. \ref{Fig: 2Q-EMDs}, which processes input combinations of only the same-sign signals, i.e., ON-ON and OFF-OFF contrast. 
In addition to that, the 6-Q model has a more complex structure (see Fig. \ref{Fig: 6Q-EMDs}), which argues that either the ON/OFF channel conveys motion information with both positive (onset) and negative (offset) contrast changes. 
Our proposed method leverages them with the 2-Q model's simpler computational structure as well as the 6-Q model's edge selectivity prior to the ON and OFF channels.

\section{Formulation of the Proposed Model}
\label{Sec: formulation}

\begin{figure}[t!]
	\begin{center}
		\includegraphics[width=\textwidth]{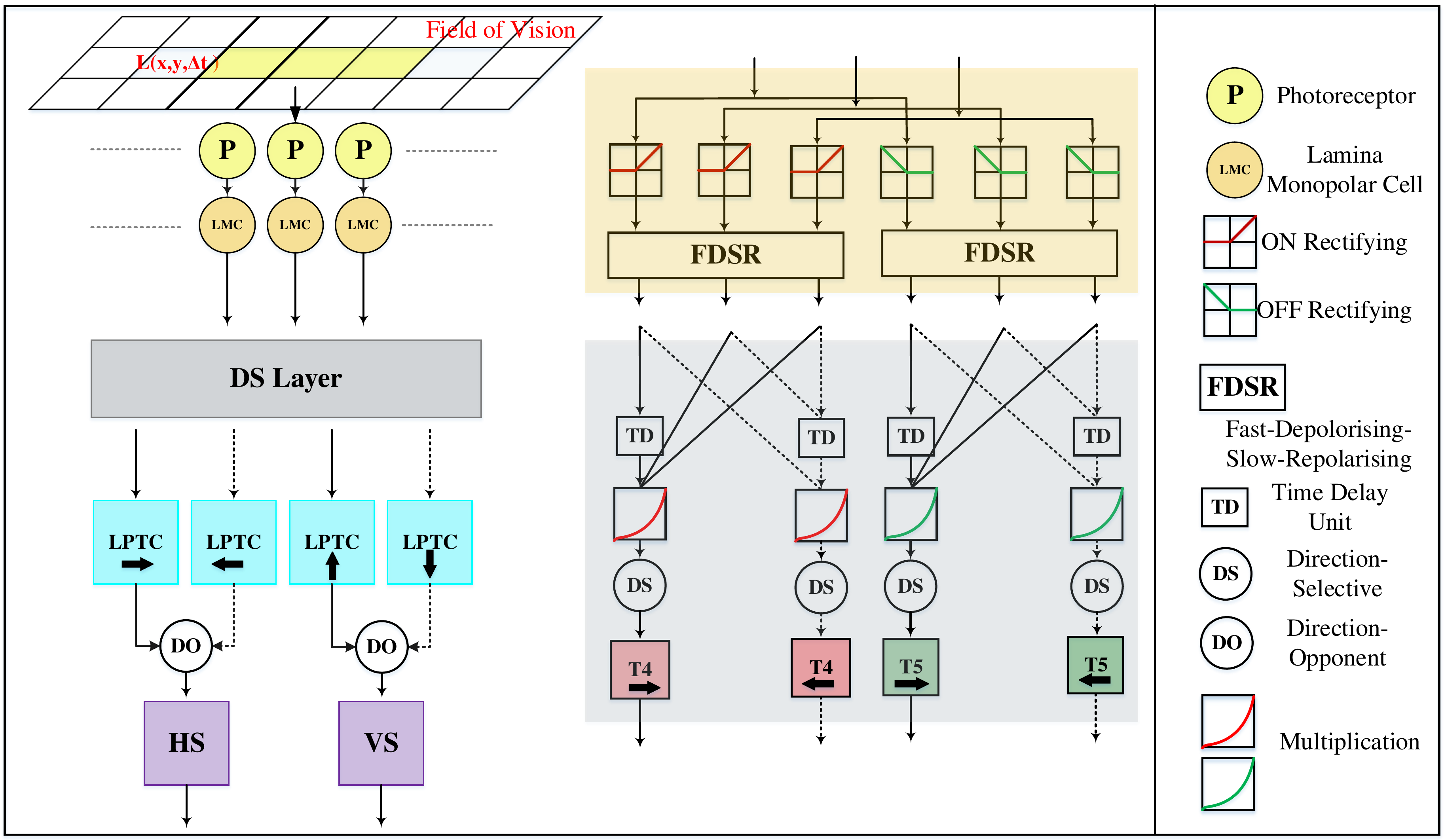}
	\end{center}
	\caption{
		Schematic illustration of the proposed model consisting of the ON and OFF motion pathways throughout several computational layers mimicking the \textit{Drosophila} physiology in Fig. \ref{Fig: neuro-layers}. 
		The DS layer exemplifies the processing of each interneuron interacting with two horizontal neighbour cells in the medulla and lobula layers, where the dashed lines with respect to the solid ones indicate the generation and transmission of opposing DS motion.
	}
	\label{Fig: model}
\end{figure}

Within this section, we present the formulation of the proposed visual system model. 
Fig. \ref{Fig: model} depicts the schematic of model structure. 
Generally speaking, for mimicking the \textit{Drosophila} physiology in Fig. \ref{Fig: neuro-layers}, the model consists of mainly five computational neuropile layers with the HS and VS systems. 
The forming of DS and DO responses in the proposed model resembles the revealed \textit{Drosophila} visual processing in a feed-forward manner \cite{Badwan-Dynamic-nonlinearities-EMD}. 
Compared to previous related methods, we highlight the following mechanisms: 
\begin{enumerate}
	\item The model combines bio-plausible spatiotemporal pre-filtering methods to remove redundant background motion to a large extent, and achieve edge selectivity, which include firstly a variant of `Difference of Gaussians' (vDoG) mechanism with ON and OFF contrast selectivity, spatially, and then a fast-depolarising-slow-repolarising (FDSR) mechanism, temporally (see Fig. \ref{Fig: key-structures}).
	\item To improve the dynamic response and alleviate the impact by temporal frequency of visual stimuli, we propose a novel structure representing ensembles of motion correlators for each interneuron inside the ON and OFF pathways to produce the DS responses in horizontal and vertical directions, i.e., multi-connected and same-polarity cells possess dynamic latency corresponding to the sampling distance between each pairwise detectors (see Fig. \ref{Fig: key-structures}).
	\item The HS and VS systems integrate the local DS responses from stratified LPTCs with inhibitions from adjacent LPi interneurons representing the DO responses to form global membrane potential. 
	Accordingly, the PD or ND translating motion is indicated by the positive or negative membrane potential of both systems.
\end{enumerate}

\subsection{Computational Retina Layer}

In the first retina layer, there are photoreceptors (see P in Fig. \ref{Fig: model}) that capture single-channel luminance (green-channel or grey-scale in our case), at ommatidia (grouped local optical units) or local pixel level from images, with respect to time. 
Let $L(x,y,t) \in \R^3$ denote the input image streams, where $x$, $y$ and $t$ are spatial and temporal positions. 
The calculation of motion signals is as follows: 
\begin{equation}
P(x,y,t) = L(x,y,t) - L(x,y,t-1) + \sum_{i=1}^{n_p}a_i \cdot P(x,y,t-i).
\label{Eq: retina-photoreceptors}
\end{equation}
The change of brightness could continue and decay for a short while of $n_p$ number of frames. 
The decay coefficient $a_i$ is computed by 
\begin{equation}
a_i = \left(1 + e^{i}\right)^{-1}.
\label{Eq: retina-decay}
\end{equation}

\subsection{Computational Lamina Layer}

As illustrated in Fig. \ref{Fig: neuro-layers} and \ref{Fig: model}, in the second lamina layer, there are LMCs that split motion signals from the retina layer into parallel ON and OFF channels encoding light-on (onset) and light-off (offset) responses, respectively. 
For enhancing the motion edge selectivity and maximising the transmission of useful information from visually cluttered environments, we propose a bio-plausible spatial mechanism, named the ``vDoG", simulating the functions of LMCs. 
Compared to the traditional DoG mechanism, it also demonstrates the ON and OFF contrast selectivity to fit with the following processing in the ON and OFF channels. 
The vDoG depicts a centre-surrounding antagonism with centre-positive and surrounding-negative Gaussians representing excitatory and inhibitory fields in space. 
That is, 
\begin{equation}
\begin{aligned}
&P_e(x,y,t) = \sum_{u=-2}^{2}\sum_{v=-2}^{2} P(x-u,y-v,t) \cdot G_{\sigma_e}(u,v),\\
&P_i(x,y,t) = \sum_{u=-4}^{4}\sum_{v=-4}^{4} P(x-u,y-v,t) \cdot G_{\sigma_i}(u,v),
\end{aligned}
\label{Eq: lamina-dog}
\end{equation}
\begin{equation}
G_{\sigma_e}(u,v) = \frac{1}{2\pi \sigma_e^2} \text{exp}\left( -\frac{u^2 + v^2}{2 \sigma_e^2} \right),\ 
G_{\sigma_i}(u,v) = \frac{1}{2\pi \sigma_i^2} \text{exp}\left( -\frac{u^2 + v^2}{2 \sigma_i^2} \right),
\label{Eq: lamina-dog-gaussian}
\end{equation}
where $\sigma_e$ and $\sigma_i$ indicate the excitatory and inhibitory standard deviations. 
The outer convolution kernel $G$ is with twice the radius of the inner one. 
Accordingly, the broader inhibitory Gaussian is subtracted from the narrower excitatory one with the polarity selectivity. 
That is, 
\begin{equation}
LA(x,y,t) = \left\{
\begin{aligned}
|P_{e}(x,y,t) - P_{i}(x,y,t)|,\ &\text{if}\ P_{e}(x,y,t) \geq 0\ \&\ P_{i}(x,y,t) \geq 0\\
-|P_{e}(x,y,t) - P_{i}(x,y,t)|,\ &\text{if}\ P_{e}(x,y,t) < 0\ \&\ P_{i}(x,y,t) < 0
\end{aligned}
\right..
\label{Eq: lamina-dog-subtraction}
\end{equation}

After that, there are ON and OFF half-wave rectifying mechanisms splitting motion information into two parallel pathways, via filtering out negative and positive inputs for the ON and OFF pathways, respectively (see Fig. \ref{Fig: model}). 
In addition to that, the negative inputs to the OFF pathway are sign-inverted. 
The calculations are expressed as follows: 
\begin{equation}
L1(x,y,t) = [LA(x,y,t)]^{+},\ L2(x,y,t) = -[LA(x,y,t)]^{-}.
\label{Eq: lamina-halfwave-rectifying}
\end{equation}
$[x]^{+}$ and $[x]^{-}$ denote $\text{max}(0,x)$ and $\text{min}(x,0)$, respectively. 
$L1$ and $L2$ indicate the LMCs in the lamina layer, i.e., L1 in the ON channels, L2 and L3 in the OFF channels (see Fig. \ref{Fig: neuro-layers}).

\begin{figure}[t!]
	\begin{center}
		\includegraphics[width=\textwidth]{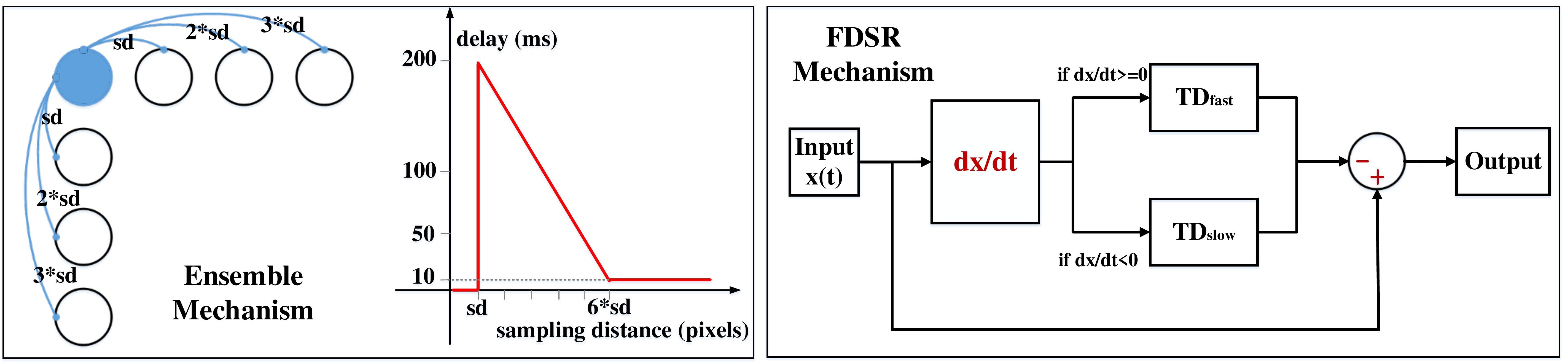}
	\end{center}
	\caption{
		Illustrations of the proposed mechanisms of spatiotemporal dynamics inside the ON and OFF pathways. 
		The left sub-figure exemplifies the ensembles of local motion detectors in two directions, where the latency is dynamic depending on the sampling distance (sd) between each pair-wise correlators. 
		The right sub-figure depicts the temporal FDSR mechanism.
	}
	\label{Fig: key-structures}
\end{figure}

For each interneuron in the lamina layer, an `adaptation state' is formed by the bio-plausible FDSR mechanism which matches the neural characteristic of `fast onset and slow decay' phenomenons. 
As depicted in Fig. \ref{Fig: key-structures}, we first check the derivative of inputs from the ON and OFF channels along the time $t$. 
As digital signals do not have continuous derivatives, we compare neuronal responses between every two successive frames to get the change:
\begin{equation}
\Delta L1(x,y) = L1(x,y,t) - L1(x,y,t-1),\ \Delta L2(x,y) = L2(x,y,t) - L2(x,y,t-1).
\label{Eq: lamina-fdsr-gradient}
\end{equation}
Subsequently, the input signals from the ON and OFF channels are delayed with two different latency constants $\tau_1, \tau_2$ in milliseconds, and $\tau_1<\tau_2$ representing the `fast depolarising' with non-negative change and the `slow repolarising' with negative change, respectively. 
That is, 
\begin{equation}
\begin{aligned}
&\hat{L1}(x,y,t) = \left\{
\begin{aligned}
&\alpha_1 L1(x,y,t)+(1-\alpha_1) L1(x,y,t-1),\ \text{if}\ \Delta L1(x,y) \ge 0,\\
&\alpha_2 L1(x,y,t)+(1-\alpha_2) L1(x,y,t-1),\ \text{if}\ \Delta L1(x,y) < 0.
\end{aligned}
\right.\\
&\hat{L2}(x,y,t) = \left\{
\begin{aligned}
&\alpha_1 L2(x,y,t)+(1-\alpha_1) L2(x,y,t-1),\ \text{if}\ \Delta L2(x,y) \ge 0,\\
&\alpha_2 L2(x,y,t)+(1-\alpha_2) L2(x,y,t-1),\ \text{if}\ \Delta L2(x,y) < 0.
\end{aligned}
\right.
\end{aligned}
\label{Eq: lamina-fdsr-delay}
\end{equation}
\begin{equation}
\alpha_1 = \tau_i / (\tau_1 + \tau_i),\ \alpha_2 = \tau_i / (\tau_2 + \tau_i).
\label{Eq: lamina-fdsr-delay-coe}
\end{equation}
$\tau_i$ is the discrete time interval in milliseconds, between frames. 
Notably, in the FDSR mechanism, the delayed signal is subtracted from the original one (see Fig. \ref{Fig: key-structures}) as 
\begin{equation}
M1(x,y,t) = L1(x,y,t) - \hat{L1}(x,y,t),\ M2(x,y,t) = L2(x,y,t) - \hat{L2}(x,y,t).
\label{Eq: lamina-fdsr-subtraction}
\end{equation}
$M1$ and $M2$ denote interneurons in the medulla layer including Mi1, Tm3 in the ON channels, and Tm1, Tm2, Tm4, Tm9 in the OFF channels (see Fig. \ref{Fig: neuro-layers}). 
Such a temporal mechanism contributes significantly to filter out irrelevant background OF and visual flickers, like the windblown vegetation in natural environments.

\subsection{Computational Medulla and Lobula Layers}

Next, both the medulla and lobula layers constitute the DS layer in Fig. \ref{Fig: model}, in order to generate the specific DS responses to four cardinal directions, where those interneurons interact with each other, non-linearly \cite{Maisak_2013(T4-T5-fly)}. 
Importantly, like the genuine T4 and T5 neurons in several individual groups sensitive to different directions (see Fig. \ref{Fig: neuro-layers}), the proposed model demonstrates the same directional tuning. 
As mentioned above, for each local cell in the DS layer, we propose the ensemble mechanism (see Fig. \ref{Fig: key-structures}) connecting same polarity motion detectors, in space, with ON-ON contrast correlation in the ON pathway and OFF-OFF contrast correlation in the OFF pathway, separately. 
Each pair-wise connection is featured by the aforementioned 2-Q correlation structure (see Fig. \ref{Fig: 2Q-EMDs}). 
In addition, as illustrated in Fig. \ref{Fig: key-structures}, the delay is dynamic, varying with respect to the sampling distance between every pair-wise detectors. 
More precisely, the combination with smaller distance has larger latency, which decreases as the space increases. 
In our preliminary modelling and bio-robotic researches \cite{Fu2017(fly-DSNs-IJCNN),Fu-ROBIO-2018}, such a multi-connected structure has demonstrated the improved dynamic response in speed tuning of translational OF perception, when challenged by a range of angular velocities. 
The computations for producing the DS responses of the T4 neurons in the medulla layer are defined as follows: 
\begin{equation}
\begin{aligned}
&T4_{r}(x,y,t) = \sum_{i=sd}^{sd \cdot n_c}\hat{M1}(x,y,t) \cdot M1(x+i,y,t),\\
&T4_{l}(x,y,t) = \sum_{i=sd}^{sd \cdot n_c}\hat{M1}(x+i,y,t) \cdot M1(x,y,t),\\
&T4_{d}(x,y,t) = \sum_{i=sd}^{sd \cdot n_c}\hat{M1}(x,y,t) \cdot M1(x,y+i,t),\\
&T4_{u}(x,y,t) = \sum_{i=sd}^{sd \cdot n_c}\hat{M1}(x,y+i,t) \cdot M1(x,y,t).
\end{aligned}
\label{Eq: medulla-ds}
\end{equation}
$\{r, l, d, u\}$ indicate the DS responses on four cardinal directions: rightward, leftward, downward and upward. 
$n_c$ and $sd$ stand for the number of correlated neighbouring cells for each original cell, and the sampling distance between each pair-wise combination, respectively. 
$\hat{M1}$ denotes the delayed signal, calculated by 
\begin{equation}
\begin{aligned}
\hat{M1}(x,y,t) = &\alpha_3 M1(x,y,t) + (1-\alpha_3) M1(x,y,t-1),\\
&\alpha_3 = \tau_i / (\tau_i + \tau_s).
\end{aligned}
\label{Eq: medulla-lobula-delay}
\end{equation}
$\tau_s$ indicates the proposed dynamic time delay, as exemplified in Fig. \ref{Fig: key-structures}. 
Similarly for the T5 neurons in the lobula layer, the forming of DS responses along four cardinal directions is expressed as follows: 
\begin{equation}
\begin{aligned}
&T5_{r}(x,y,t) = \sum_{i=sd}^{sd \cdot n_c}\hat{M2}(x,y,t) \cdot M2(x+i,y,t),\\
&T5_{l}(x,y,t) = \sum_{i=sd}^{sd \cdot n_c}\hat{M2}(x+i,y,t) \cdot M2(x,y,t),\\
&T5_{d}(x,y,t) = \sum_{i=sd}^{sd \cdot n_c}\hat{M2}(x,y,t) \cdot M2(x,y+i,t),\\
&T5_{u}(x,y,t) = \sum_{i=sd}^{sd \cdot n_c}\hat{M2}(x,y+i,t) \cdot M2(x,y,t).
\end{aligned}
\label{Eq: lobula-ds}
\end{equation}
The delay computation of $\hat{M2}$ conforms to Eq. \ref{Eq: medulla-lobula-delay}, which is not restated here. 

Importantly, a latest biological research has revealed that the distinct DS responses are all generated in a feed-forward manner when arriving the T4 and T5 neurons, each group of which demonstrates the specific direction selectivity \cite{Badwan-Dynamic-nonlinearities-EMD}. 
As introduced in Section \ref{Sec: literature}, though the different mechanisms forming such DS responses, with either PD motion enhancement or ND motion suppression, are still in debate, the proposed visual system model reconciles well the generation of DS responses with feed-forward signal processing.

\subsection{Computational Lobula Plate Layer}

After that, as illustrated in Fig. \ref{Fig: neuro-layers}, the LPTCs in four stratified sub-layers integrate the DS responses from different groups of T4 and T5 neurons each with specific PD motion tuning, where the same DS responses converge at an identical sub-layer of the lobula plate. 
That is, 
\begin{equation}
\begin{aligned}
&LP_r(t) = \sum_{x=1}^{C}\sum_{y=1}^{R}T4_{r}(x,y,t) + T5_{r}(x,y,t),\\ 
&LP_l(t) = \sum_{x=1}^{C}\sum_{y=1}^{R}T4_{l}(x,y,t) + T5_{l}(x,y,t),\\
&LP_d(t) = \sum_{x=1}^{C}\sum_{y=1}^{R}T4_{d}(x,y,t) + T5_{d}(x,y,t),\\ 
&LP_u(t) = \sum_{x=1}^{C}\sum_{y=1}^{R}T4_{u}(x,y,t) + T5_{u}(x,y,t).
\end{aligned}
\label{Eq: lptcs}
\end{equation}
$C$ and $R$ indicate columns and rows of the two-dimensional visual field. 
In addition to that, the proposed DO responses by opposing motions are generated via a sign-inverting operation representing the functionality of LPi interneurons inhibiting the LPTCs in neighbouring sub-layer (see Fig. \ref{Fig: neuro-layers} and \ref{Fig: model}), which are pooled by the HS and VS systems as the following: 
\begin{equation}
HS(t) = LP_r(t) - LP_l(t),\ VS(t) = LP_d(t) - LP_u(t).
\label{Eq: hs-vs}
\end{equation}
With regard to the non-linear and symmetric mapping in each combination of local ON-ON or OFF-OFF motion correlators inside the ON and OFF pathways, the model response is tuned to be positive by the PD (rightward and downward) motions, while negative by the ND (leftward and upward) motions.

Finally, the global membrane potential of either the HS/VS system is activated by a sigmoid function. 
Let the $HS(t)$ or $VS(t)$ be $x$, the function is defined as 
\begin{equation}
f(x) = 2 \cdot \operatorname{sgn}(x) \cdot ((1 + e^{-|x| \cdot (C \cdot R \cdot k)^{-1}})^{-1} - 0.5),
\label{Eq: smp}
\end{equation}
where $k$ is a scale coefficient. 
Accordingly, the output of proposed model is regulated within $[0,1)$ for the positive input, and $(-1,0]$ for the negative input.

\subsection{Setting Model Parameters}

\begin{table}[t!]
	\centering
	\begin{tabular}{l|l|l}
		\toprule
		\textbf{Parameter}&\textbf{Description}&\textbf{Value}\\
		\hline
		$n_p$&number of persistent frames&$0\sim 2$\\
		$\tau_i$&time interval constant between frames&$1000/30$ ms\\
		$\tau_1$&fast depolarising time constant&$1$ ms\\
		$\tau_2$&slow repolarising time constant&$100$ ms\\
		$sd$&sampling distance between each combination&$4$\\
		$\{\sigma_e,\sigma_i\}$&standard deviations in vDoG mechanism&$\{2,4\}$\\
		$n_c$&number of correlating cells&$4$\\
		$\tau_s$&dynamic delay in ensembles of correlators&$10\sim 200$ ms\\
		$\{C,R\}$&columns and rows of the visual field&adaptable\\
		$k$&coefficient in sigmoid function&$0.01$\\
		\bottomrule
	\end{tabular}
	\caption{Setting parameters of the proposed visual system model}
	\label{Tab: Tab2}
\end{table}

The parameters of the proposed model are given in Table \ref{Tab: Tab2}. 
All of them are decided, empirically, with considerations of the functionality of the \textit{Drosophila} visual system, and based on our previous modelling and experimenting experience in \cite{Fu2017(fly-DSNs-IJCNN),Fu-2017(ROBIO-fixation),Fu-ROBIO-2018}. 
In particular, the parameters of the vDoG mechanism correspond to the sampling distance between local motion correlators ($sd=\sigma_i$). 
It is also worthwhile pointing out that since each local cell inside the ON and OFF channels correlates with multiple neighbouring cells in horizontal and vertical directions, increasing the number of correlating cells ($n_c$) could further improve the dynamic response to translating stimuli, at the cost though of more computational consumption.

\section{Experimental Setting}
\label{Sec: setting}

In this section, we introduce the experimental settings. 
Generally speaking, all the experiments can be categorised into two types of tests. 
In the first type of tests, we aim at demonstrating the robust DS and DO responses, as the basic characteristic of the proposed neural system model, challenged by visual stimuli against various backgrounds, from simple to complex. 
More specifically, the visual inputs are with $320 \times 180$ pixels, at 30 frames per second (fps) for the clean and real-world scenarios. 
After that, more systematic experiments are carried on with two cluttered moving backgrounds. 
We simulate a bar with $25 \times 120$ pixels in size, at three certain grey levels (white, moderate, dark), translating rightward at three individual angular velocities of 9, 18, 27 degrees per second (degrees/s), in front of a cluttered moving background. 
Note that both the two backgrounds shift in an opposite direction (leftward) relative to the foreground translating targets, at a range of velocities (-5, -10, -20, -30, -40 degrees/s), respectively. 
The visual inputs are with $700 \times 180$ pixels, at 30 fps.

\begin{figure}[t!]
	\begin{center}
		\subfloat{\includegraphics[width=0.2\textwidth]{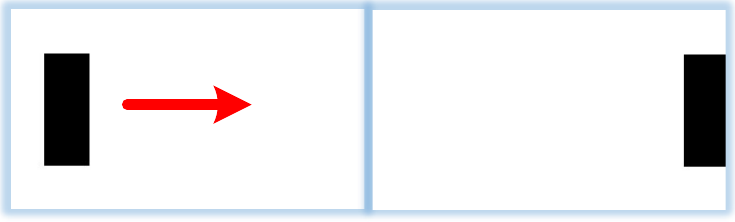}}
		\hfill
		\subfloat{\includegraphics[width=0.2\textwidth]{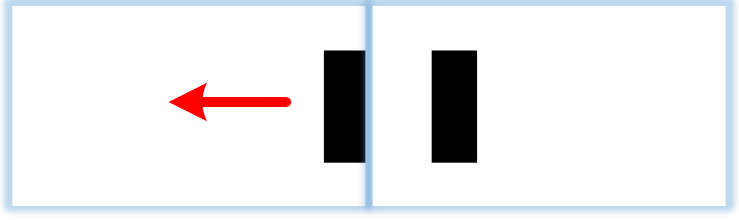}}
		\hfill
		\subfloat{\includegraphics[width=0.2\textwidth]{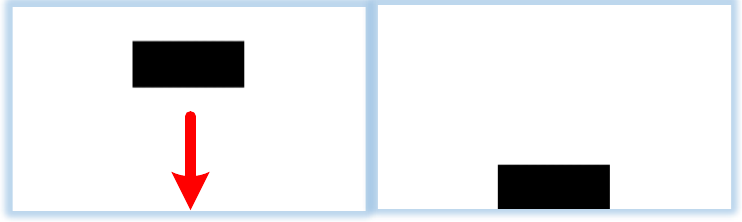}}
		\hfill
		\subfloat{\includegraphics[width=0.2\textwidth]{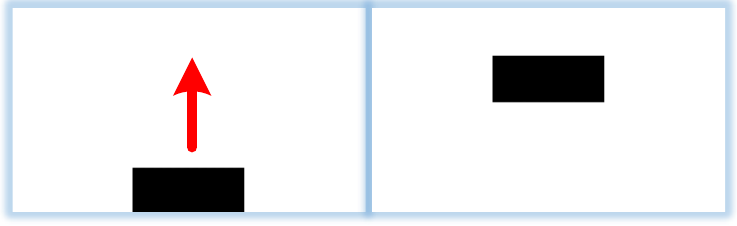}}
		\vfil
		\vspace{-10pt}
		\subfloat{\includegraphics[width=0.24\textwidth]{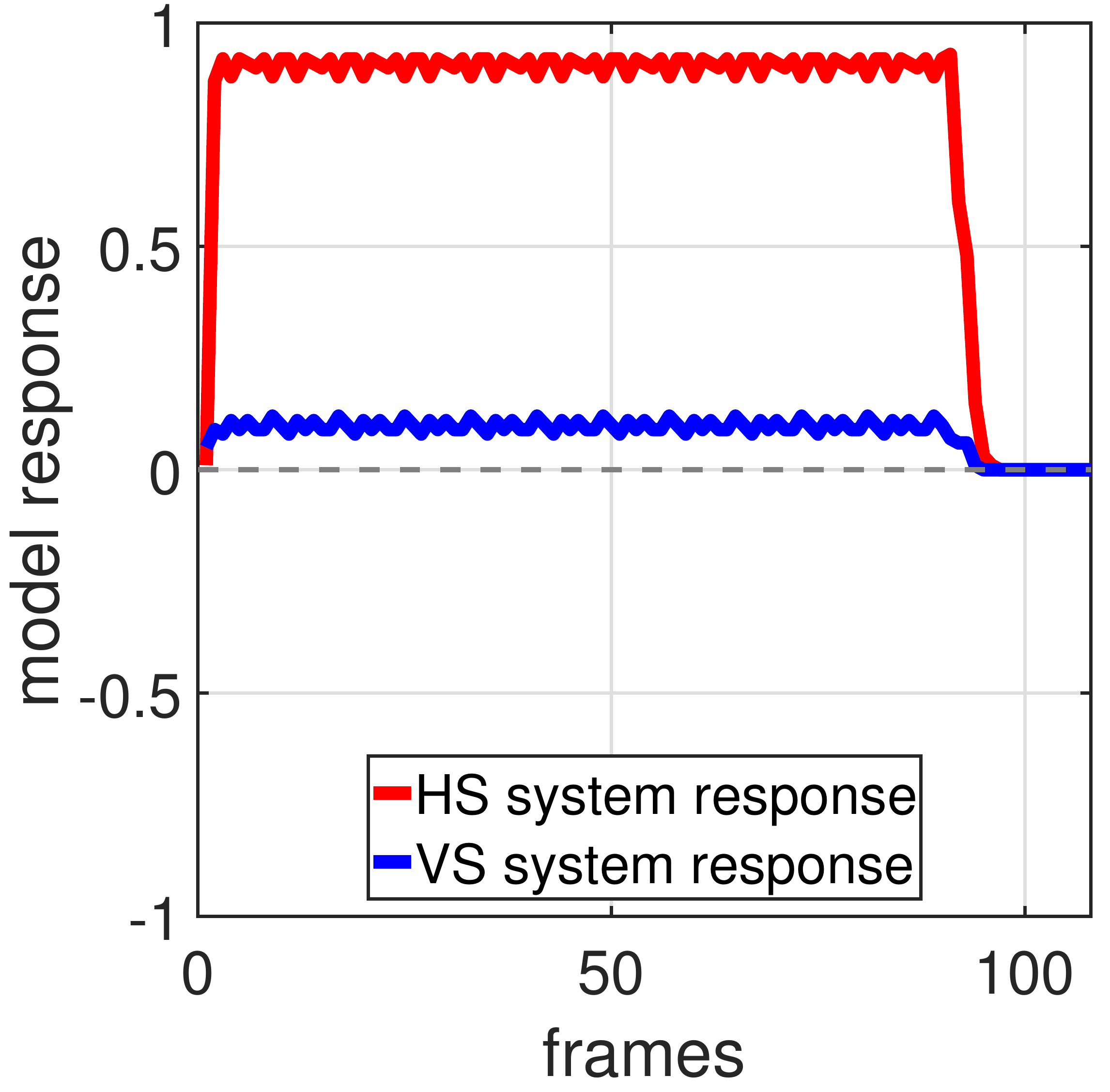}}
		\hfill
		\subfloat{\includegraphics[width=0.24\textwidth]{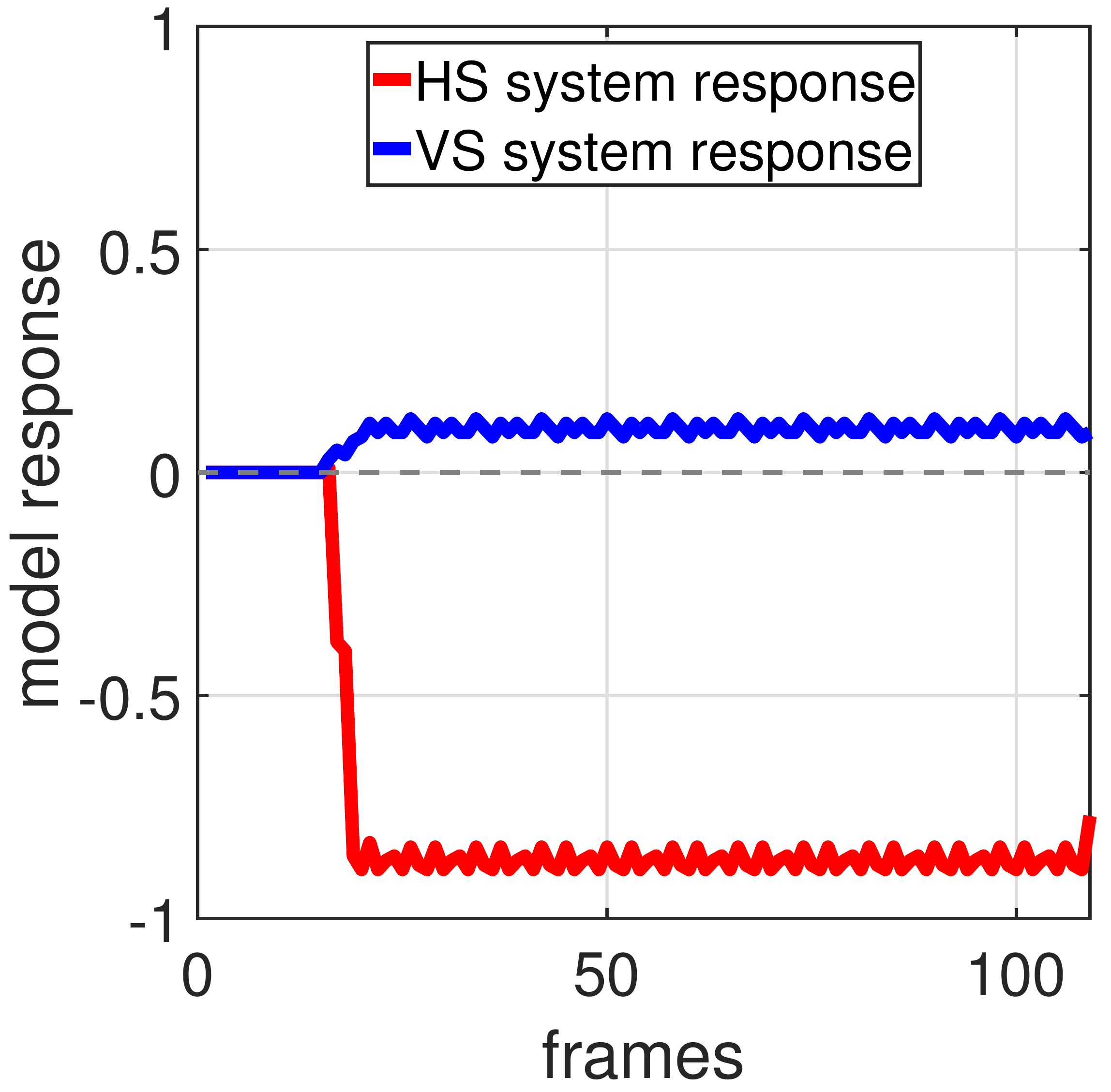}}
		\hfill
		\subfloat{\includegraphics[width=0.24\textwidth]{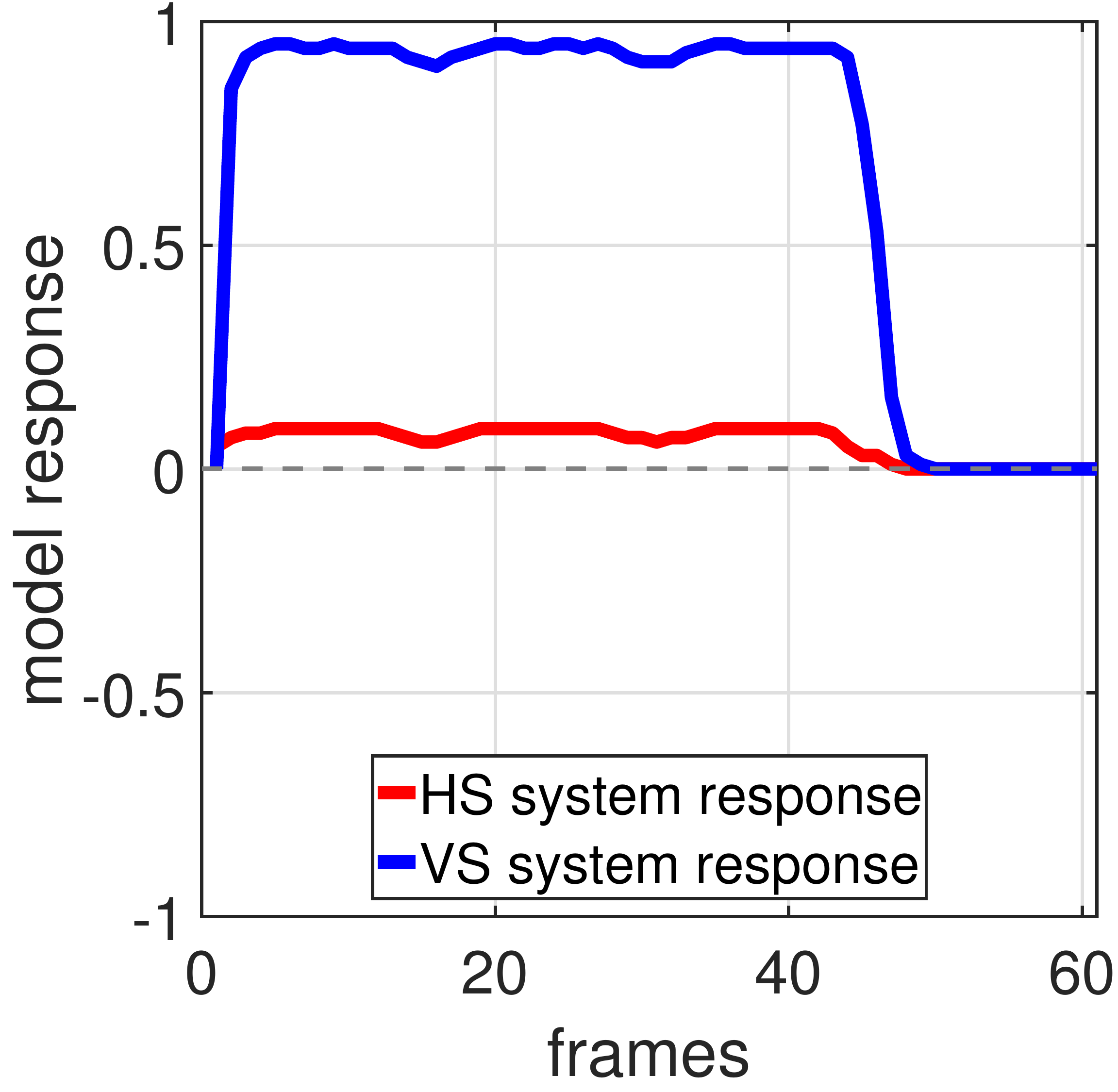}}
		\hfill
		\subfloat{\includegraphics[width=0.24\textwidth]{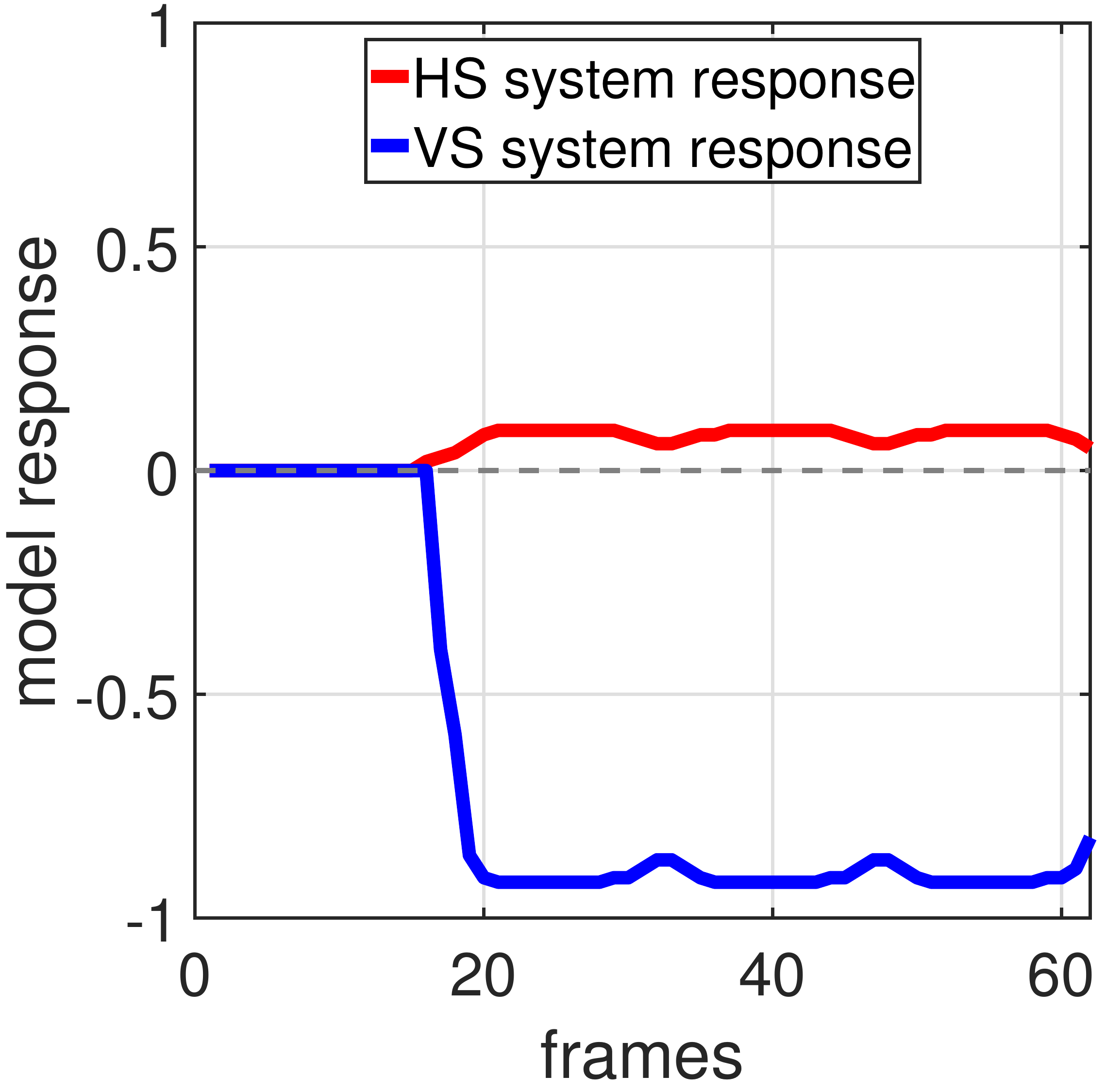}}
		\vfil
		\subfloat{\includegraphics[width=0.2\textwidth]{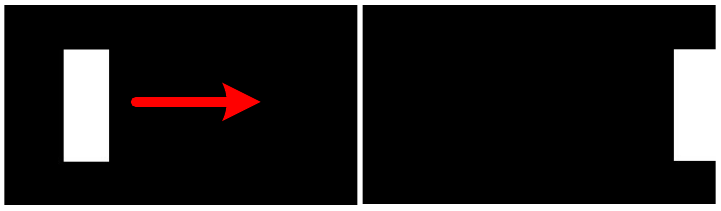}}
		\hfill
		\subfloat{\includegraphics[width=0.2\textwidth]{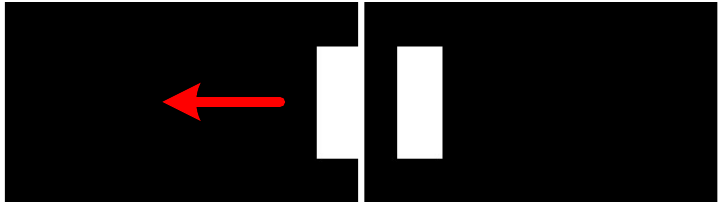}}
		\hfill
		\subfloat{\includegraphics[width=0.2\textwidth]{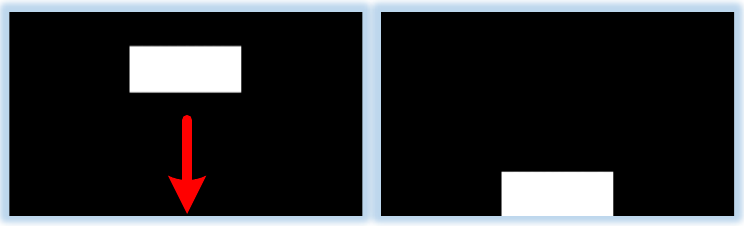}}
		\hfill
		\subfloat{\includegraphics[width=0.2\textwidth]{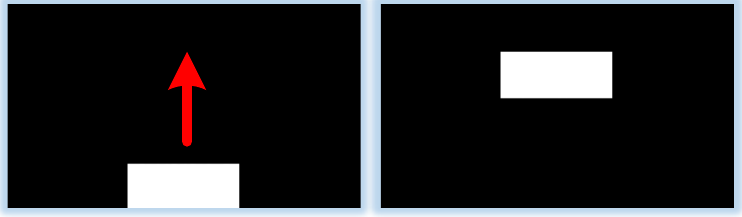}}
		\vfil
		\vspace{-10pt}
		\subfloat{\includegraphics[width=0.24\textwidth]{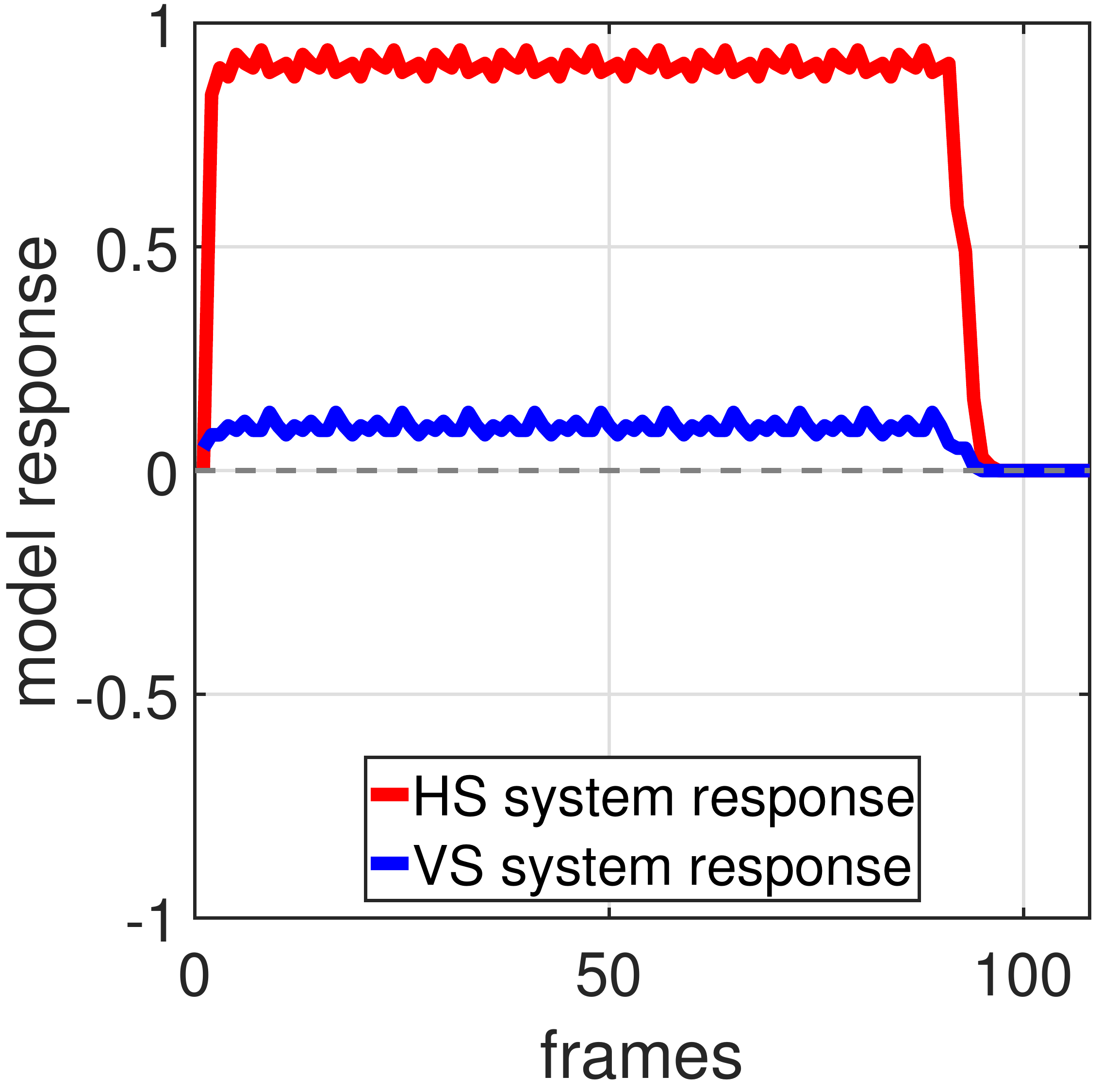}}
		\hfill
		\subfloat{\includegraphics[width=0.24\textwidth]{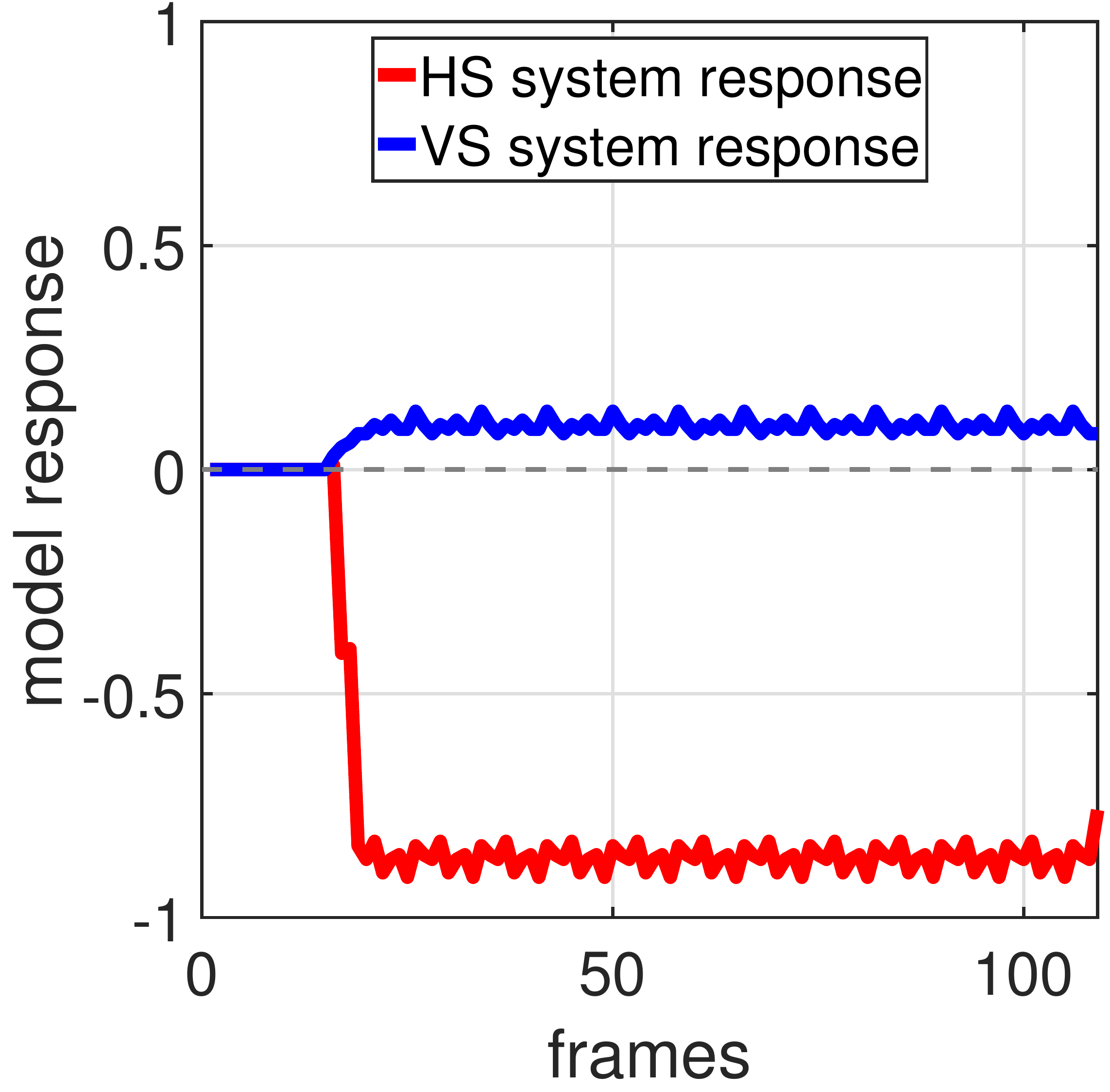}}
		\hfill
		\subfloat{\includegraphics[width=0.24\textwidth]{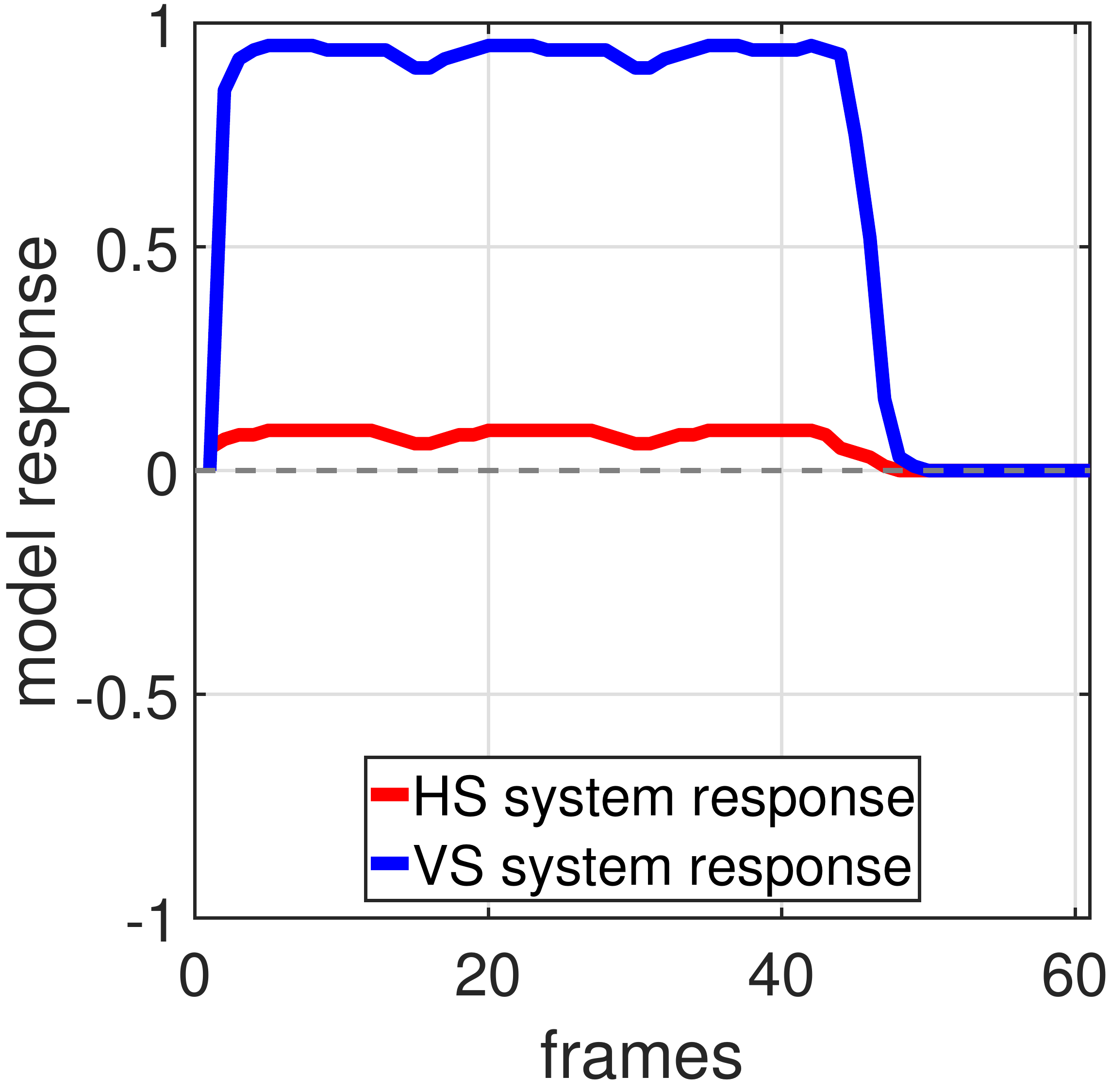}}
		\hfill
		\subfloat{\includegraphics[width=0.24\textwidth]{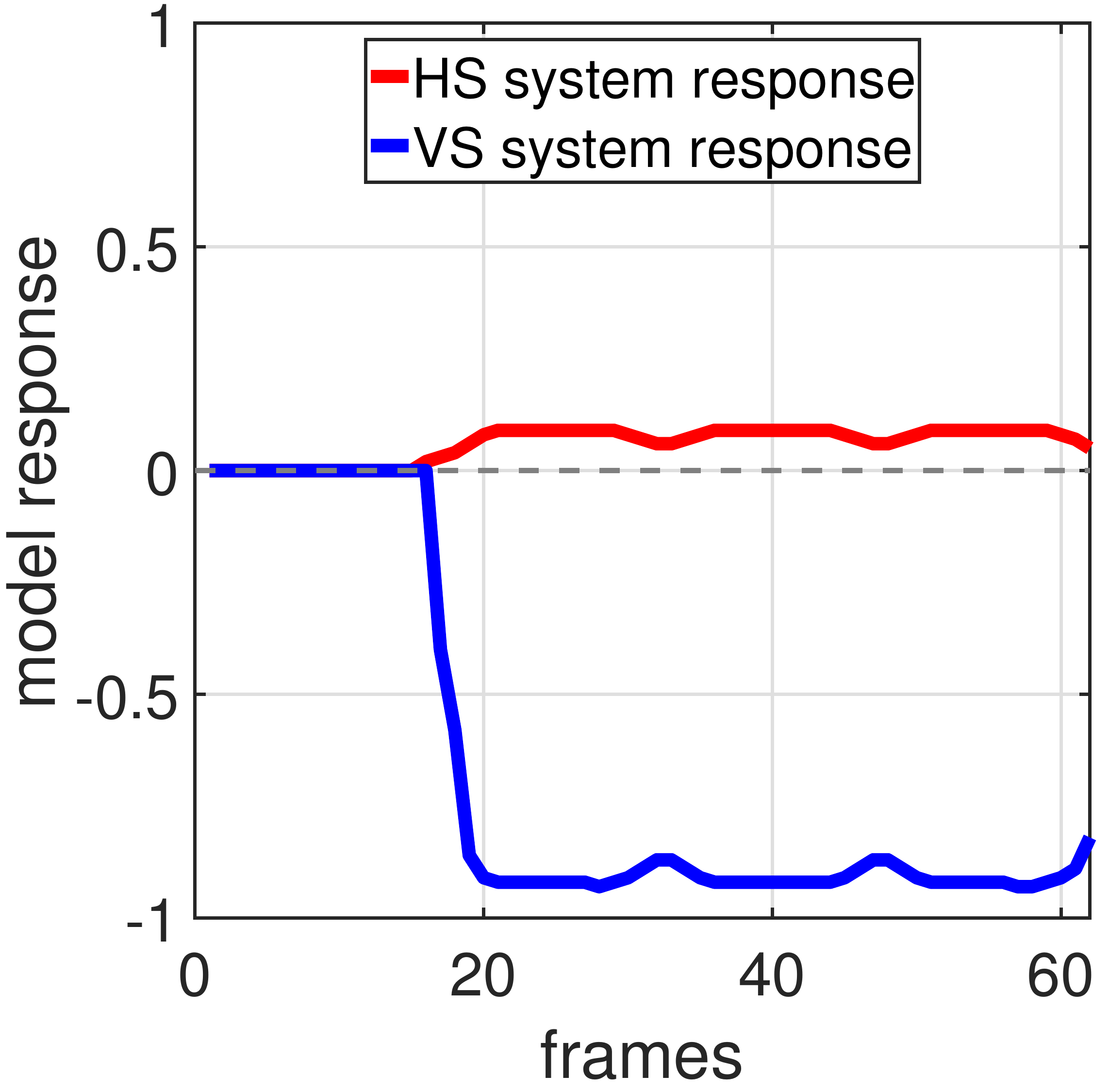}}
	\end{center}
	\caption{
		Proposed model responses challenged by dark and white objects translating constantly in four cardinal directions, against clean backgrounds.
	}
	\label{Fig: Fig-offline-clean-background-translating-tests}
\end{figure}

\begin{figure}[h!]
	\begin{center}
		\subfloat{\includegraphics[width=0.2\textwidth]{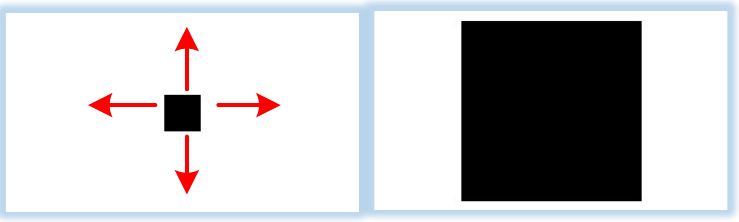}}
		\hfill
		\subfloat{\includegraphics[width=0.2\textwidth]{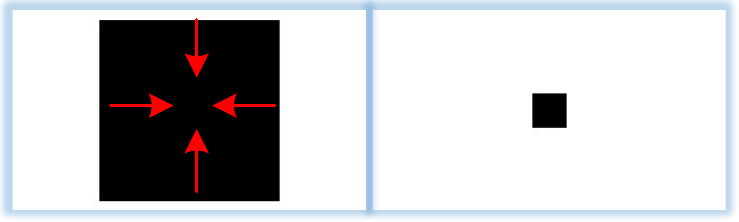}}
		\hfill
		\subfloat{\includegraphics[width=0.2\textwidth]{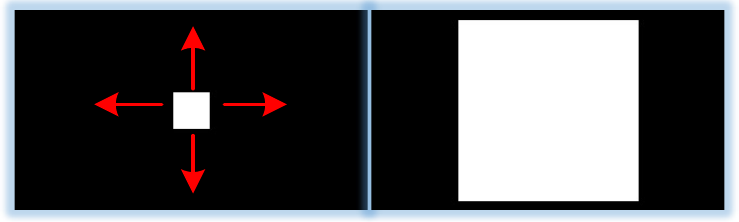}}
		\hfill
		\subfloat{\includegraphics[width=0.2\textwidth]{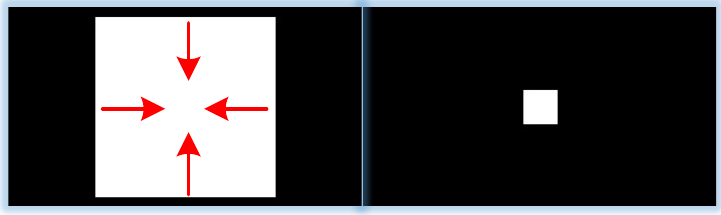}}
		\vfil
		\vspace{-10pt}
		\subfloat{\includegraphics[width=0.24\textwidth]{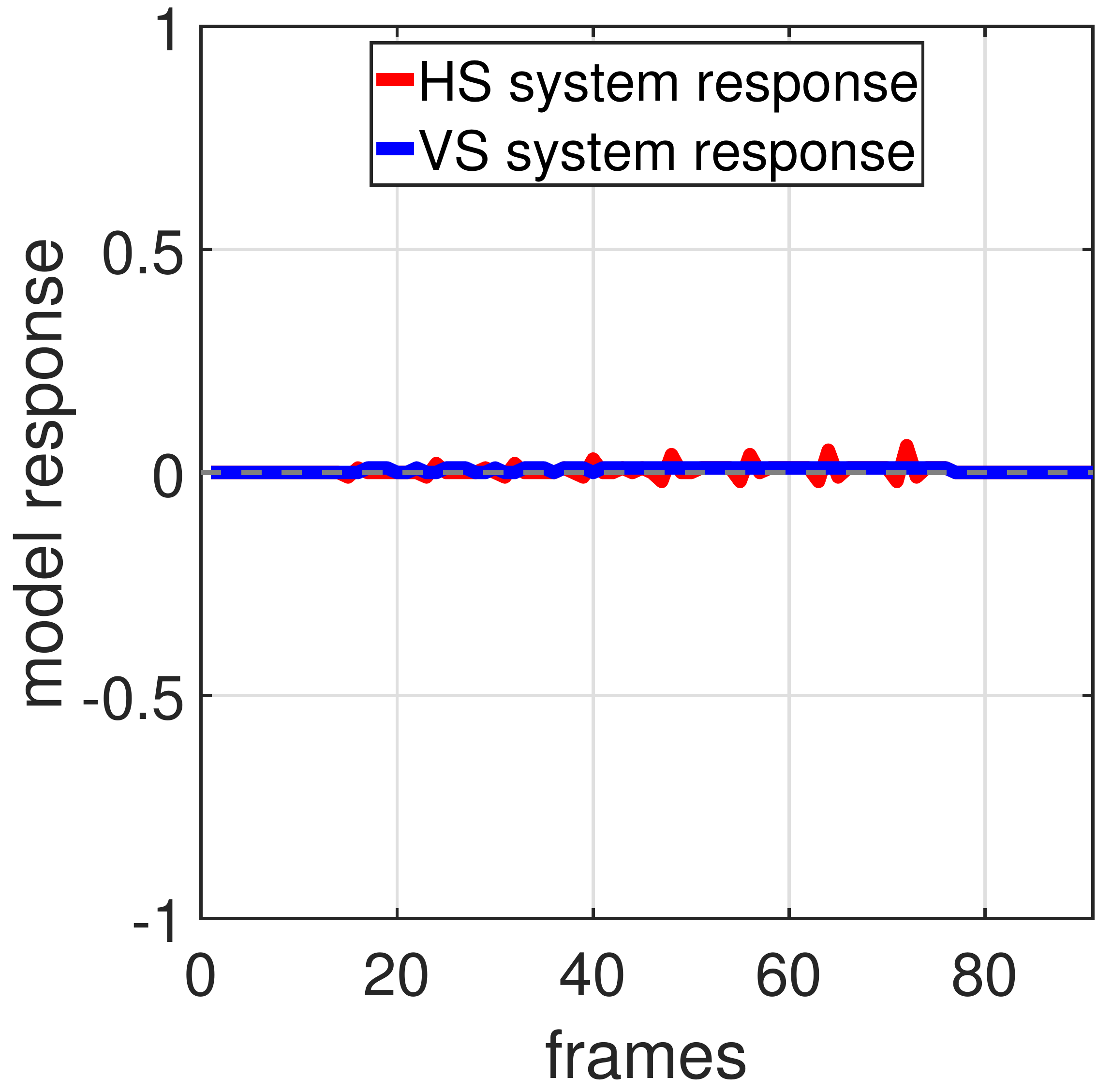}}
		\hfill
		\subfloat{\includegraphics[width=0.24\textwidth]{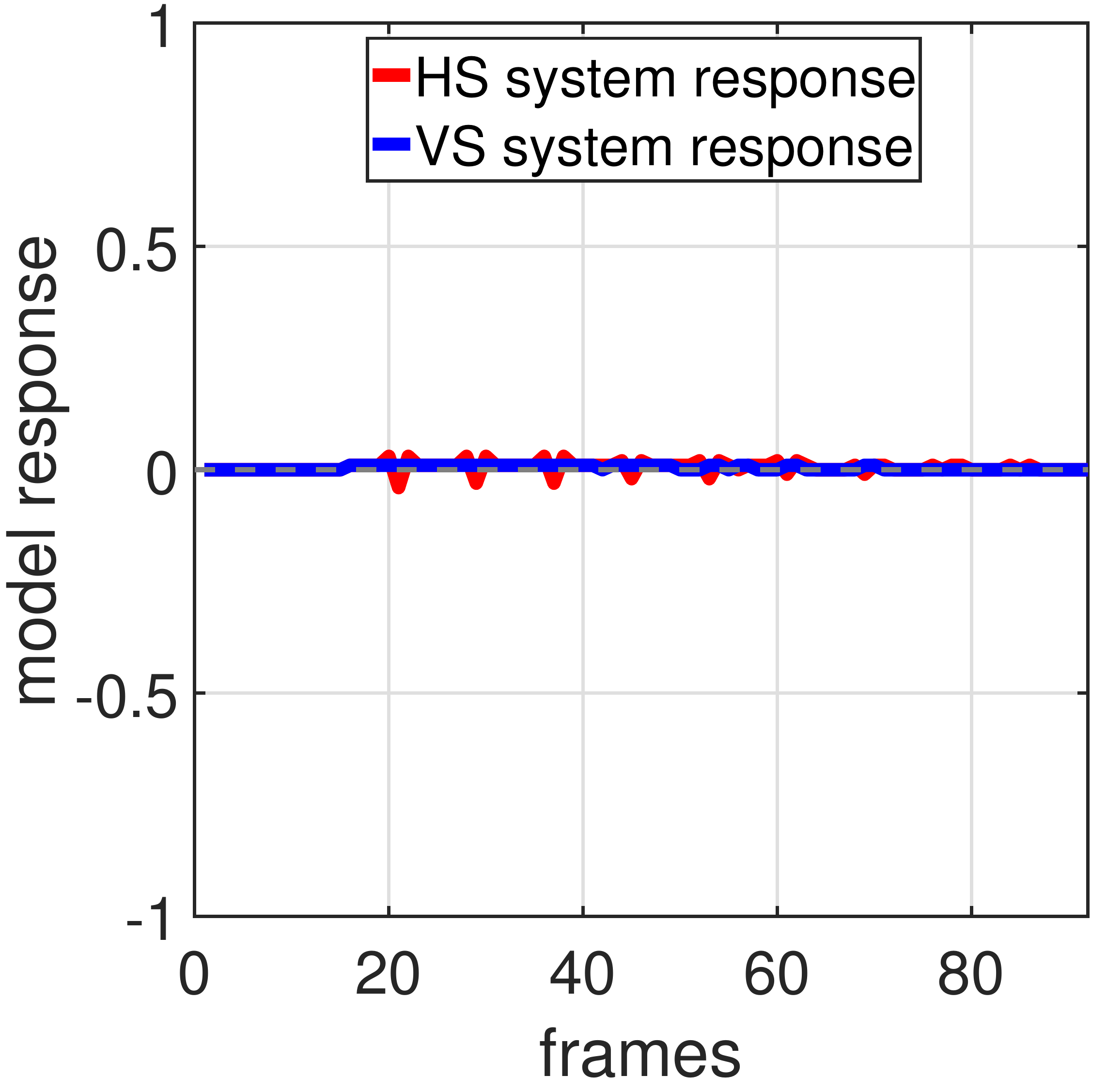}}
		\hfill
		\subfloat{\includegraphics[width=0.24\textwidth]{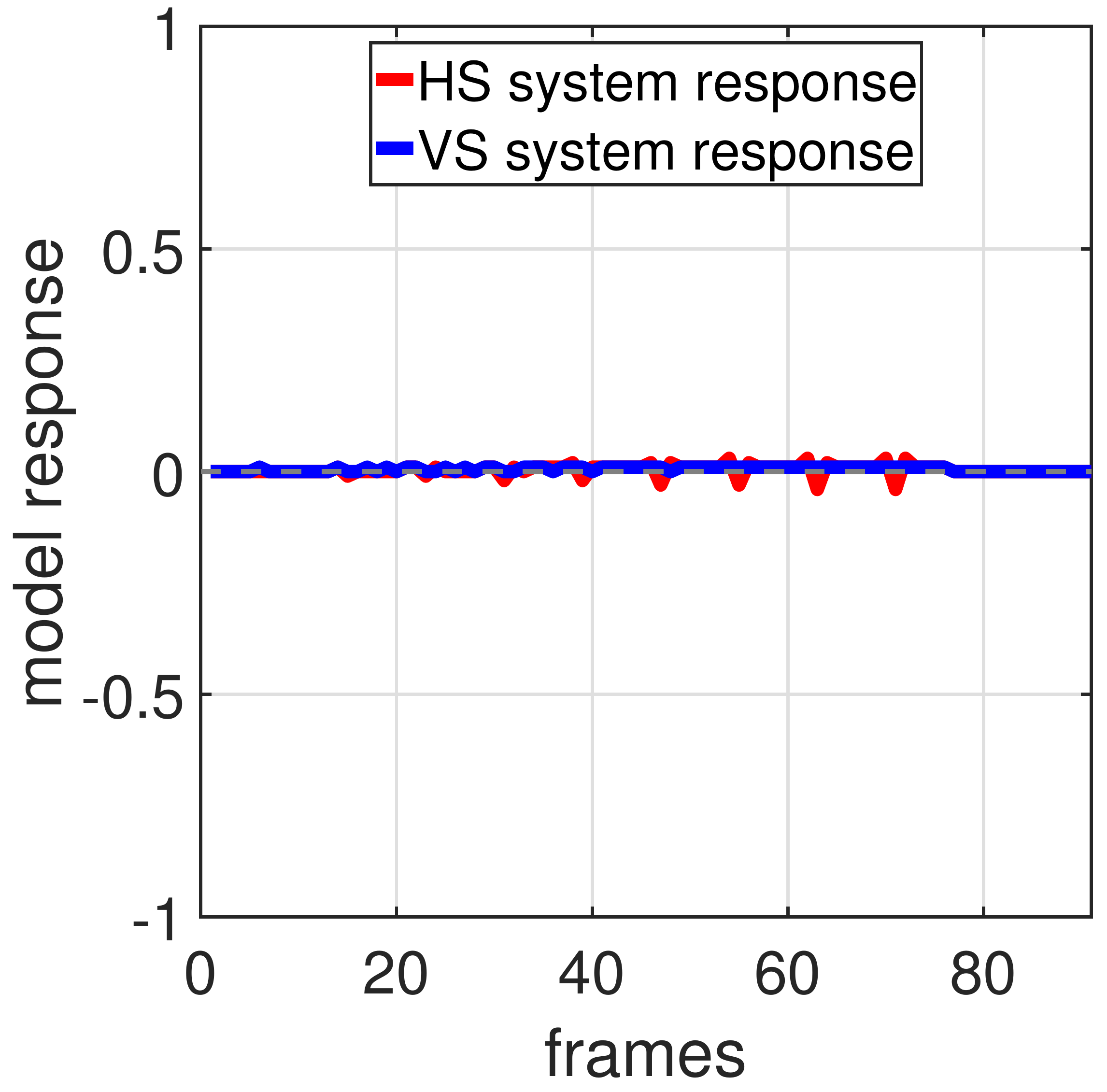}}
		\hfill
		\subfloat{\includegraphics[width=0.24\textwidth]{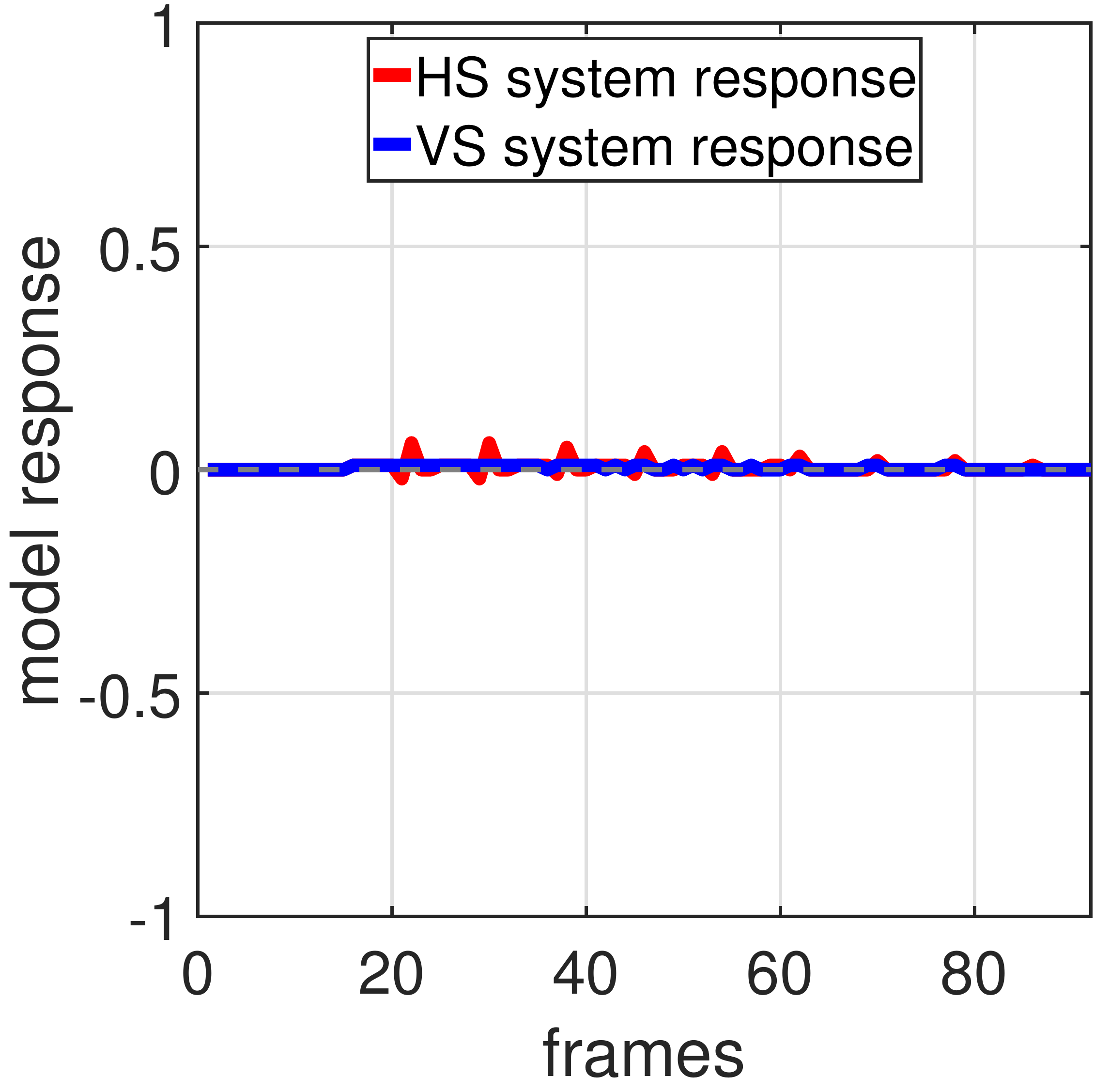}}
	\end{center}
	\caption{
		Proposed model responses by dark and white objects approaching and receding, against clean backgrounds.
	}
	\label{Fig: Fig-offline-clean-background-approaching-tests}
\end{figure}

In the second type of tests, we look deeper into the properties of both  model and stimuli considering the effects of proposed spatiotemporal dynamics on decoding the direction of translating objects in front of cluttered moving backgrounds. 
We compare the performance of models with and without the investigated mechanisms including the coordination of motion pre-filtering vDoG and FDSR, and the parameter $n_c$ (number of correlating detectors) in the ensembles of local ON-ON/OFF-OFF motion correlators. 
Moreover, we also give insight into the model performance on perception of different sized targets ($10 \times 10$, $25 \times 25$, $50 \times 50$ and $100 \times 100$ pixels), and also against very high-speed moving backgrounds (-50, -100, -200 degrees/s). 
The main objectives are 
1) demonstrating the significance of proposed new structures and mechanisms for detecting the foreground translating objects and reproducing the DS/DO response; 
2) revealing the model's responsive preferences.

\begin{figure}[H]
	\begin{center}
		\subfloat[a pedestrian moving rightward]{\includegraphics[width=0.49\textwidth]{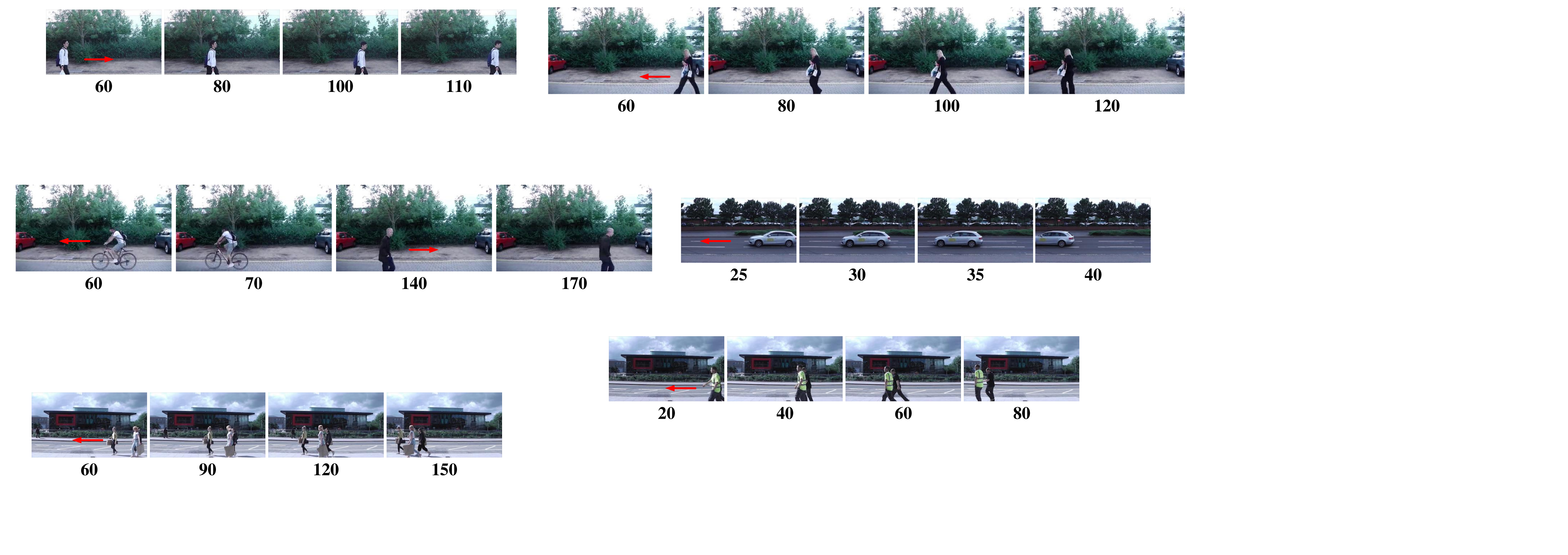}}
		\hfil
		\subfloat[a pedestrian moving leftward]{\includegraphics[width=0.49\textwidth]{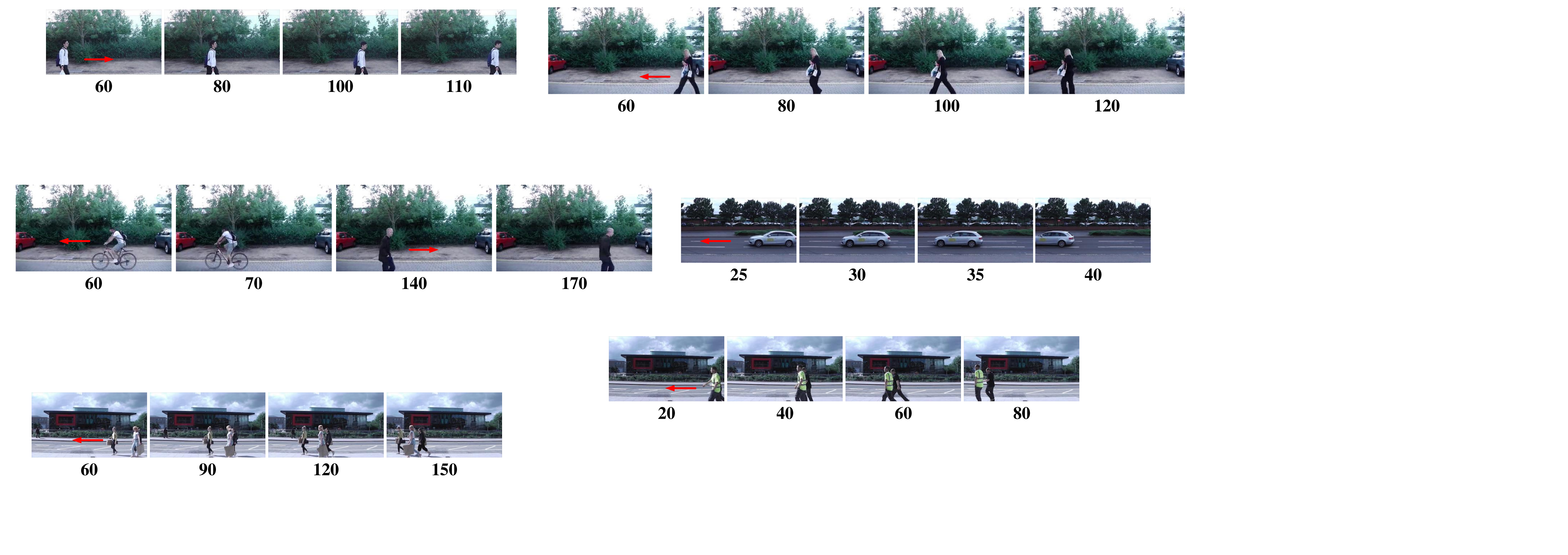}}
		\vfill
%		\vspace{-10pt}
		\subfloat[model outputs]{\includegraphics[width=0.49\textwidth]{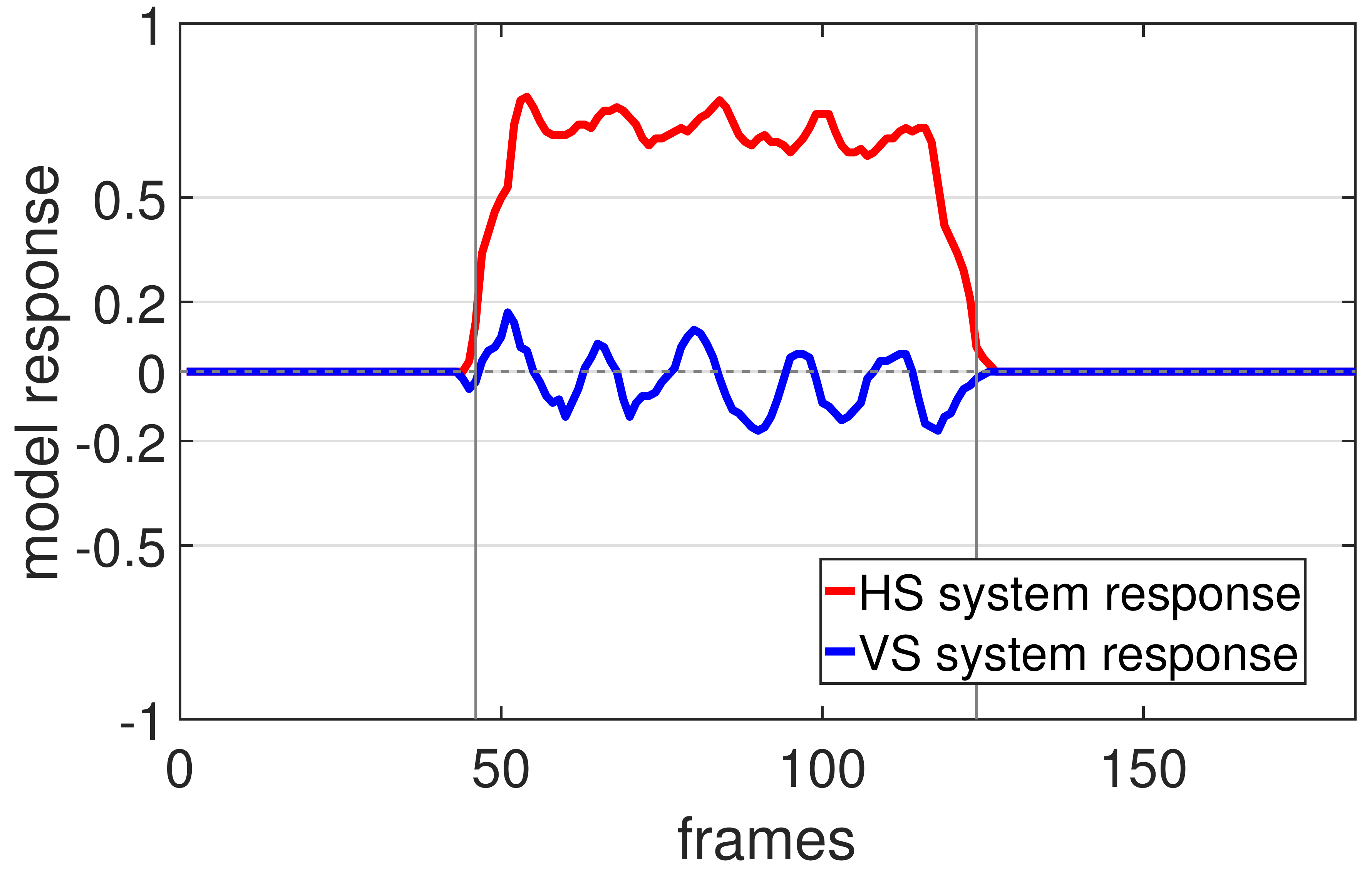}}
		\hfil
		\subfloat[model outputs]{\includegraphics[width=0.49\textwidth]{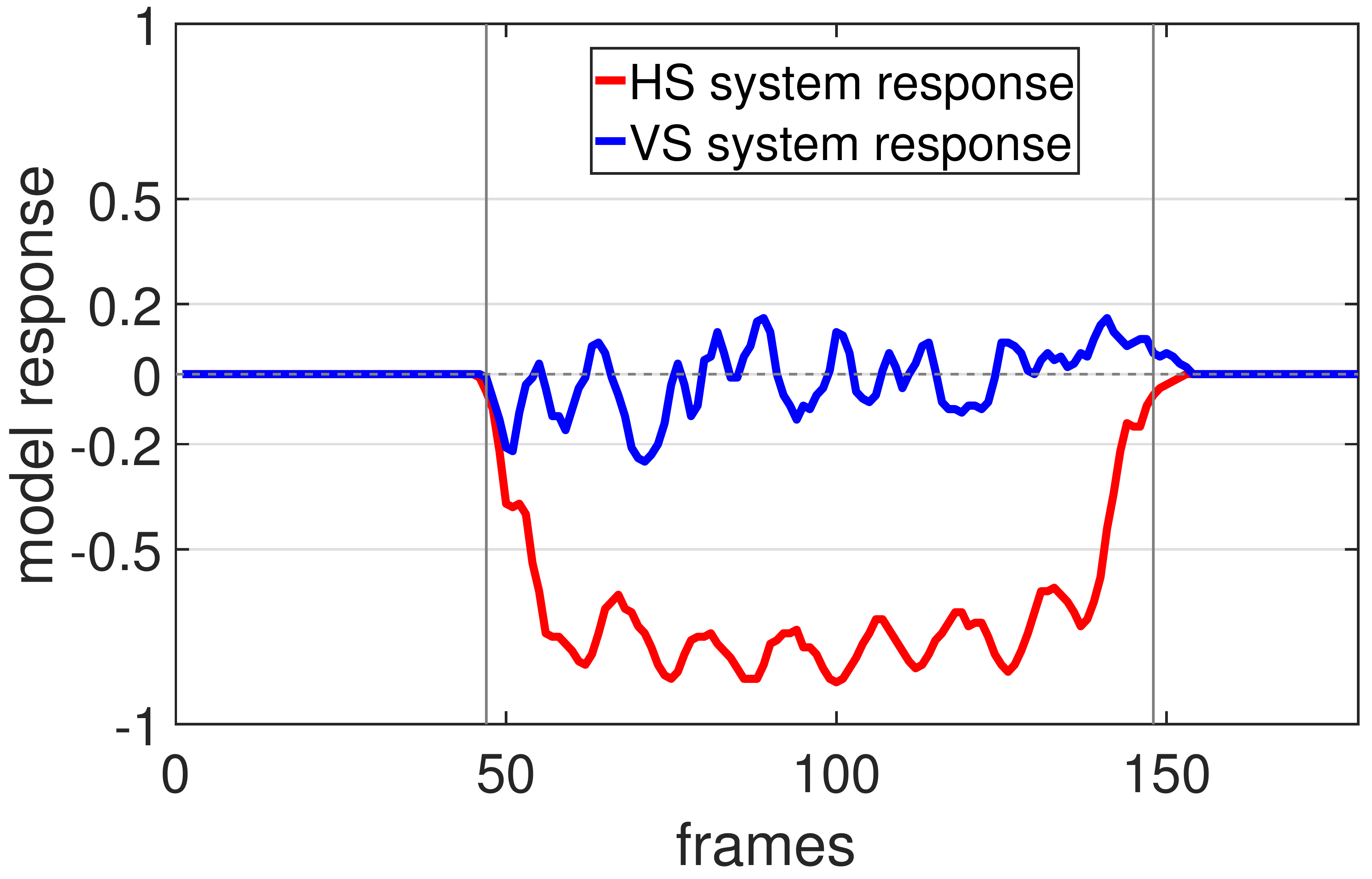}}
		\vfill
		\subfloat[a leftward translation followed by a rightward one]{\includegraphics[width=0.49\textwidth]{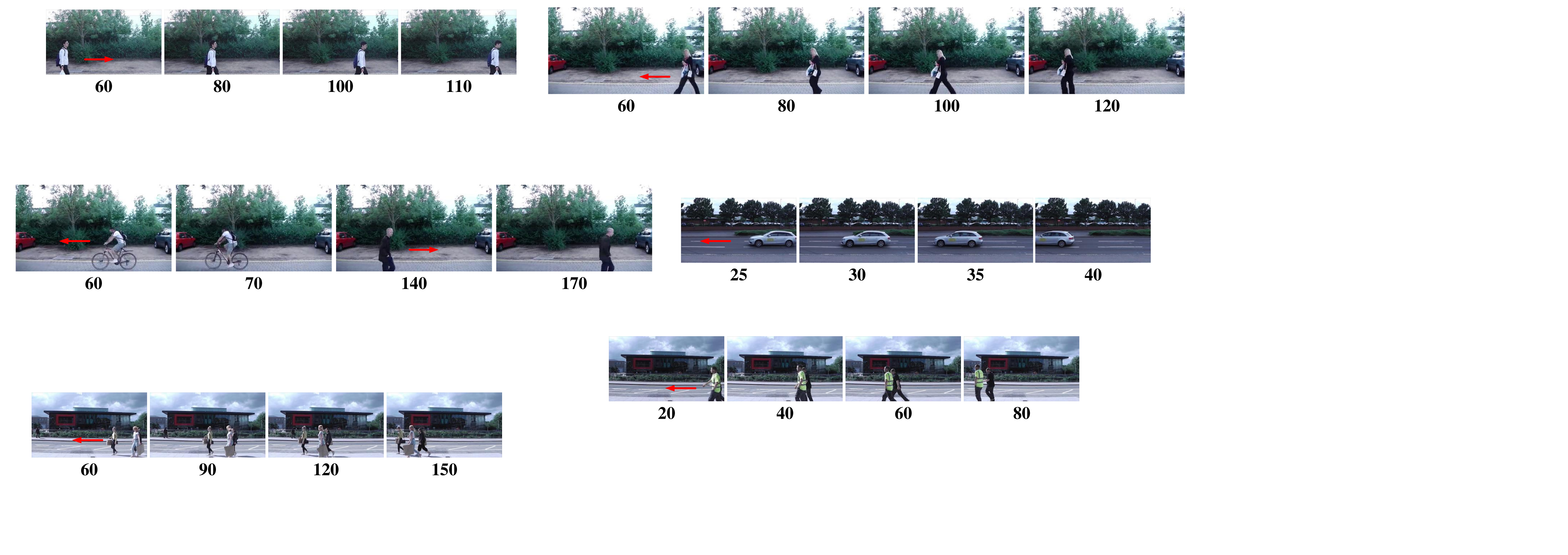}}
		\hfil
		\subfloat[a vehicle crossing leftward]{\includegraphics[width=0.49\textwidth]{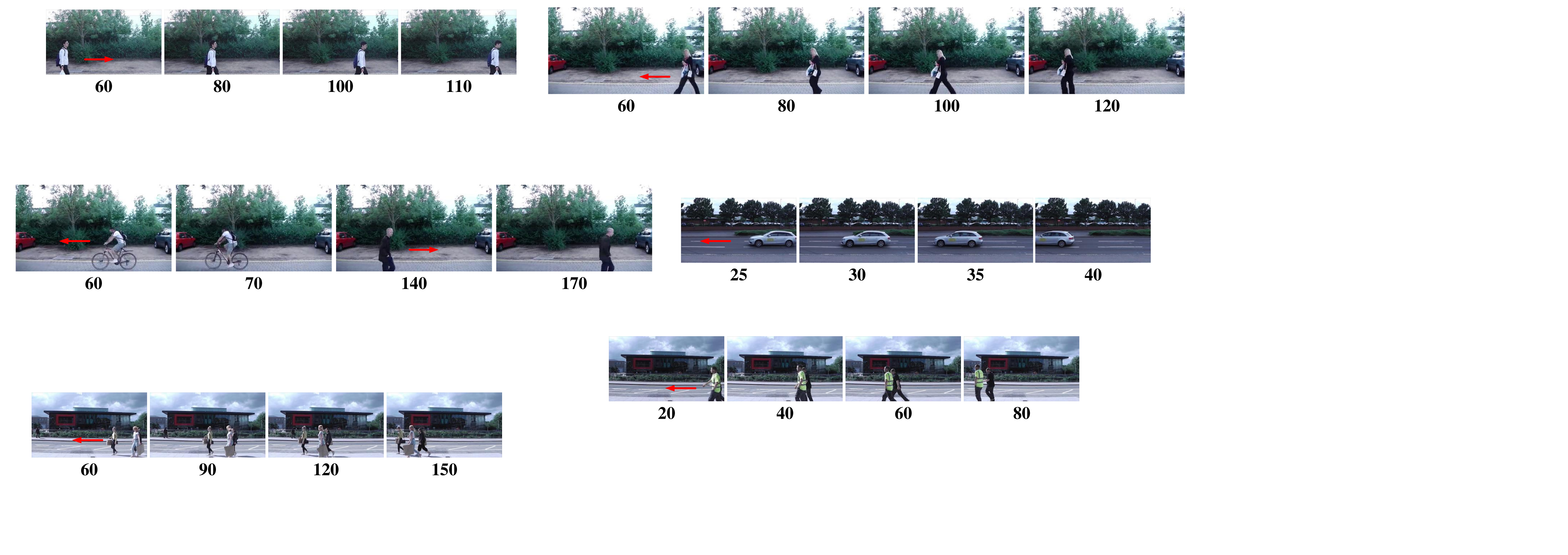}}
		\vfill
%		\vspace{-10pt}
		\subfloat[model outputs]{\includegraphics[width=0.49\textwidth]{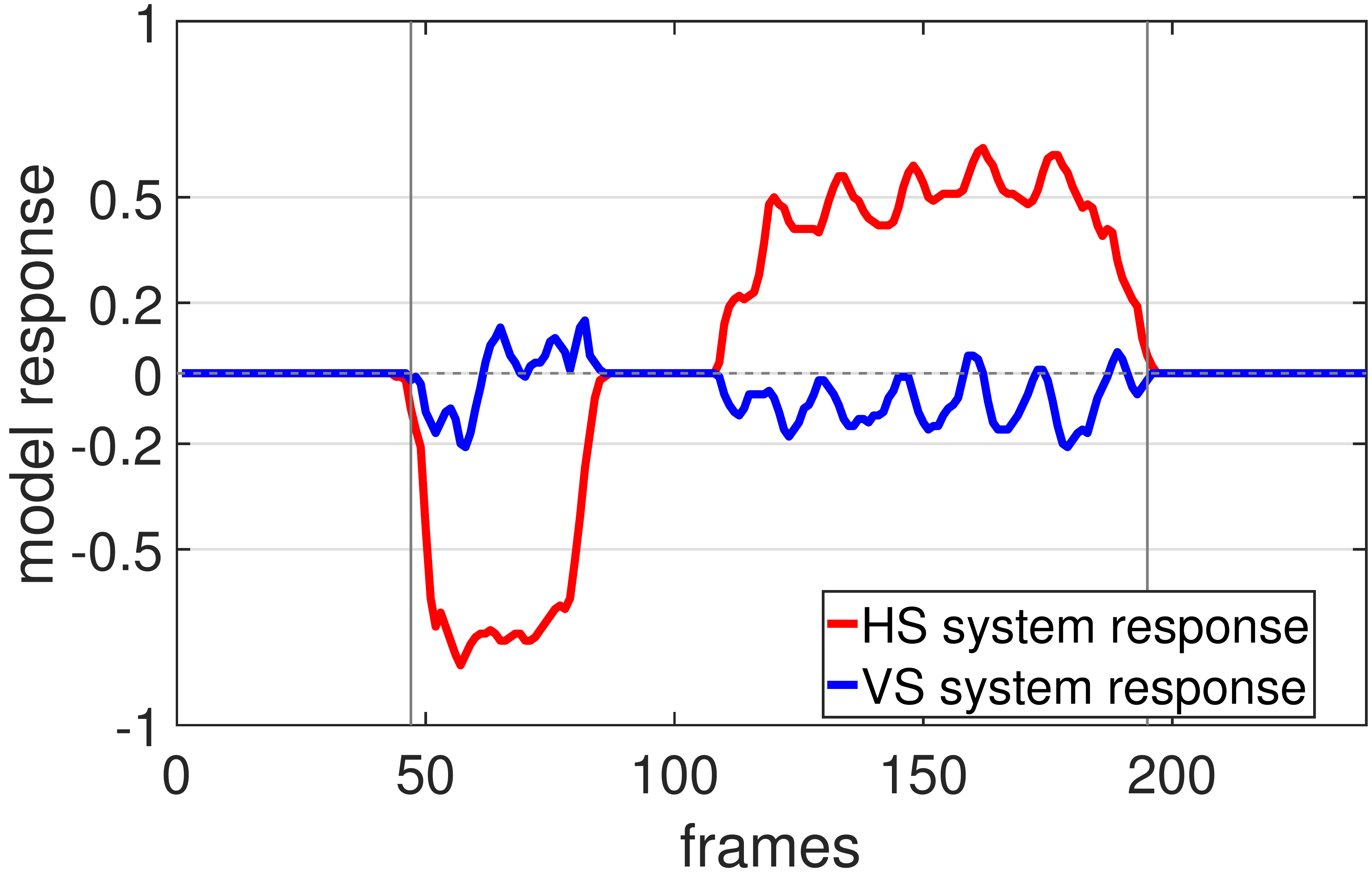}}
		\hfil
		\subfloat[model outputs]{\includegraphics[width=0.49\textwidth]{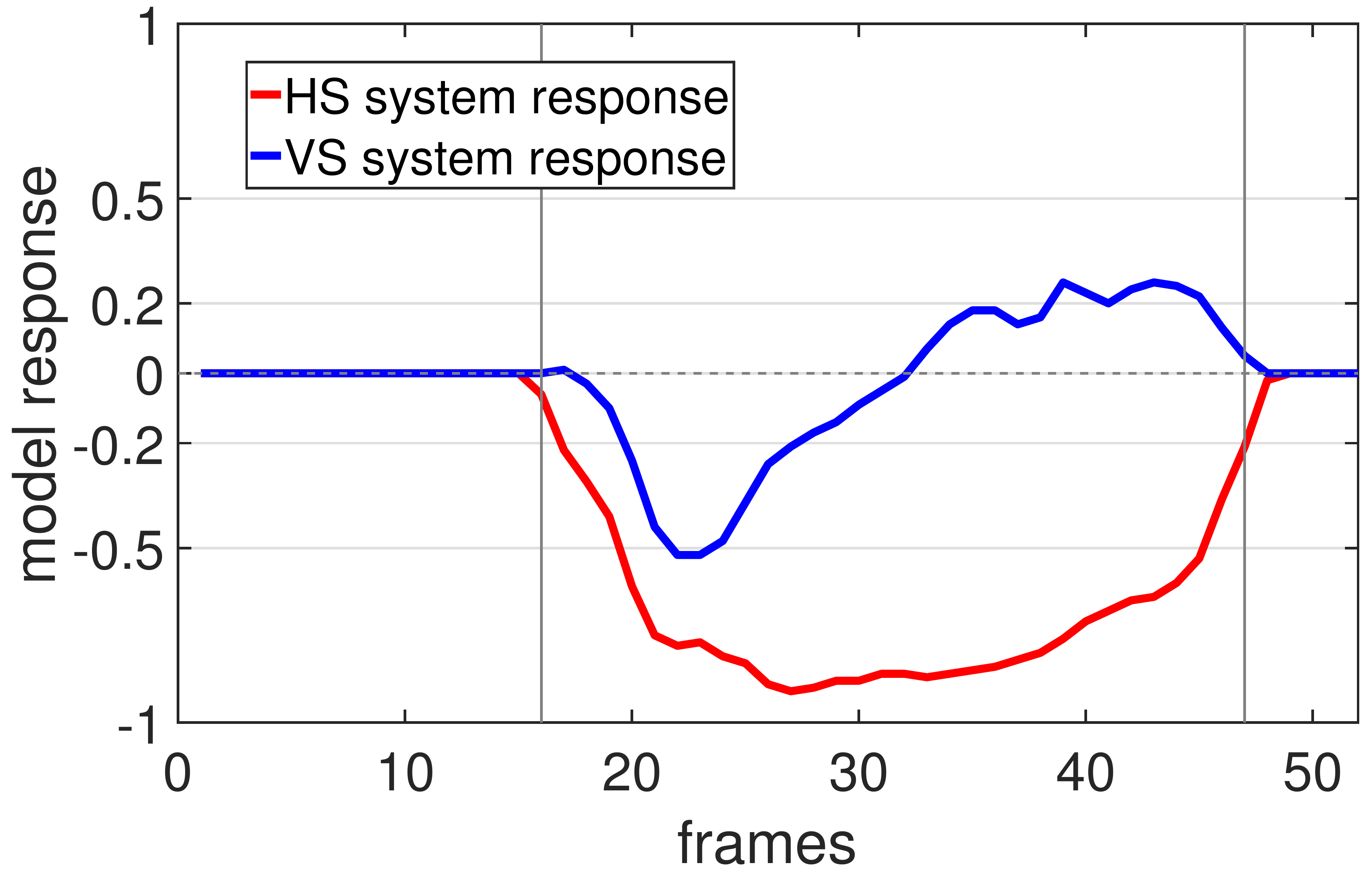}}
		\vfill
		\subfloat[grouped pedestrians moving leftward]{\includegraphics[width=0.49\textwidth]{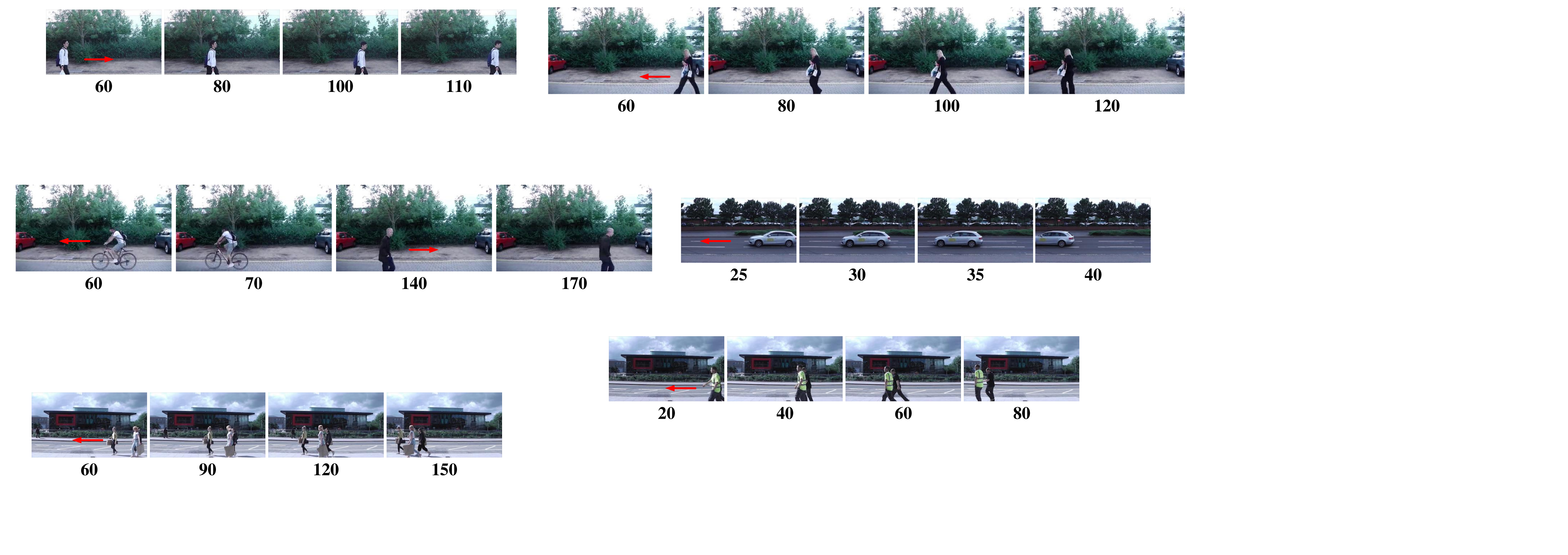}}
		\hfil
		\subfloat[grouped pedestrians moving leftward]{\includegraphics[width=0.49\textwidth]{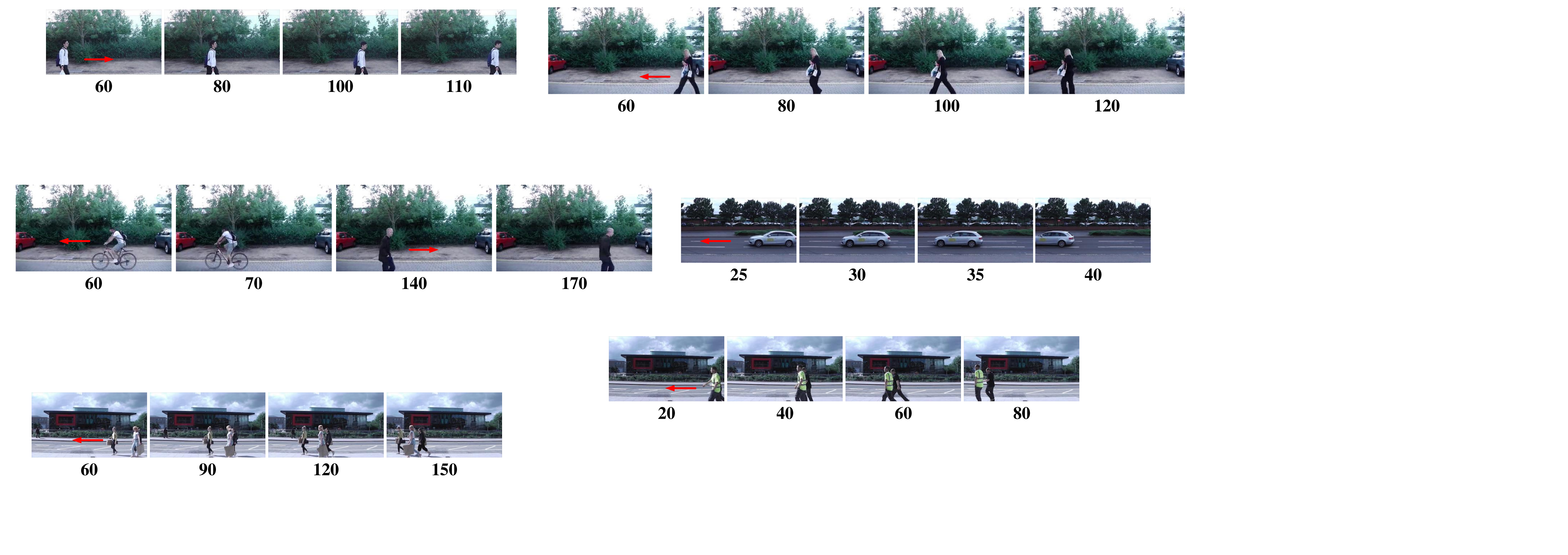}}
		\vfill
%		\vspace{-10pt}
		\subfloat[model outputs]{\includegraphics[width=0.49\textwidth]{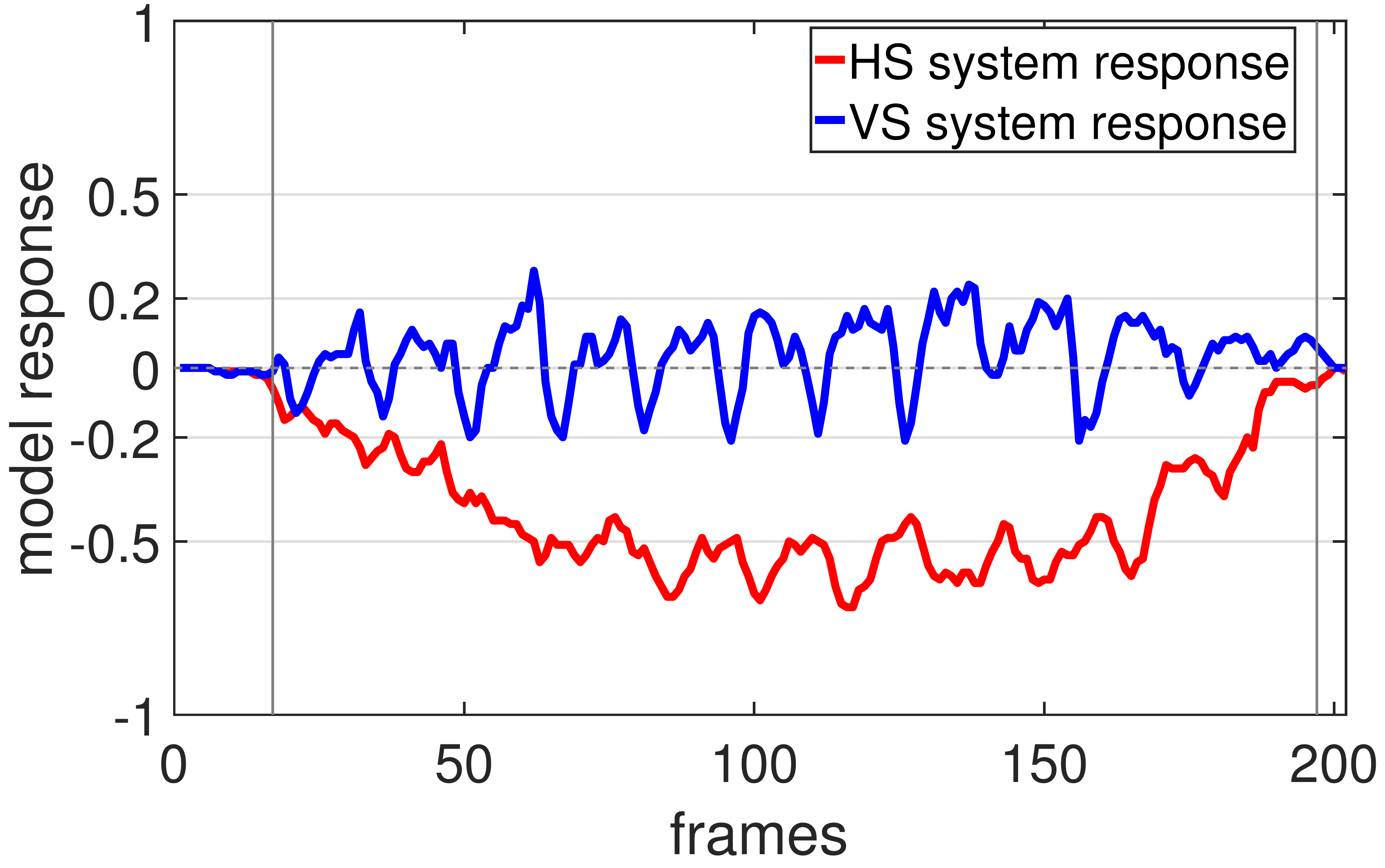}}
		\hfil
		\subfloat[model outputs]{\includegraphics[width=0.49\textwidth]{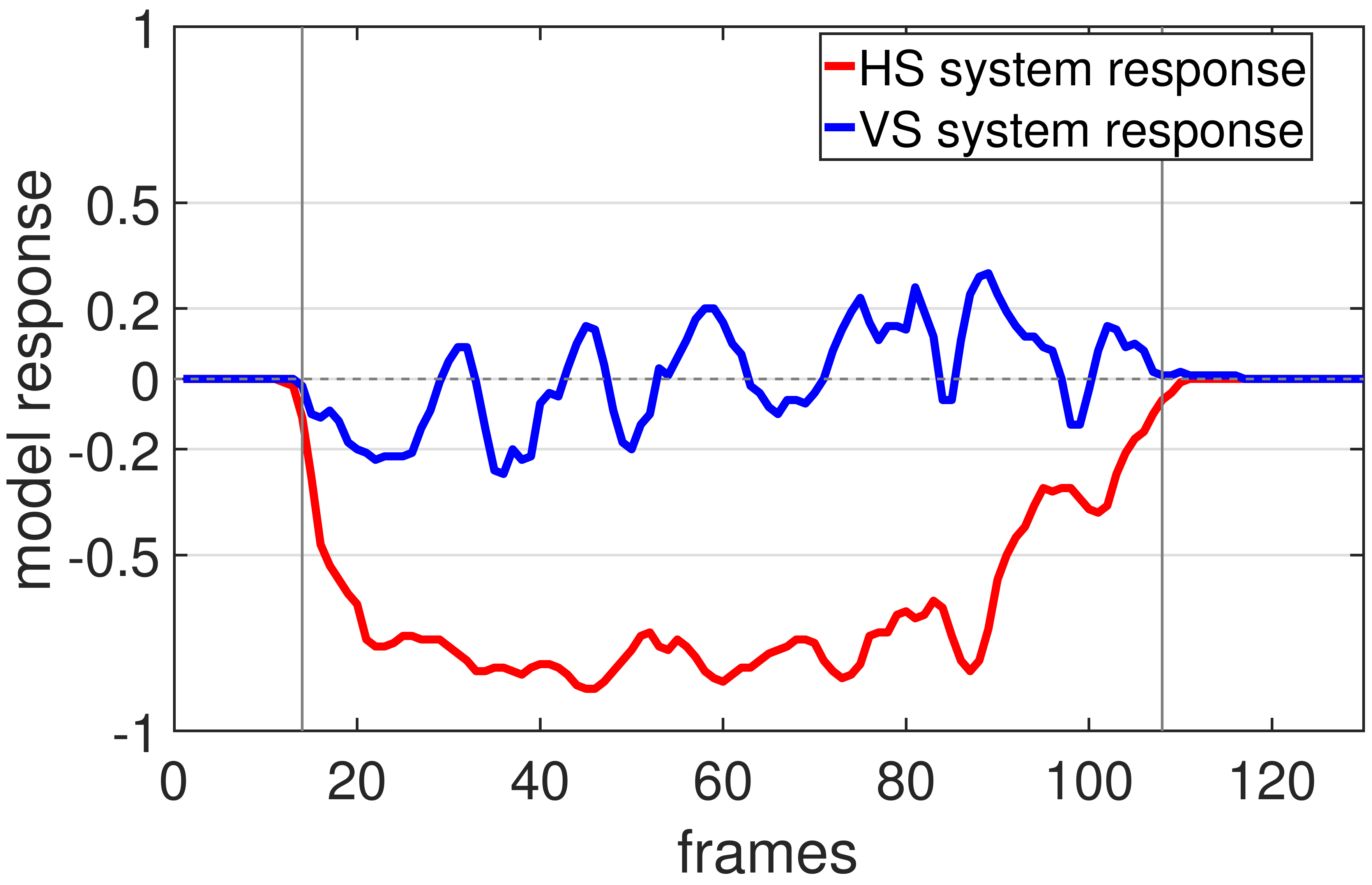}}
	\end{center}
	\caption{
		Proposed model responses challenged by foreground translating stimuli in real physical scenes. 
		The frame number is labelled at the bottom of each snapshot. 
		The red arrow in snapshots denotes the ground truth of primary direction of foreground translational OF. 
		The two vertical lines in each result indicate the time window of the appearance of foreground translational OF.
	}
	\label{Fig: Fig-offline-real-world-translating-tests}
\end{figure}

In all the experiments, the proposed visual system model was set up in Visual Studio (Microsoft Corporation). 
The synthetic visual stimuli were generated by a Python open-source library, i.e., the Vision Egg \cite{Straw-VisionEgg}. 
The real world stimuli were recorded by a camera. 
Data analysis and visualisations were implemented in MATLAB (The MathWorks, Inc., Natick, MA, USA).

\section{Results}
\label{Sec: result}

\subsection{Demonstrations of the Specific Direction Selectivity}

In this kind of tests, we systematically examine the specific direction selectivity of the proposed \textit{Drosophila} visual system model. 
Firstly, the model is challenged by a few typical motion patterns with clean backgrounds, which include darker and brighter objects translating in four cardinal directions, approaching and receding. 
The model responses are shown in Fig. \ref{Fig: Fig-offline-clean-background-translating-tests} and \ref{Fig: Fig-offline-clean-background-approaching-tests}. 
More specifically, when challenged by translating in four cardinal directions (see Fig. \ref{Fig: Fig-offline-clean-background-translating-tests}), the HS and VS systems are highly activated by horizontal and vertical movements, respectively. 
In the process of translation, the leading and trailing edges of a darker object bring about OFF and ON contrasts, respectively; while a brighter object leads to the opposite responses. 
As a result, the model with ON and OFF channels can encode both polarity contrast in separate pathways in order to generate the specific DS responses to four cardinal directions: 
the HS system is rigorously activated by merely the horizontal translational OF representing positive or negative response to its PD (rightward) or ND (leftward) motion; 
the VS system only responds to the vertical translational OF that also shows positive or negative response to its PD (downward) or ND (upward) motion.

On the other hand, when challenged against approaching and receding motion patterns, i.e., movements in depth (see Fig. \ref{Fig: Fig-offline-clean-background-approaching-tests}), both the HS and VS systems are rigorously inhibited during every entire process: the ON and OFF contrast by PD and ND motions with contracting and distracting edges are cancelled by each other. 
The results demonstrate clearly the direction selectivity of proposed visual system model to translation in four cardinal directions.

Next, the model is tested by more challenging real world scenarios. 
Compared to the computer-simulated stimuli, the real physical backgrounds are unstructured including motion distractors, such as the windblown vegetation, and etc. 
As the input visual stimuli, all the translating targets have the ground truth of primary direction in horizontal, as illustrated in Fig. \ref{Fig: Fig-offline-real-world-translating-tests}. 
Accordingly, the HS system is highly activated when translations appear in the field of vision. 
Notably, the VS system is also activated compared to the above stimuli in clean backgrounds, caused by irregular locomotion of translating targets in vertical directions and background distractors. 
Despite that, the HS system responds much more strongly to the visual stimuli, which can well indicate the principal direction of foreground moving objects. 
The PD or ND motion in horizontal is well decoded as positive or negative membrane potential of the neural system model. 
The results verify that the proposed model responds more consistently to foreground translating objects rather than irrelevant background flows that is robust to generate the DS and DO responses against more variable backgrounds.

\begin{figure}[H]
	\begin{center}
		\subfloat[$V_t=27$ degrees/s, $V_b=-20$ degrees/s]{\includegraphics[width=0.45\textwidth]{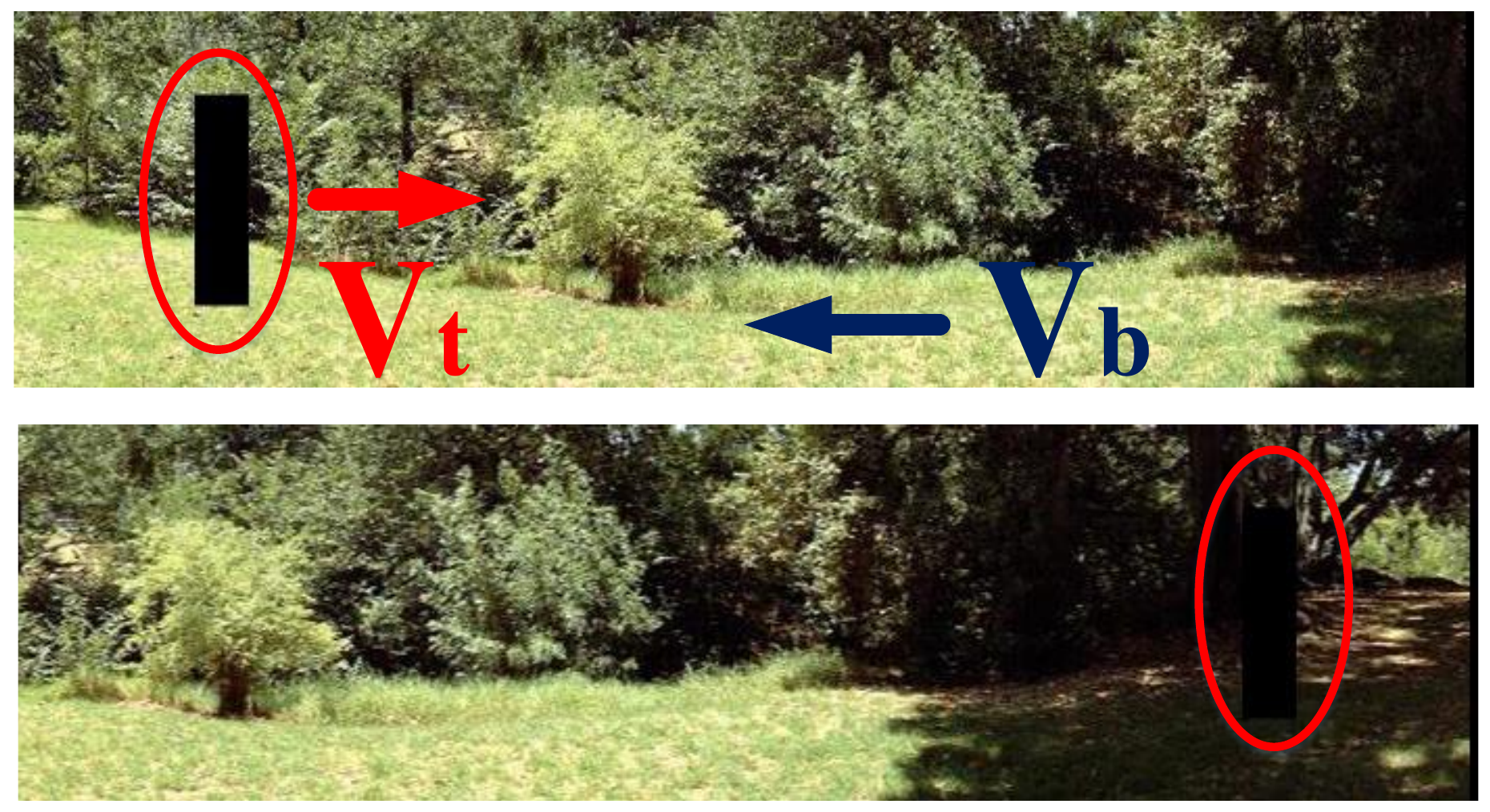}}
		\hfil
		\subfloat[$V_t=27$ degrees/s, $V_b=-20$ degrees/s]{\includegraphics[width=0.45\textwidth]{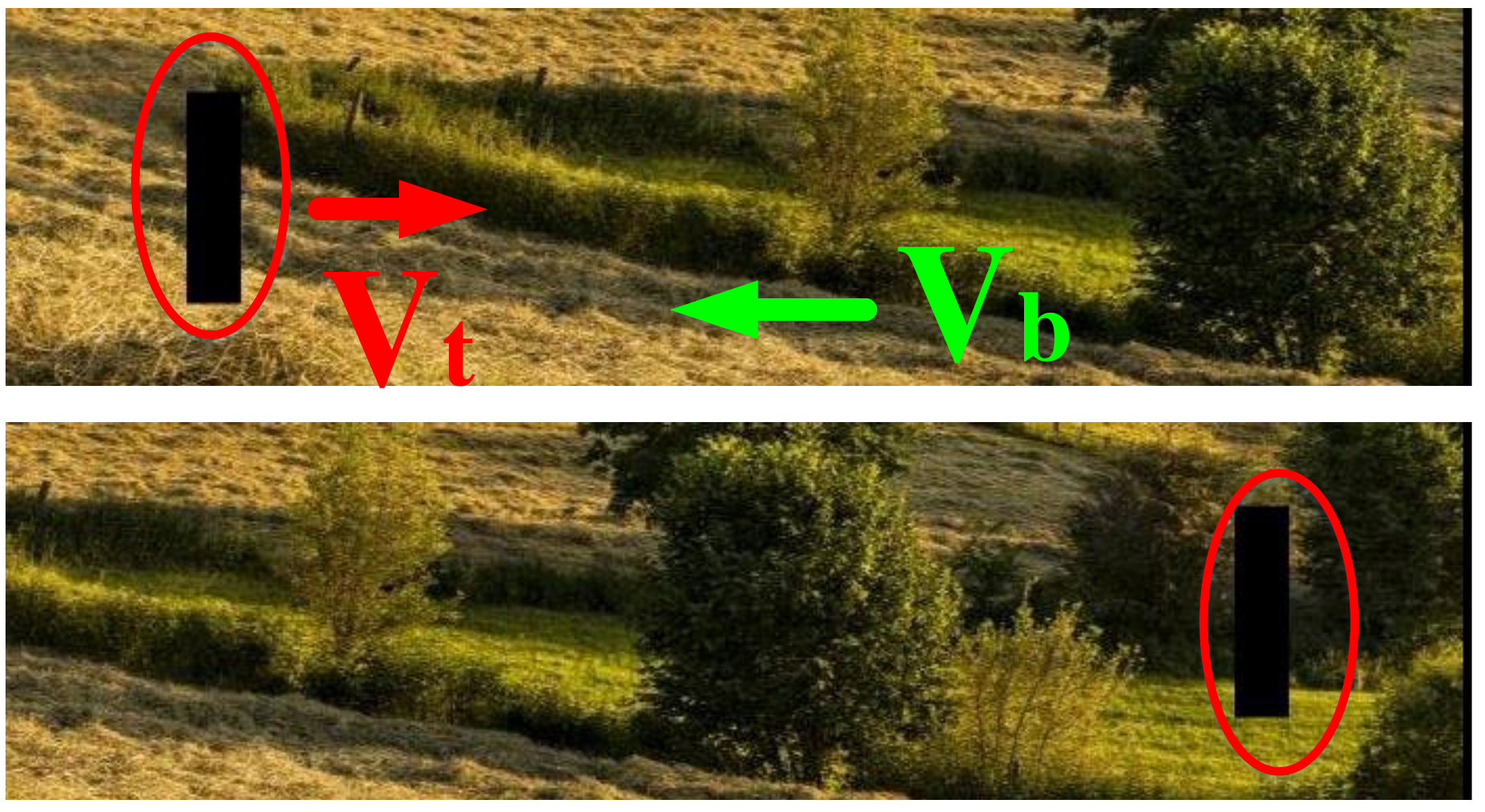}}
		\vfill
		\vspace{-10pt}
		\subfloat[model response (scene \#1)]{\includegraphics[width=0.49\textwidth]{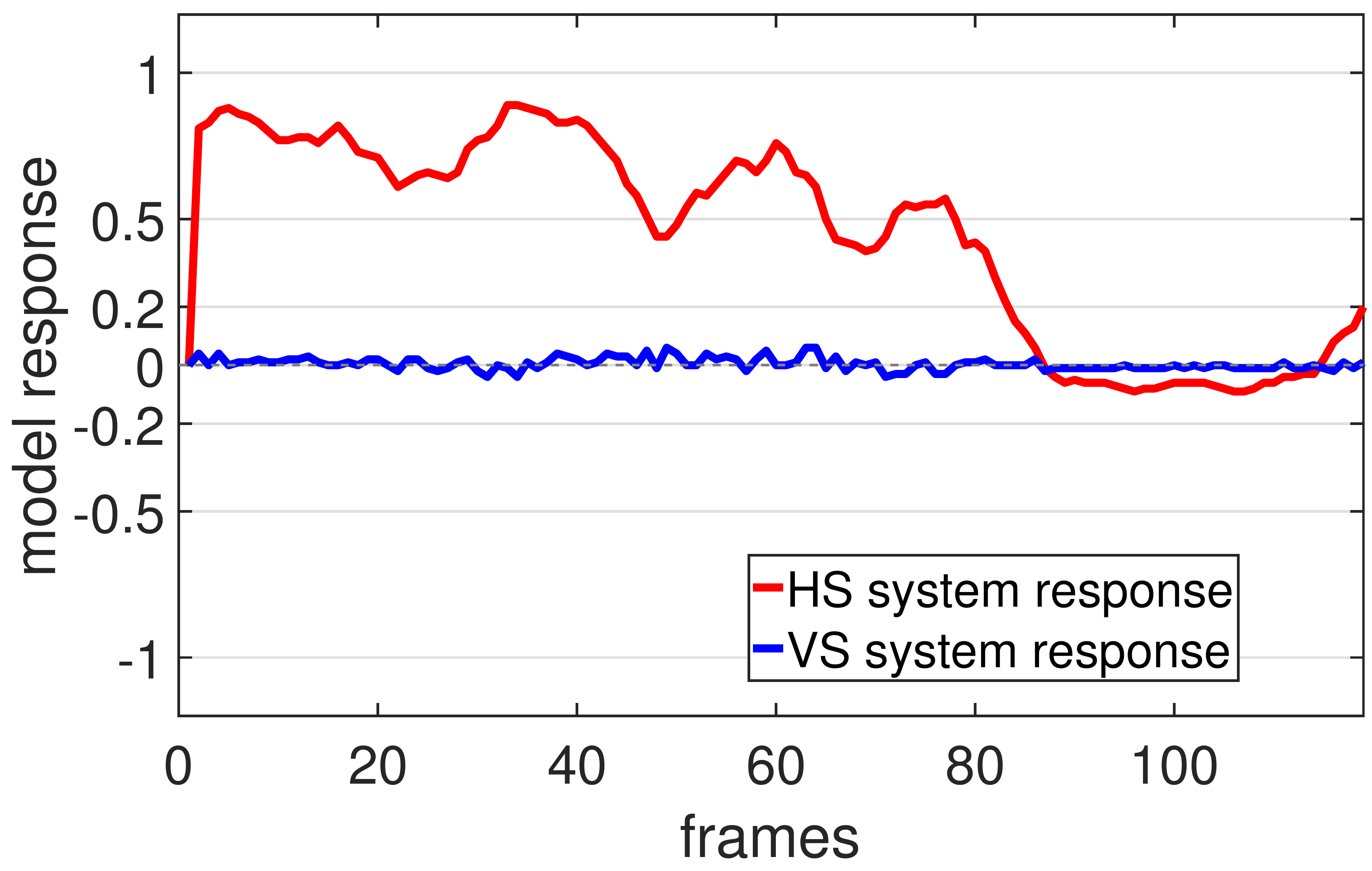}}
		\hfil
		\subfloat[model response (scene \#2)]{\includegraphics[width=0.49\textwidth]{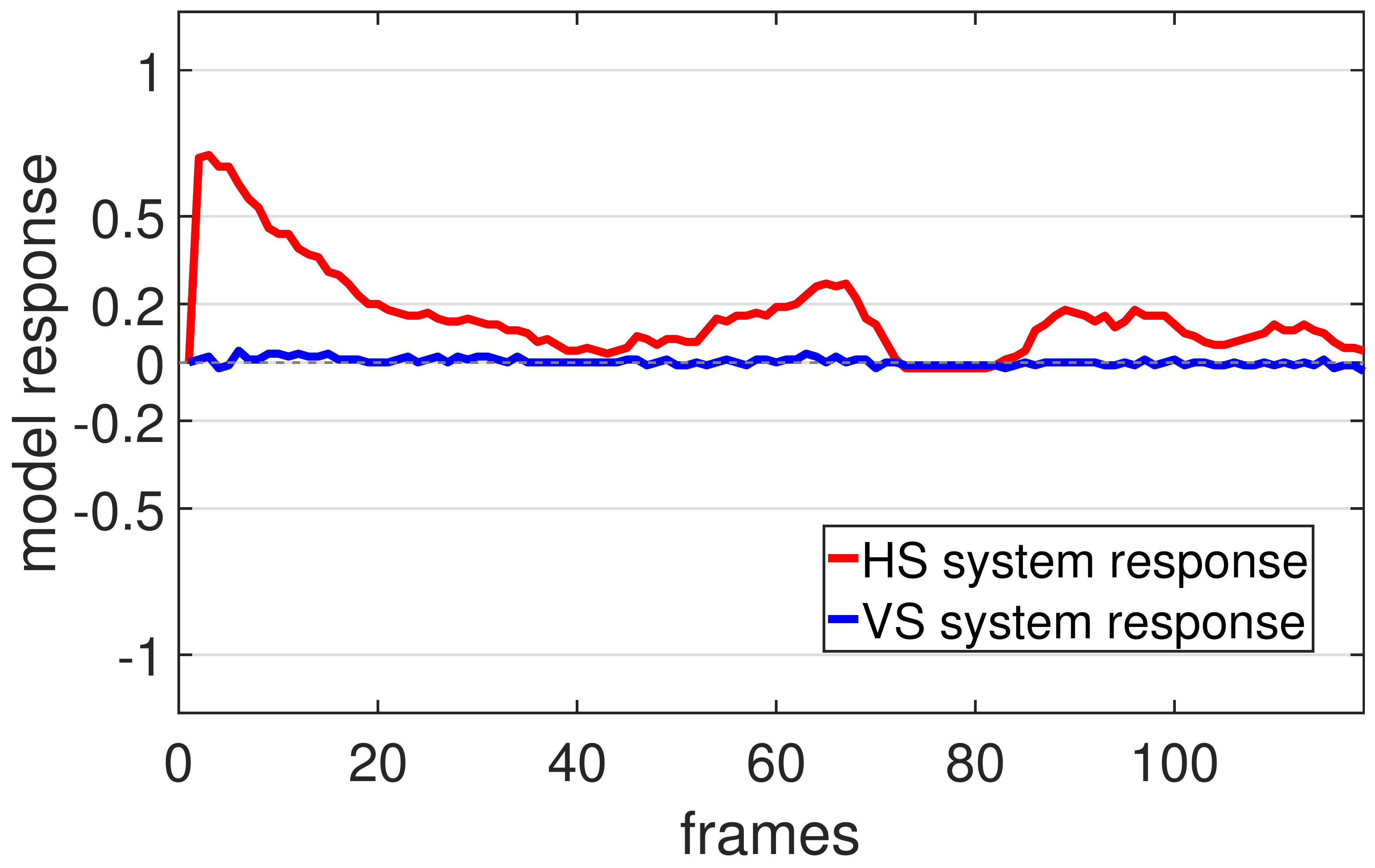}}
		\vfill
		\subfloat[$V_t=27$ degrees/s, $V_b=-20$ degrees/s]{\includegraphics[width=0.45\textwidth]{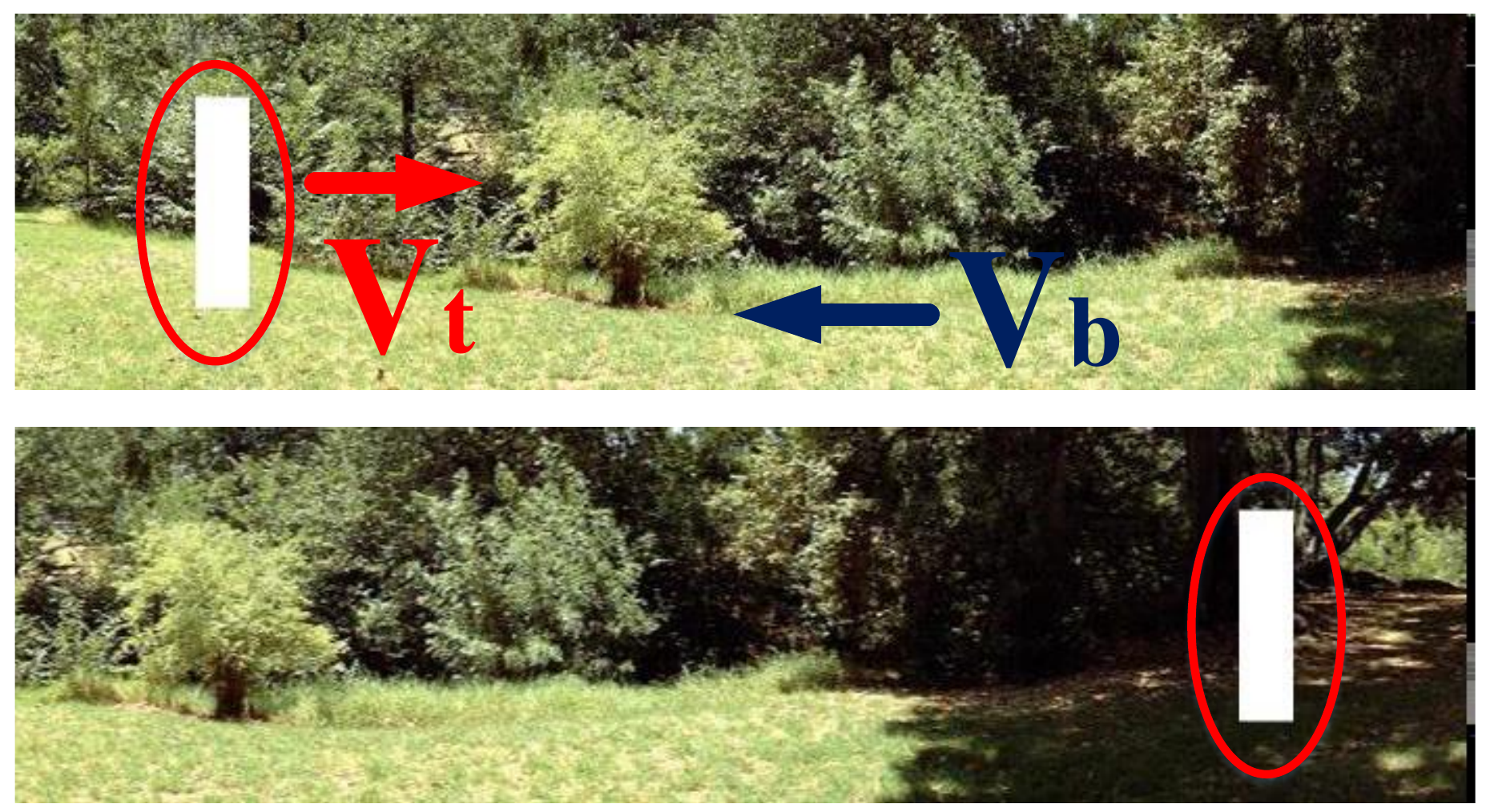}}
		\hfil
		\subfloat[$V_t=27$ degrees/s, $V_b=-20$ degrees/s]{\includegraphics[width=0.45\textwidth]{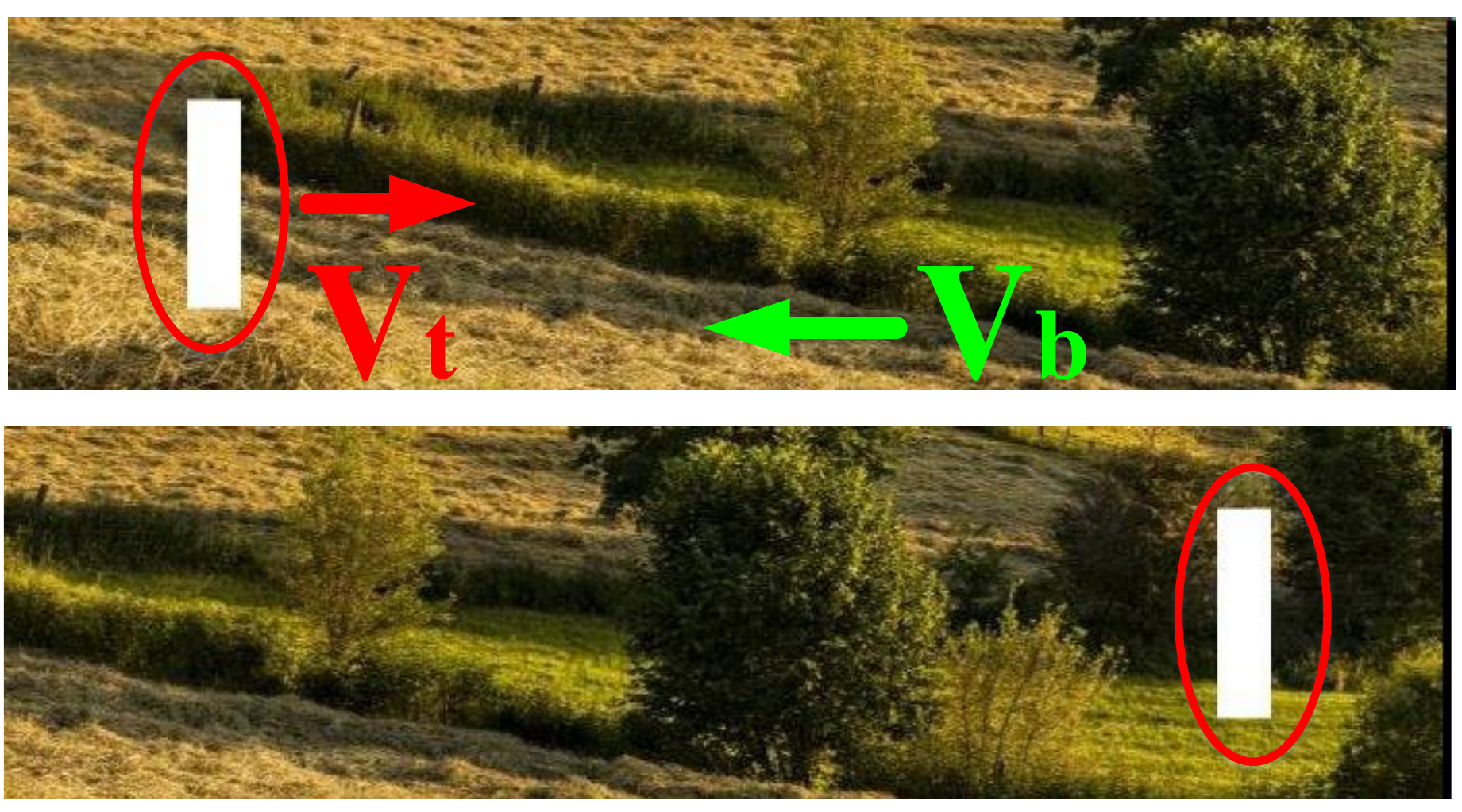}}
		\vfill
		\vspace{-10pt}
		\subfloat[model response (scene \#1)]{\includegraphics[width=0.49\textwidth]{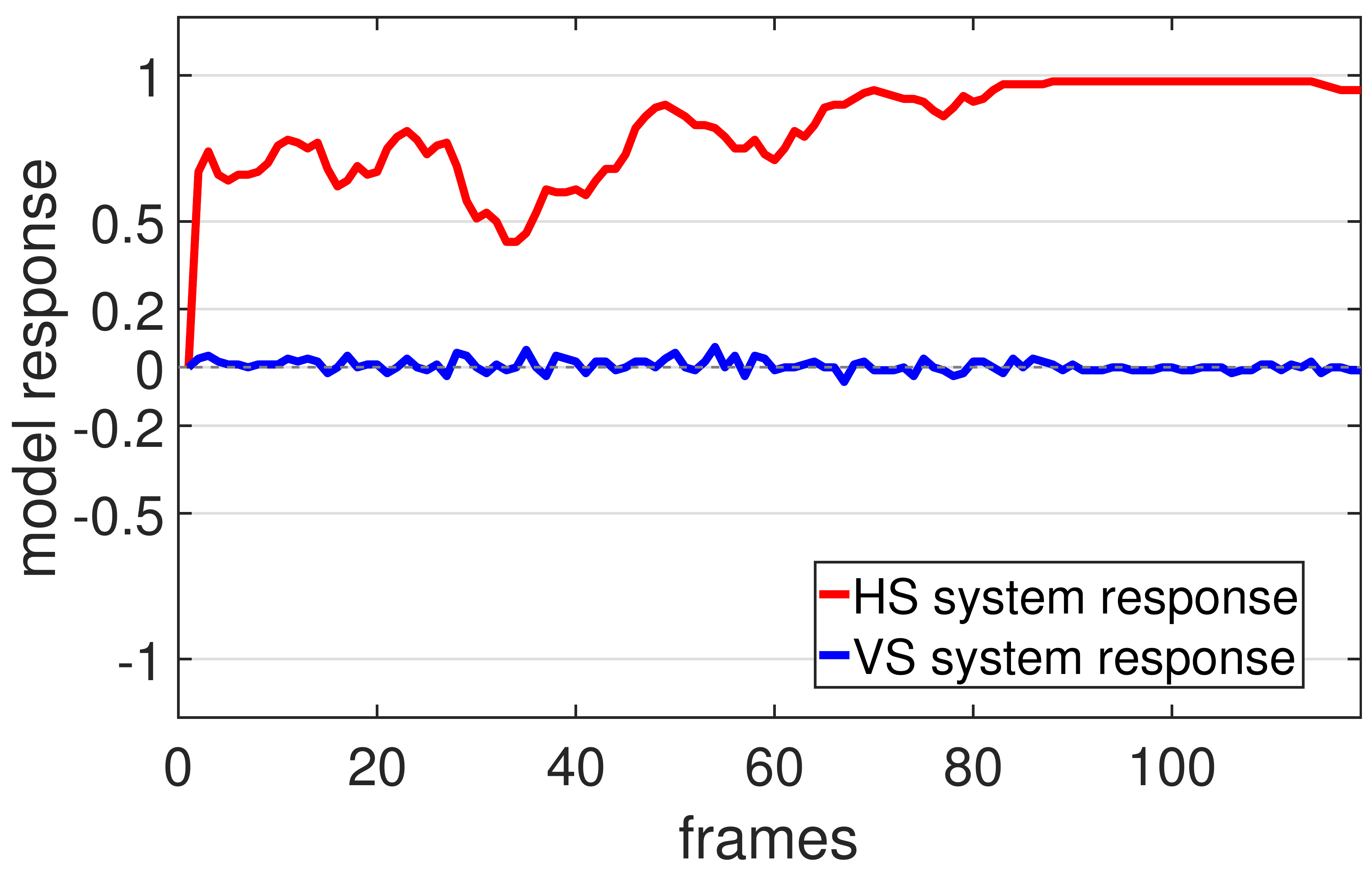}}
		\hfil
		\subfloat[model response (scene \#2)]{\includegraphics[width=0.49\textwidth]{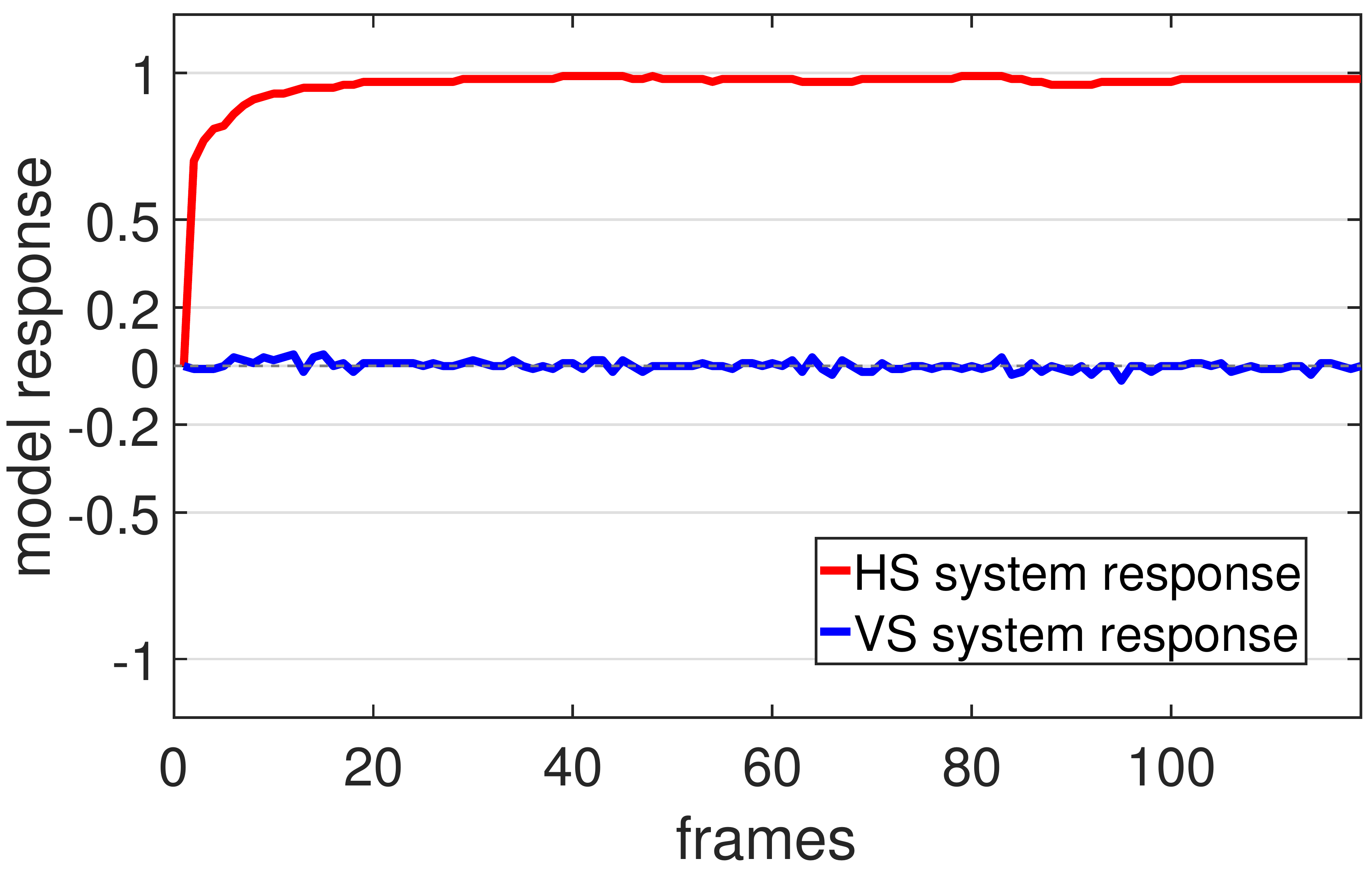}}
		\vfill
		\subfloat[$V_t=27$ degrees/s, $V_b=-20$ degrees/s]{\includegraphics[width=0.45\textwidth]{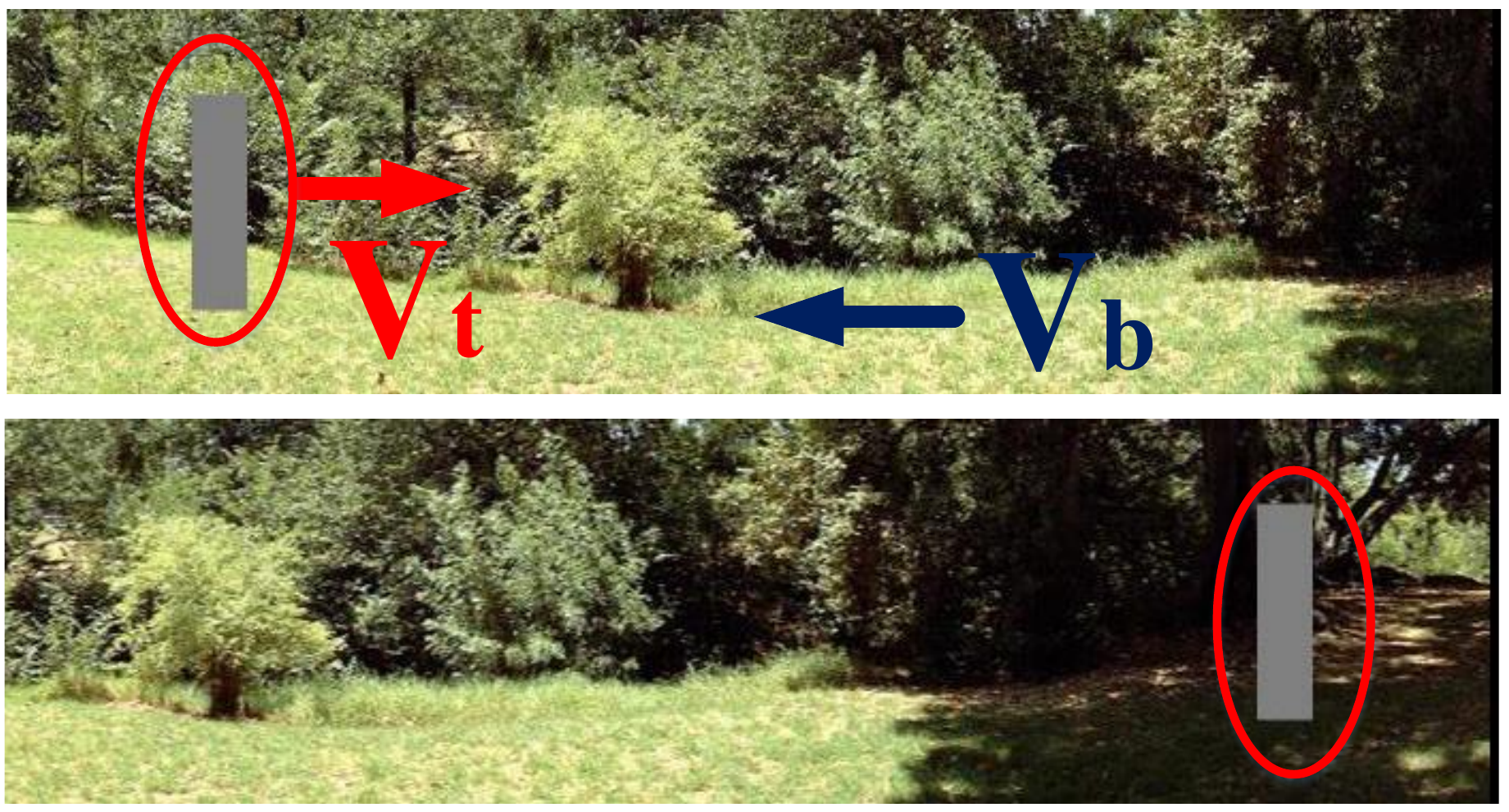}}
		\hfil
		\subfloat[$V_t=27$ degrees/s, $V_b=-20$ degrees/s]{\includegraphics[width=0.45\textwidth]{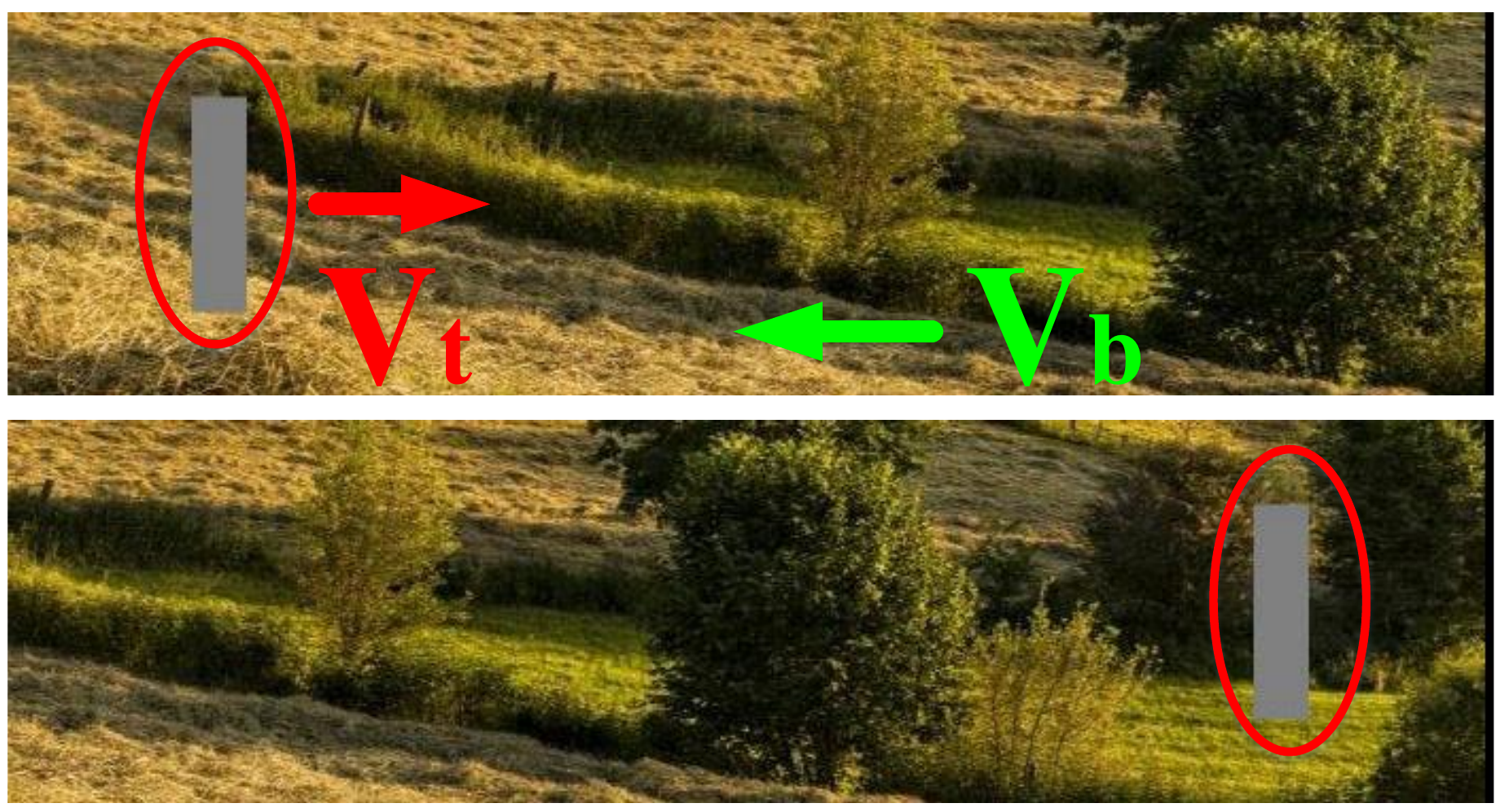}}
		\vfill
		\vspace{-10pt}
		\subfloat[model response (scene \#1)]{\includegraphics[width=0.49\textwidth]{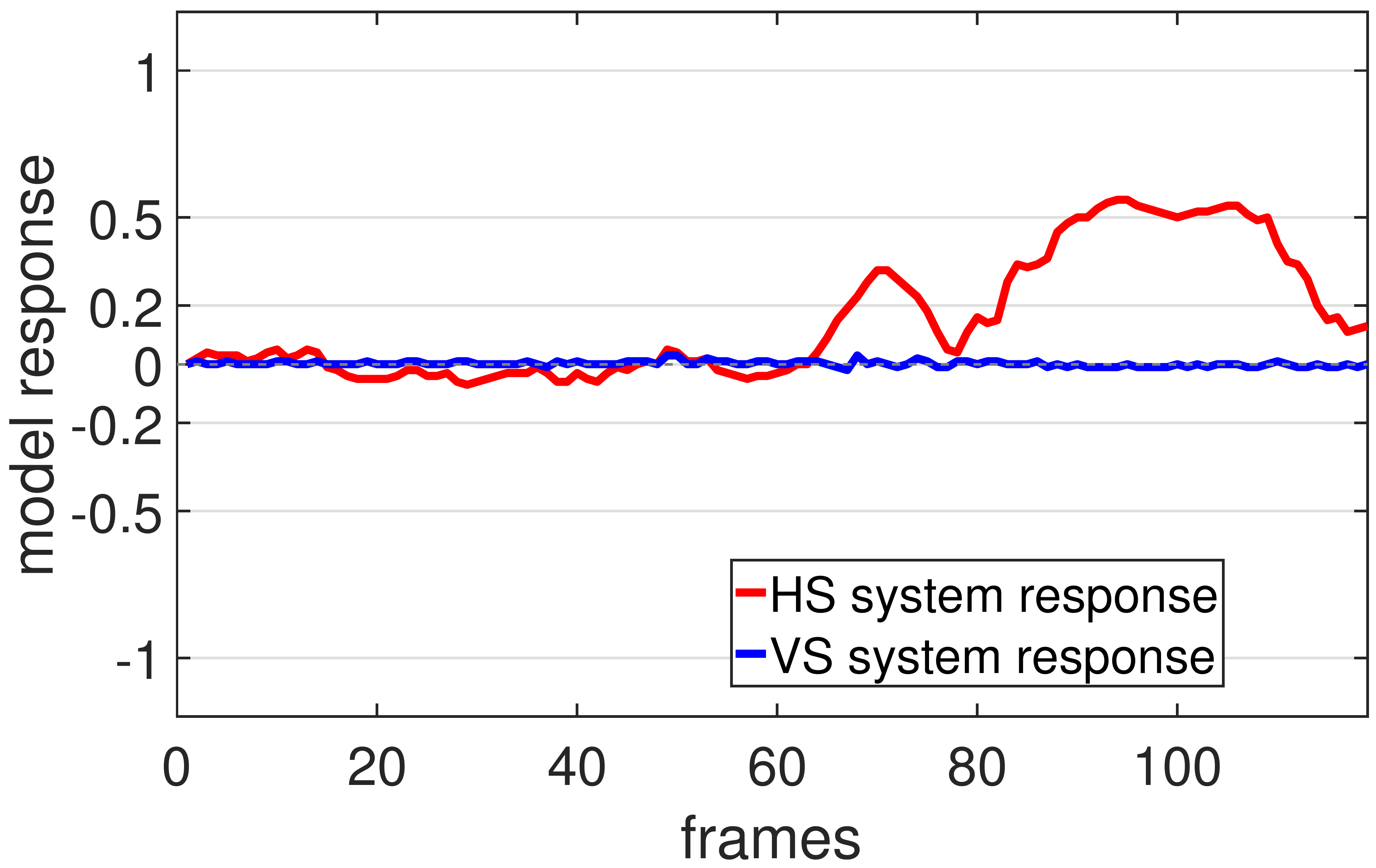}}
		\hfil
		\subfloat[model response (scene \#2)]{\includegraphics[width=0.49\textwidth]{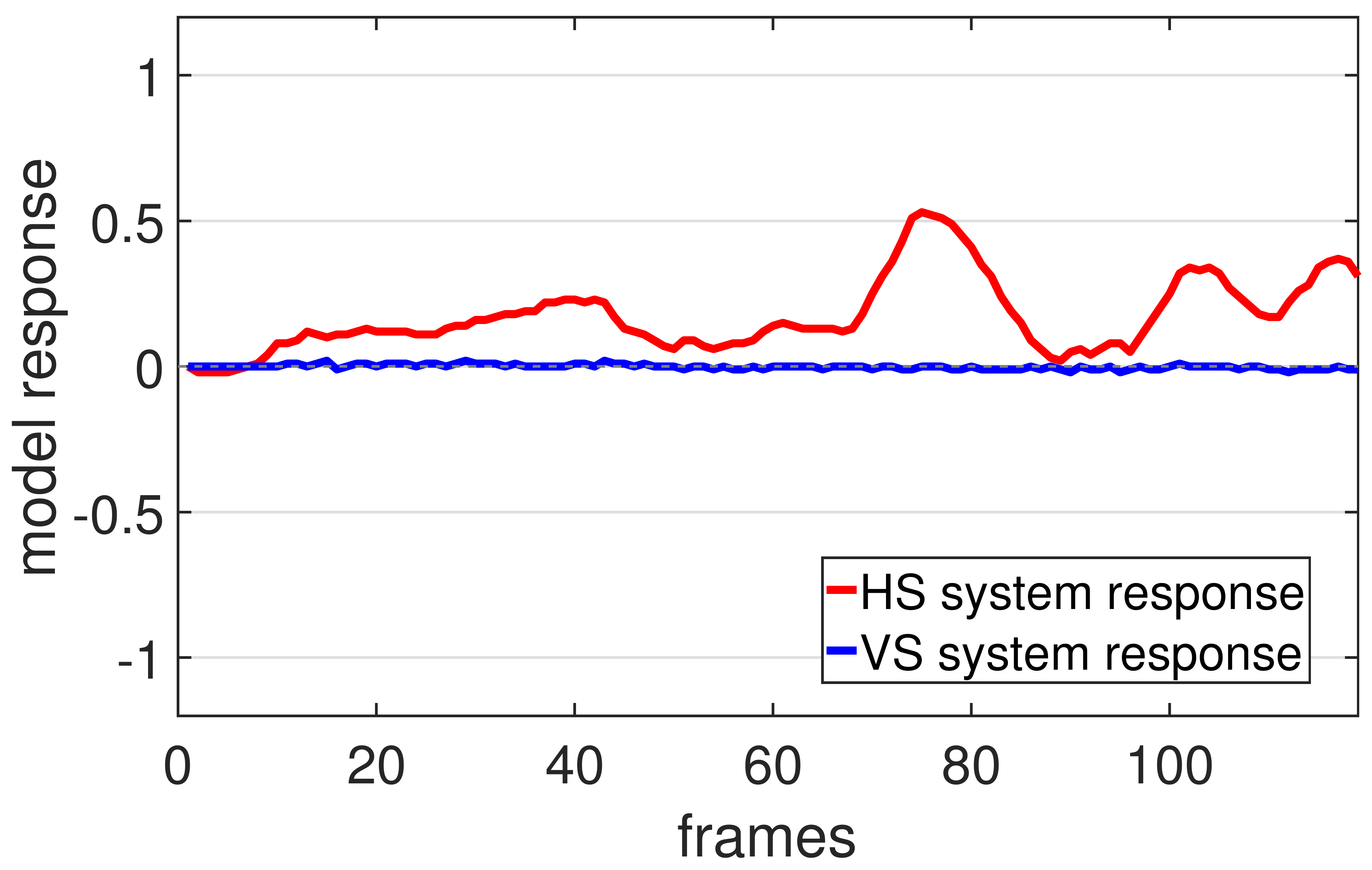}}
	\end{center}
	\caption{
		Proposed model responses challenged by two cluttered moving backgrounds. 
		The white, moderate and dark objects start from the left side and translate \textbf{rightward} at 27 degrees/second. 
		The cluttered backgrounds shift \textbf{leftward} at -20 degrees/second. 
		The red ellipses mark the start and end positions. 
		The arrows indicate the ground truth directions of foreground objects($V_t$) and moving backgrounds($V_b$).
	}
	\label{Fig: Fig-offline-moving-clutter-translating-tests}
\end{figure}

After that, more systematic experiments are carried on with cluttered moving backgrounds, in which the angular velocities of both the foreground translating objects and the shifting backgrounds are manually controlled. 
Fig. \ref{Fig: Fig-offline-moving-clutter-translating-tests} illustrates the model responses by three grey-scale objects moving in front of two shifting natural backgrounds, respectively. 
We have the following observations: 
\begin{enumerate}
	\item The proposed model is effective to decode the direction of translating objects against cluttered moving backgrounds: only the HS system is activated representing positive response to PD translational motion in front of the ND shifting backgrounds; whilst the VS system is rigorously suppressed.
	\item The model is sensitive to the contrast between translating objects and backgrounds: the white object, with relatively larger contrast to the moving backgrounds, leads to more constant and stronger responses; whilst the moderate object, with relatively smaller contrast to the moving backgrounds, brings about weaker responses. 
	Moreover, the model is not responding to the dark or moderate object translating in front of the background with little contrast, e.g., the dark object moving into the shadowed area.
\end{enumerate}

\begin{figure}[t!]
	\begin{center}
		\subfloat[scene \#1]{\includegraphics[width=0.33\textwidth]{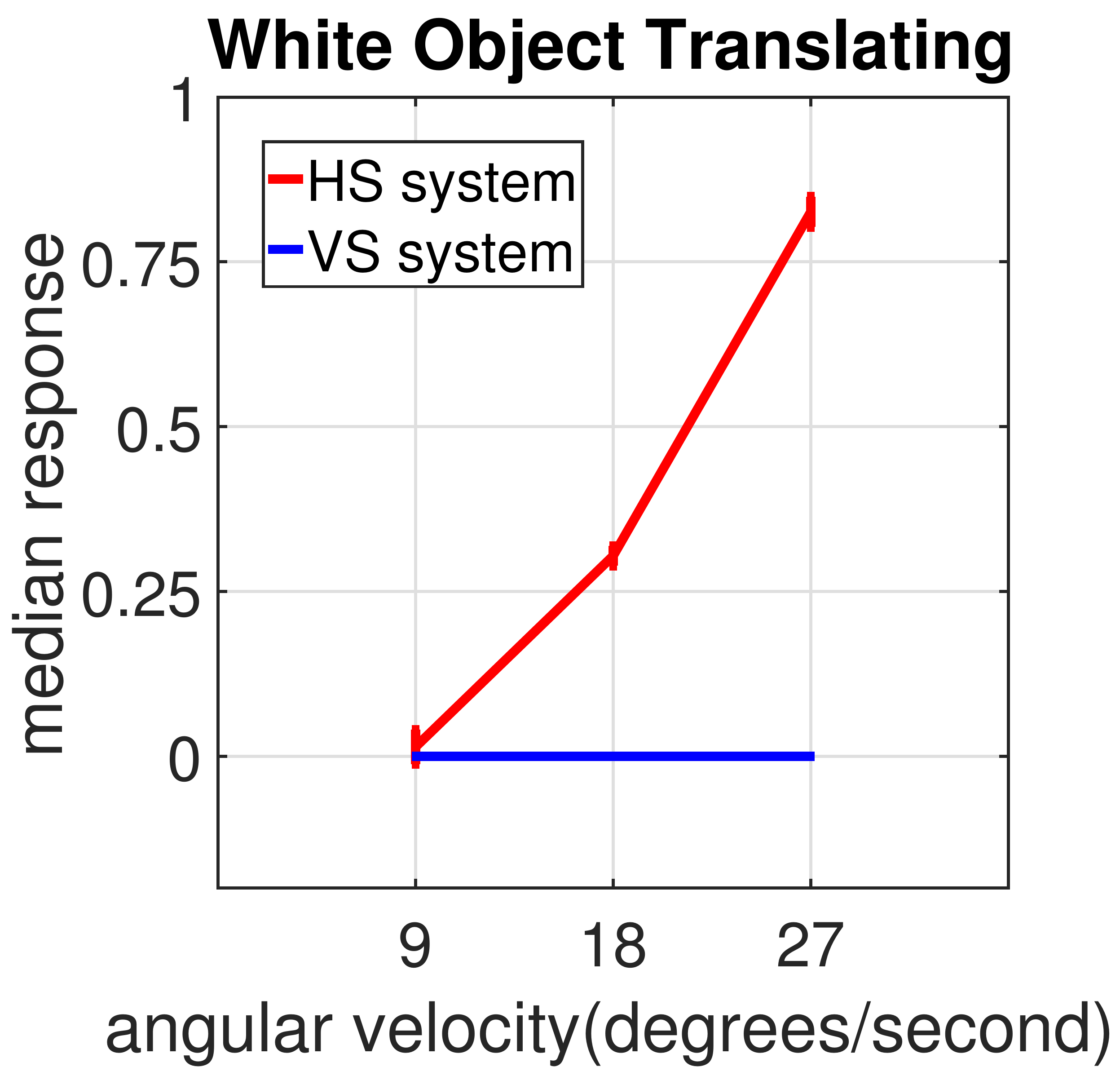}}
		\hfil
		\subfloat[scene \#1]{\includegraphics[width=0.33\textwidth]{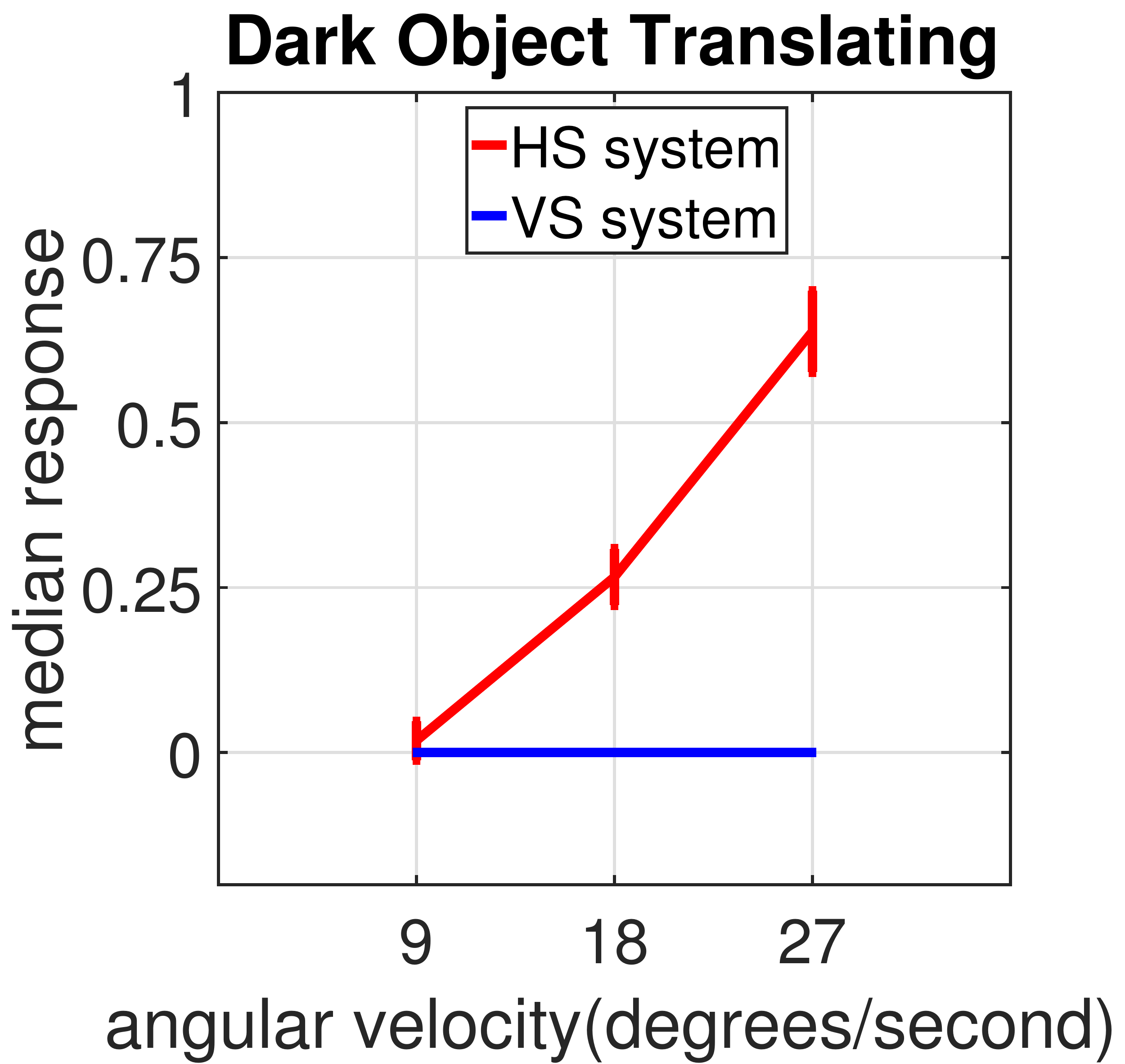}}
		\hfil
		\subfloat[scene \#1]{\includegraphics[width=0.33\textwidth]{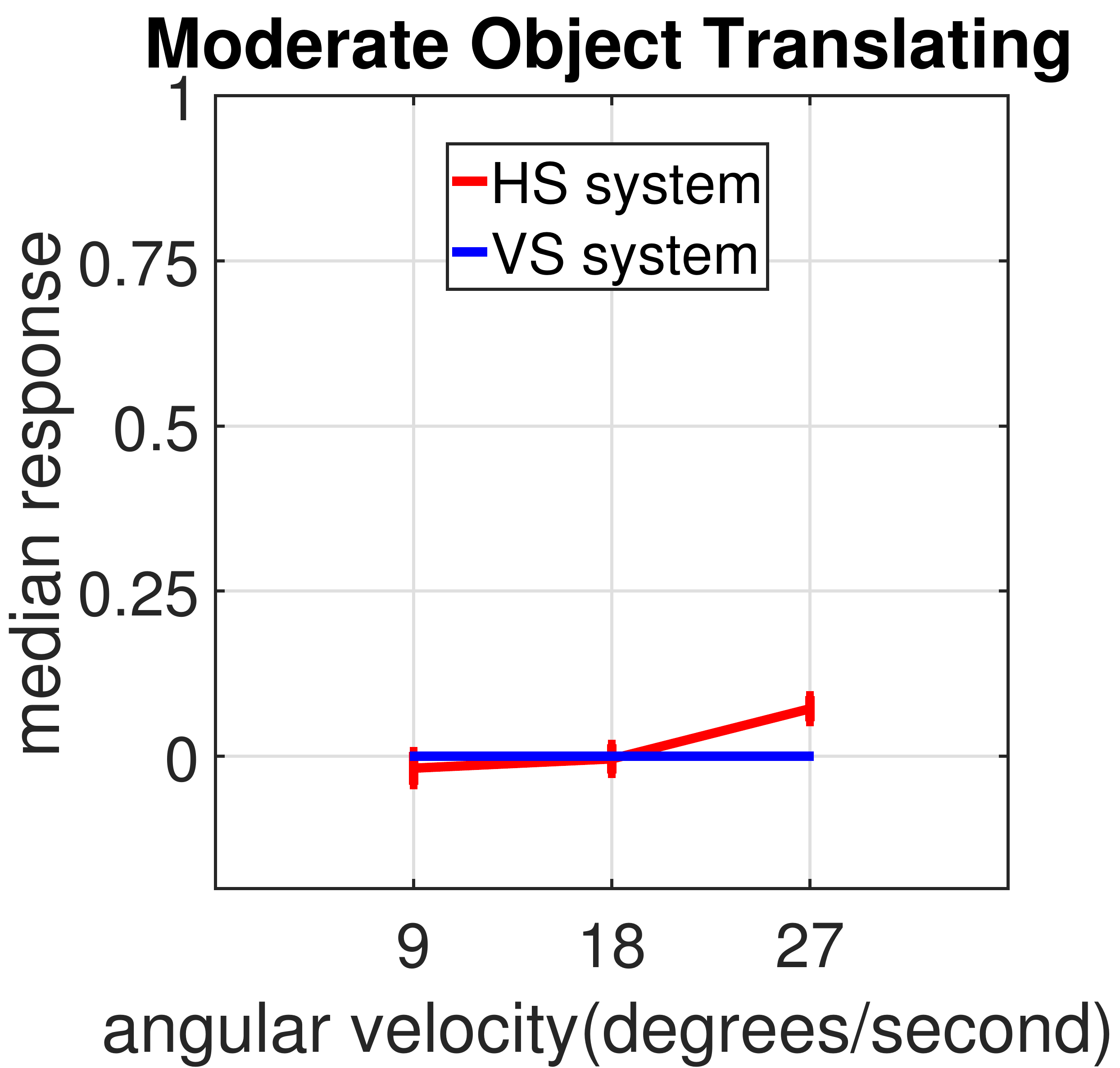}}
		\vfill
		\subfloat[scene \#2]{\includegraphics[width=0.33\textwidth]{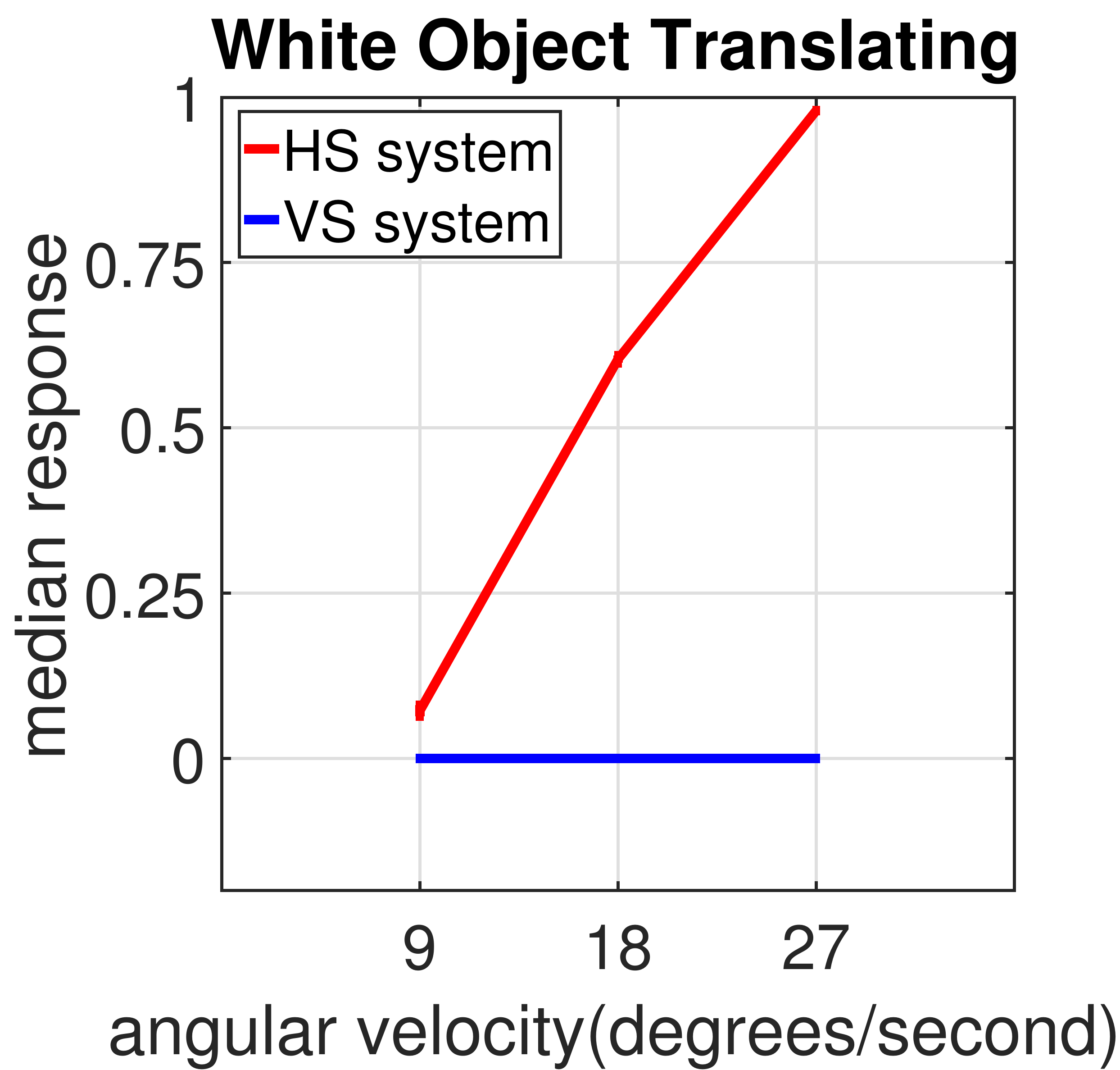}}
		\hfil
		\subfloat[scene \#2]{\includegraphics[width=0.33\textwidth]{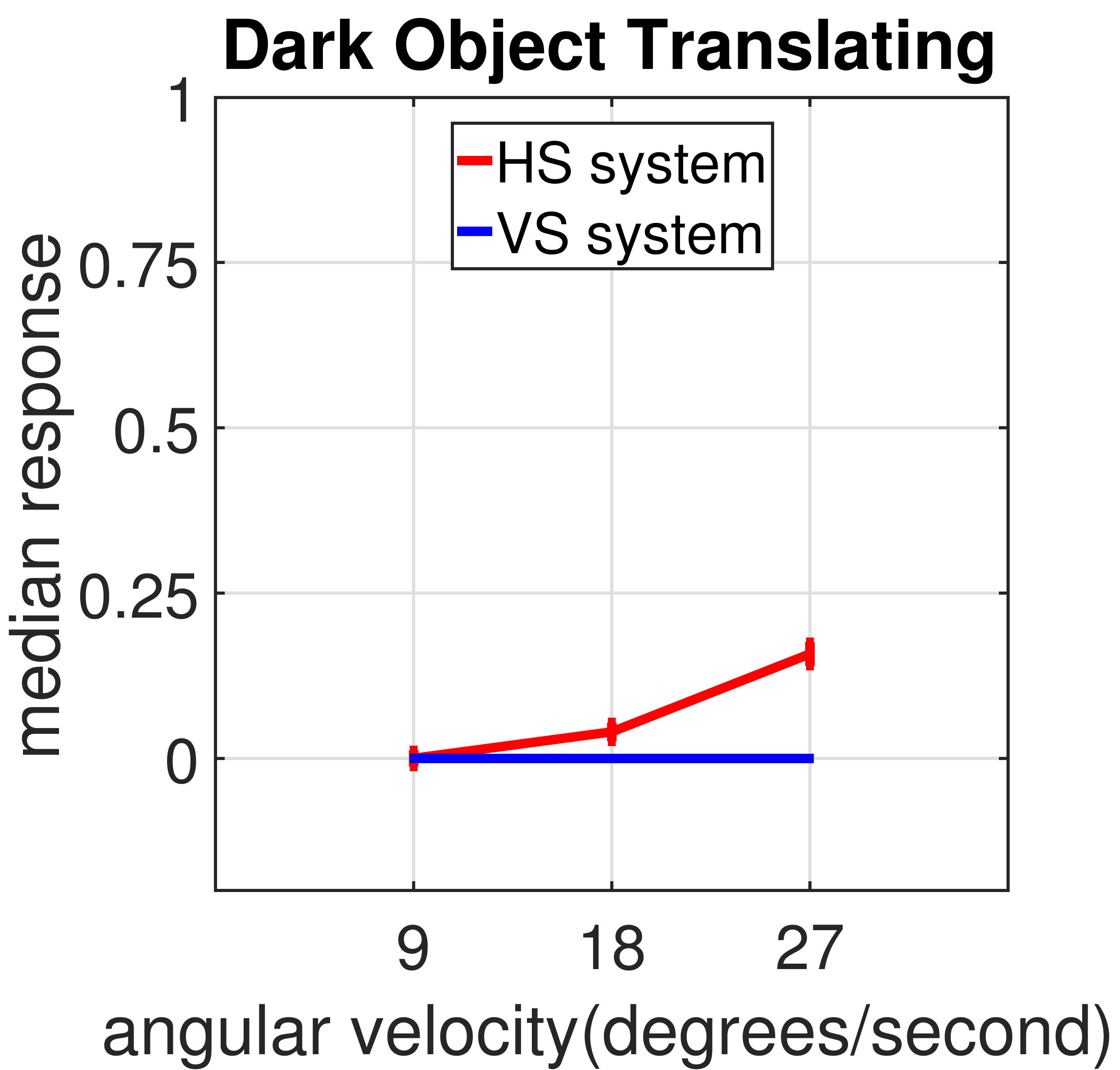}}
		\hfil
		\subfloat[scene \#2]{\includegraphics[width=0.33\textwidth]{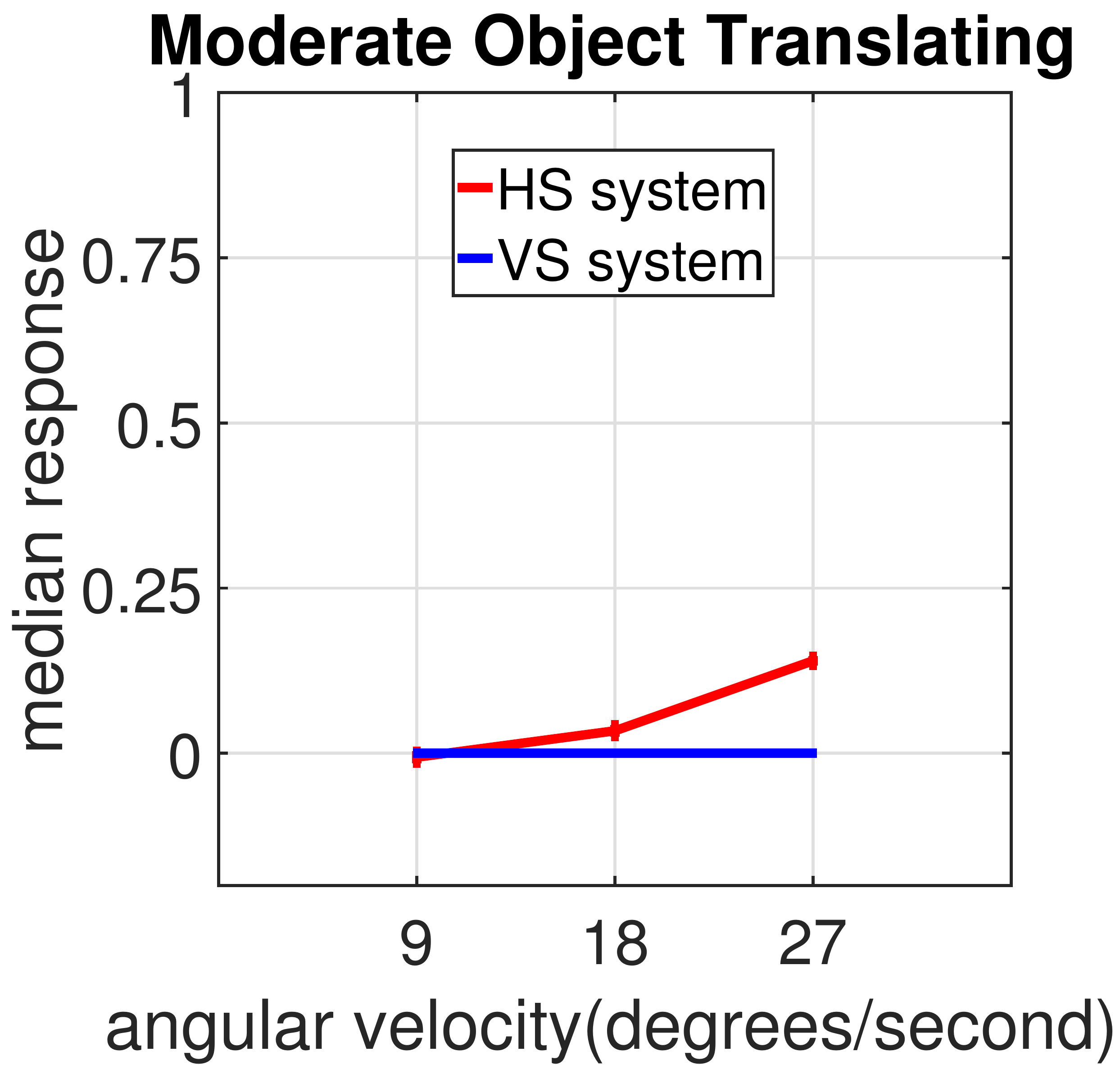}}
	\end{center}
	\caption{
		Statistical results of the median responses of HS and VS systems with variance and mean information, challenged by the three grey-scale objects translating at three individual angular velocities, in front of the two moving backgrounds each shifting at a range of angular velocities (-5, -10, -20, -30, -40 degrees/s).
	}
	\label{Fig: hs-vs-stats}
\end{figure}

\begin{figure}[t!]
	\begin{center}
		\subfloat[dark object stimuli in scene \#1]{\includegraphics[width=0.49\textwidth]{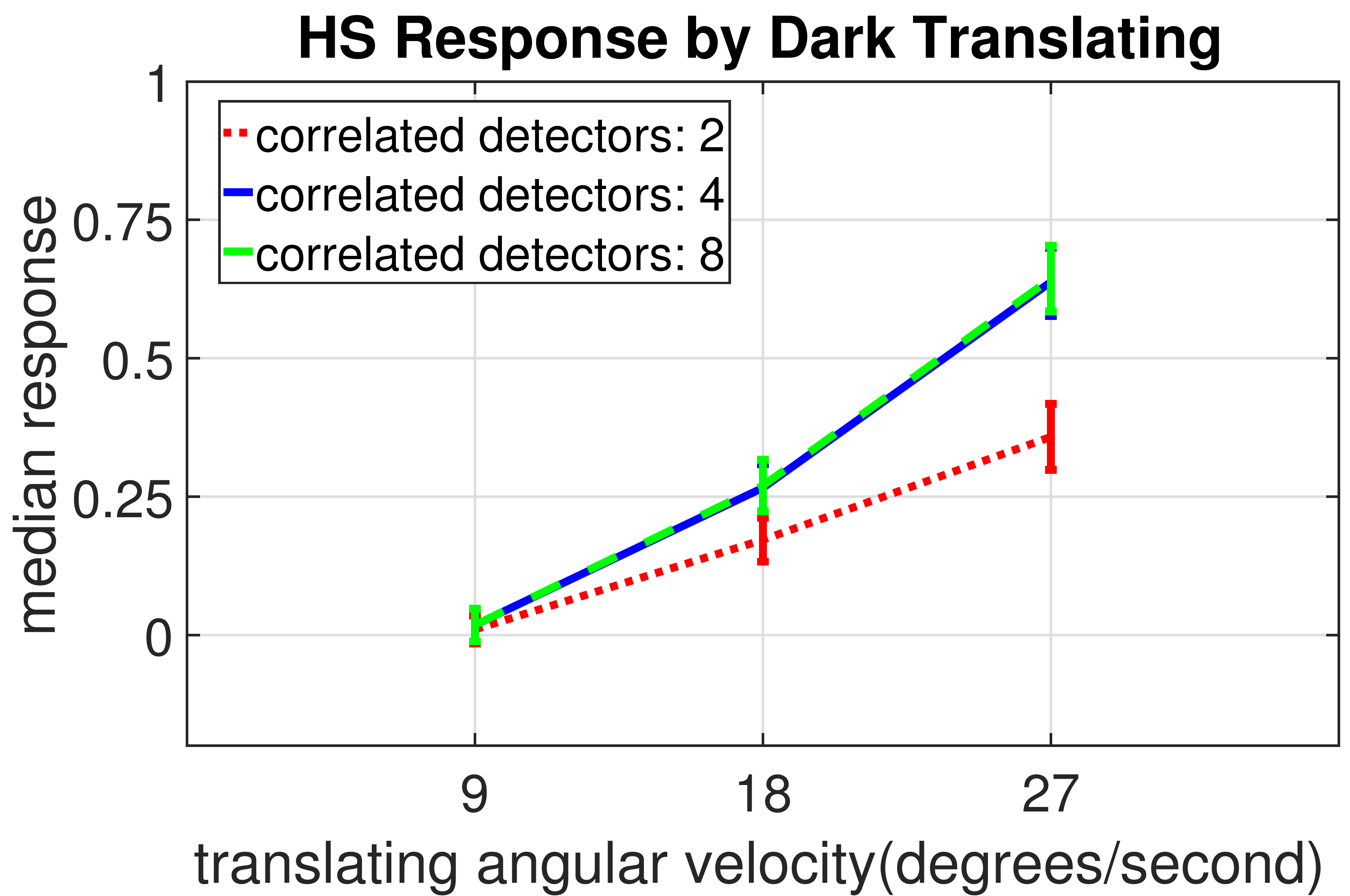}}
		\hfil
		\subfloat[white object stimuli in scene \#1]{\includegraphics[width=0.49\textwidth]{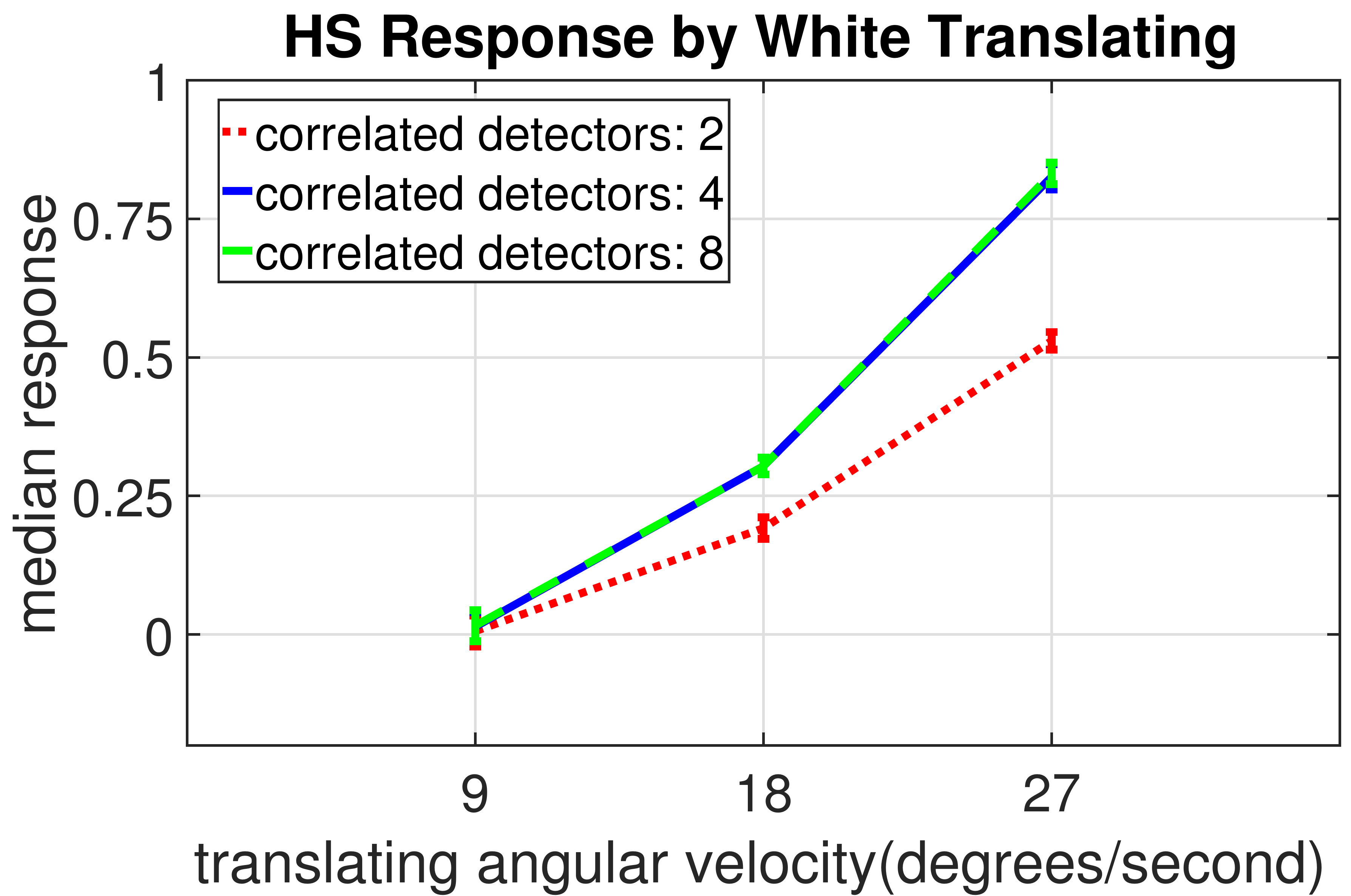}}
	\end{center}
	\caption{
		Results of investigation on the different number of correlated motion detectors inside the ON and OFF channels, under the same stimuli settings in Fig. \ref{Fig: hs-vs-stats}.
	}
	\label{Fig: ensemble-parameter-investigation}
\end{figure}

Subsequently, we compare the dynamic response in speed tuning between the three grey-scale translating objects, against the two cluttered moving backgrounds. 
Based on the experimental setting introduced in Section \ref{Sec: setting}, a range of angular velocities for both the foreground targets and the backgrounds are investigated. 
Fig. \ref{Fig: hs-vs-stats} illustrates the statistical results. 
Intuitively, the HS system responds more strongly to the PD translating stimuli at faster speeds; while the VS system maintains inactive in all the tests. 
Importantly, challenged by the ND moving backgrounds at a range of angular velocities, little variance is shown at all tested foreground translation speeds and contrasts, which indicates the proposed model performs robustly and consistently to decode the direction of translation in front of cluttered moving backgrounds. 
The irrelevant background translational OF mixed with distractors, such as the woods, have been satisfactorily suppressed, which is a significant achievement of this modelling research. 
Moreover, the model represents a broader dynamic range on the larger-contrast moving target which matches the above observations in Fig. \ref{Fig: Fig-offline-moving-clutter-translating-tests}.

\begin{figure}[t!]
	\begin{center}
		\subfloat[$V_t=27$ degrees/s, $V_b=-20$ degrees/s]{\includegraphics[width=0.45\textwidth]{Fig-moving-background-dark-scene1.pdf}}
		\hfil
		\subfloat[$V_t=27$ degrees/s, $V_b=-20$ degrees/s]{\includegraphics[width=0.45\textwidth]{Fig-moving-background-dark-scene2.pdf}}
		\vfill
		%		\vspace{-10pt}
		\subfloat[model response (scene \#1)]{\includegraphics[width=0.49\textwidth]{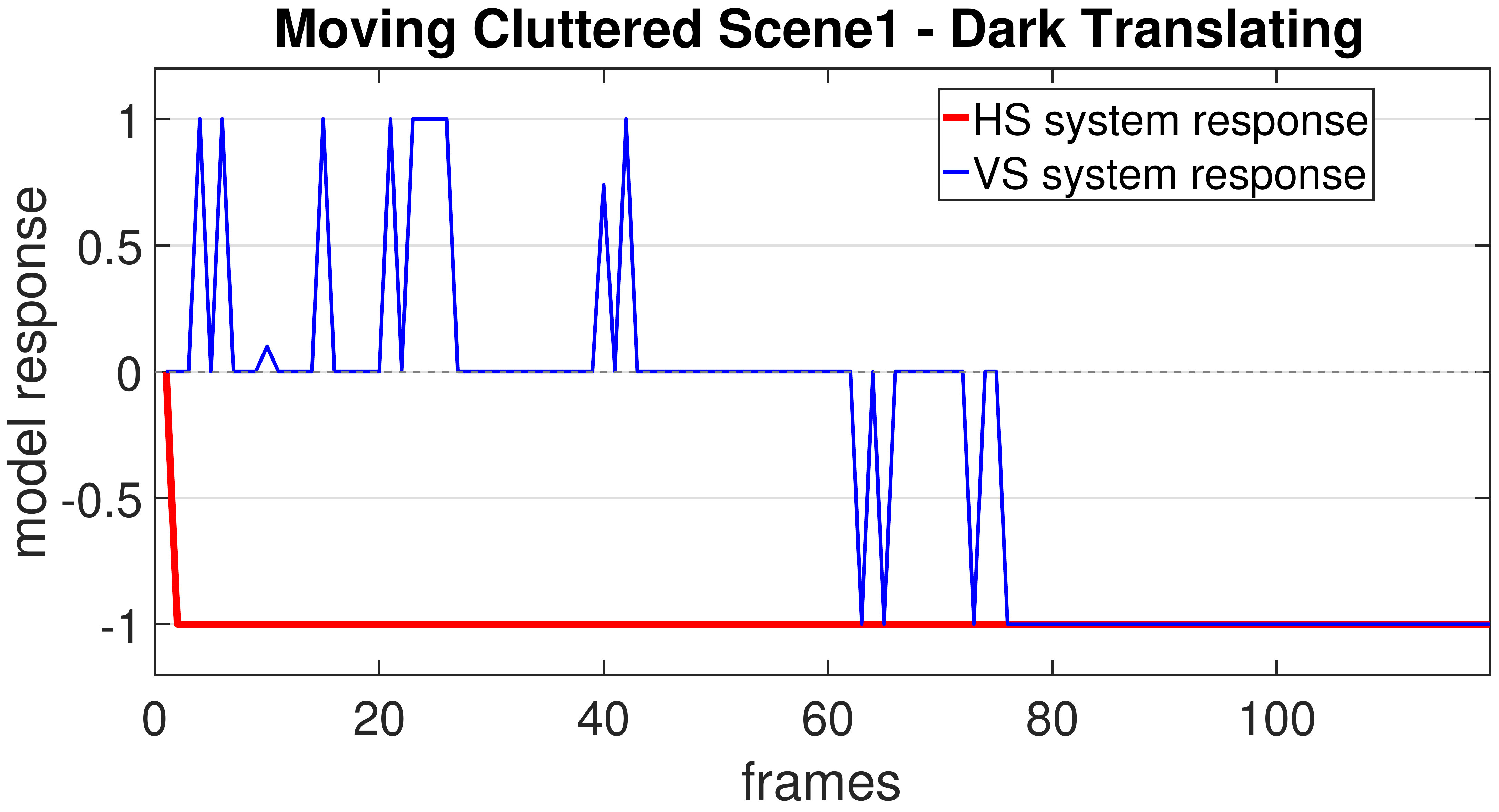}}
		\hfil
		\subfloat[model response (scene \#2)]{\includegraphics[width=0.49\textwidth]{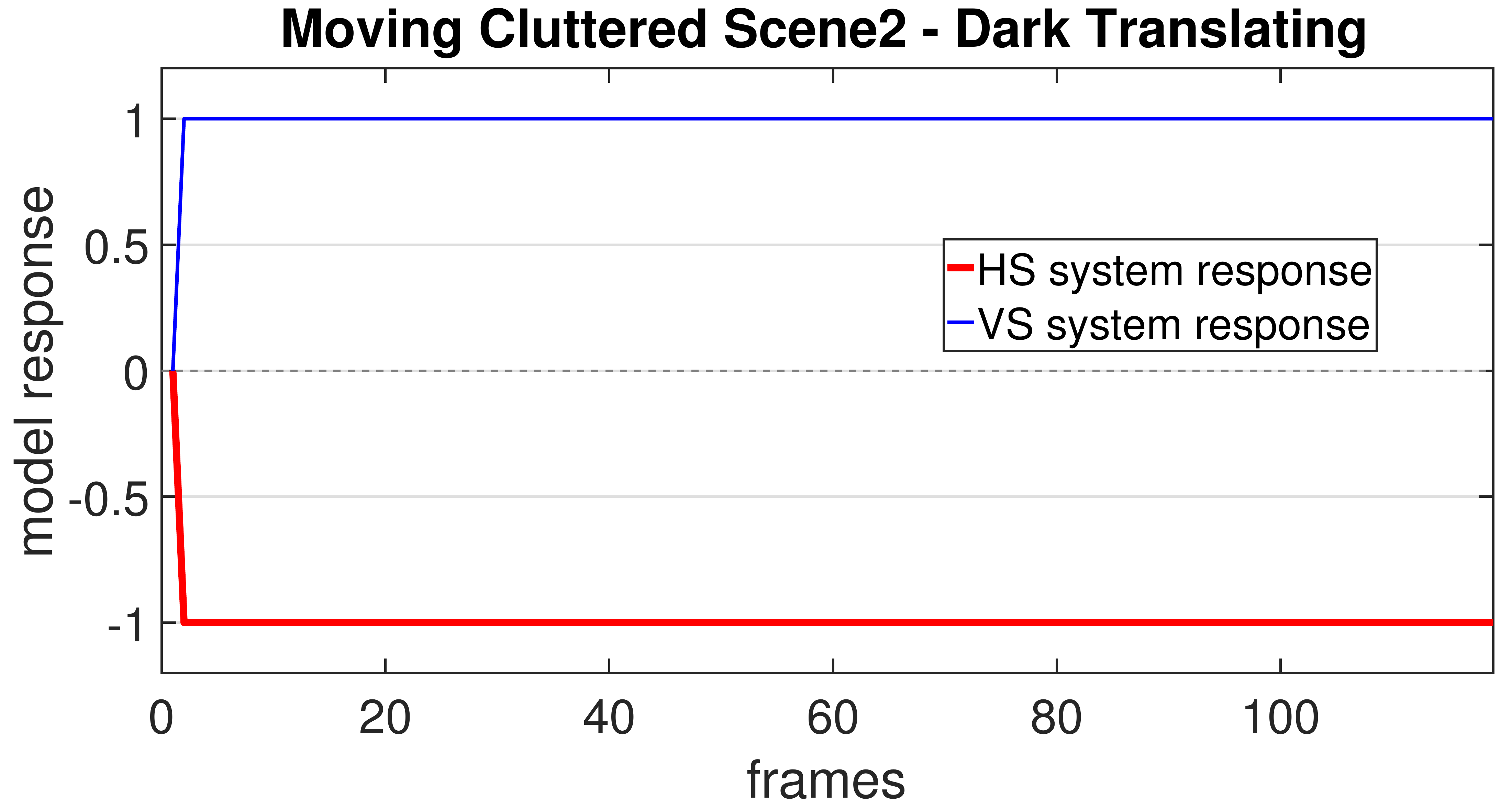}}
		\vfill
		\subfloat[$V_t=27$ degrees/s, $V_b=-20$ degrees/s]{\includegraphics[width=0.45\textwidth]{Fig-moving-background-white-scene1.pdf}}
		\hfil
		\subfloat[$V_t=27$ degrees/s, $V_b=-20$ degrees/s]{\includegraphics[width=0.45\textwidth]{Fig-moving-background-white-scene2.pdf}}
		\vfill
		%		\vspace{-10pt}
		\subfloat[model response (scene \#1)]{\includegraphics[width=0.49\textwidth]{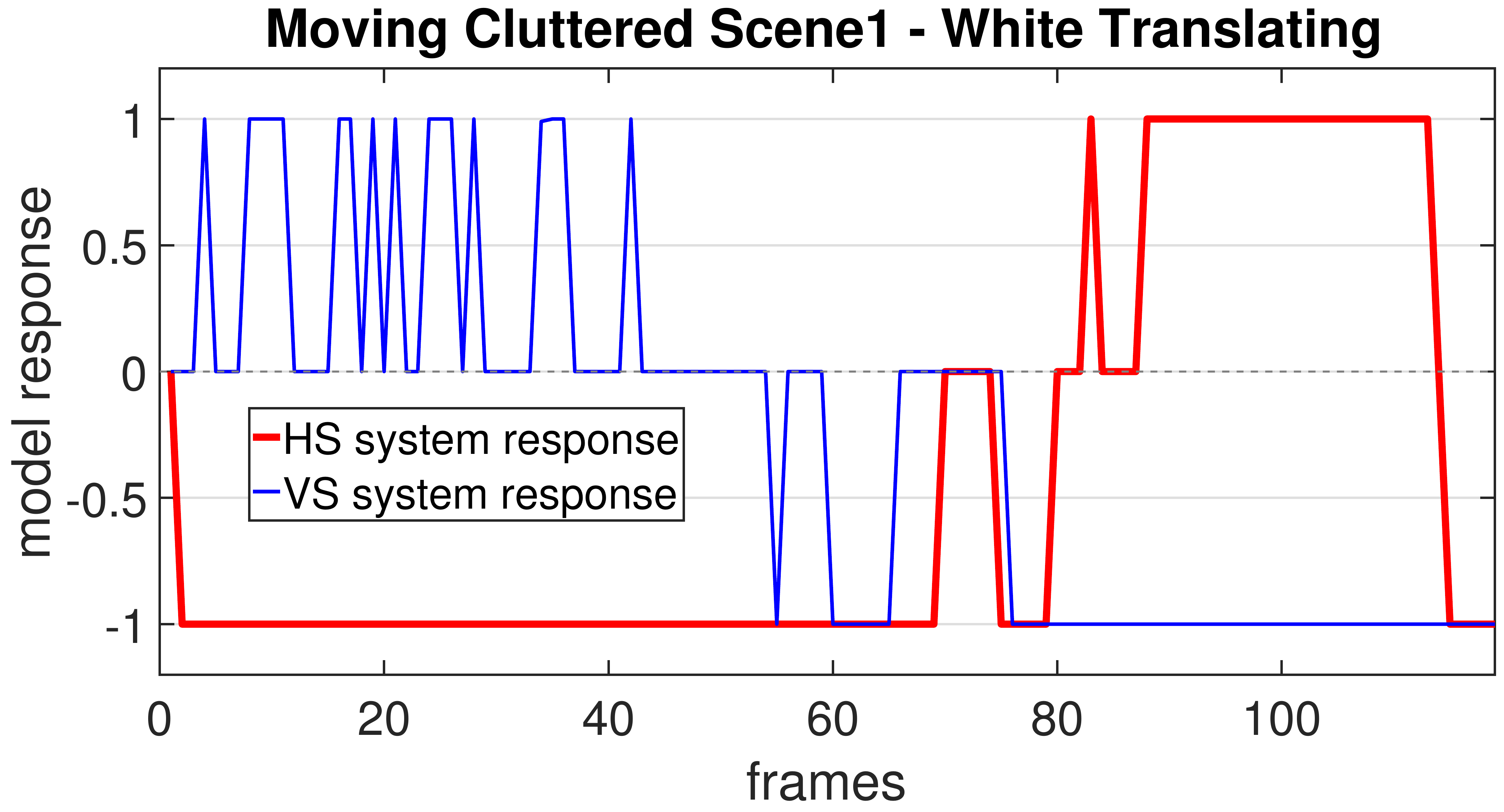}}
		\hfil
		\subfloat[model response (scene \#2)]{\includegraphics[width=0.49\textwidth]{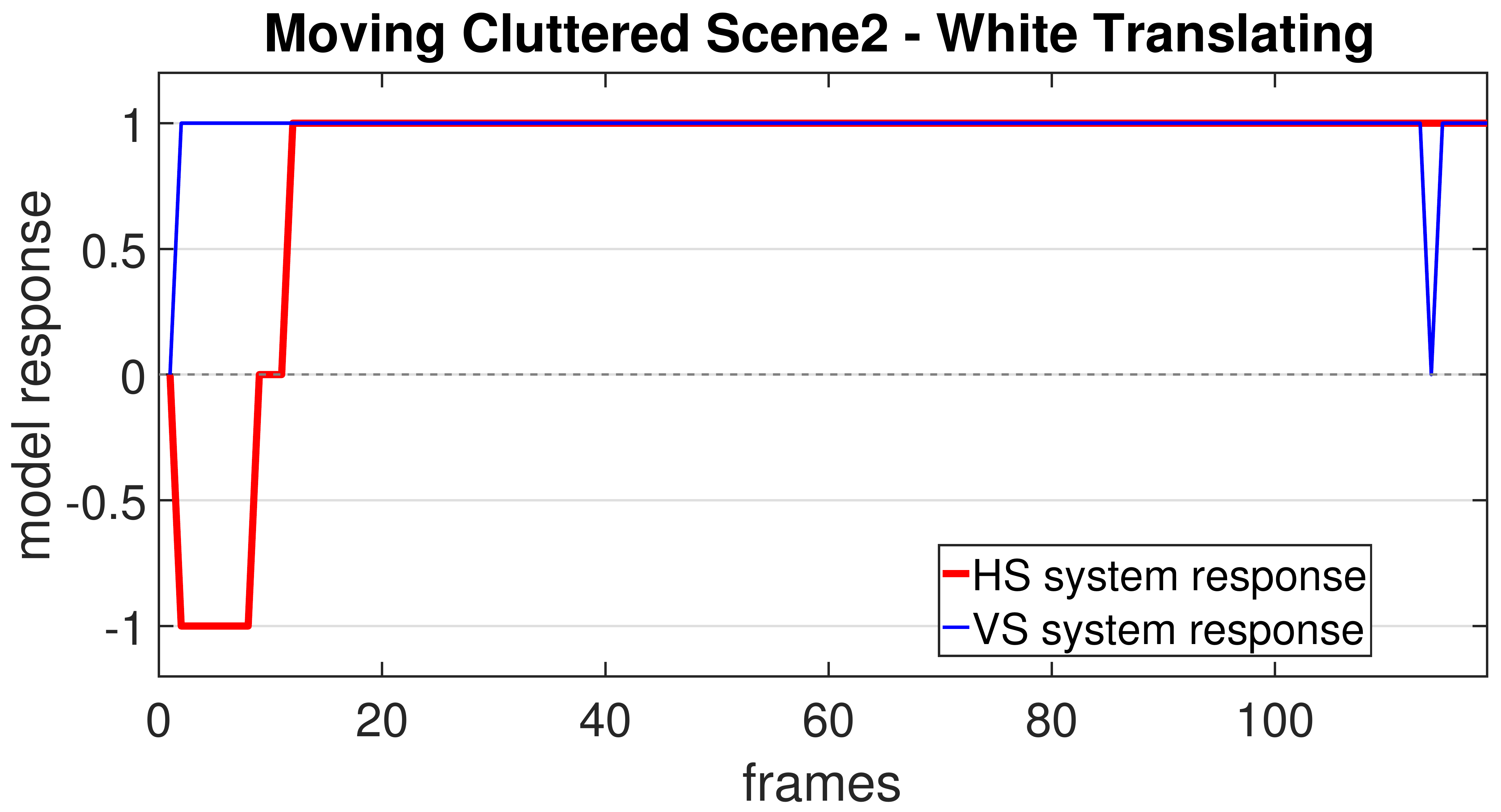}}
	\end{center}
	\caption{
		Proposed model responses without the combination of motion pre-filtering mechanisms including the vDoG and the FDSR, challenged by the two cluttered moving backgrounds. 
		The stimuli settings are in accordance with the Fig. \ref{Fig: Fig-offline-moving-clutter-translating-tests}.
	}
	\label{Fig: dynamics-investigation}
\end{figure}

\subsection{Investigations on Model Characteristics}

To provide insight into the significance of proposed new mechanisms or structures in decoding the direction of translating objects against cluttered moving backgrounds, we investigate the effectiveness of spatiotemporal dynamics in the proposed visual system model. 
Firstly, Fig. \ref{Fig: ensemble-parameter-investigation} demonstrates the effects of ensembles of ON-ON/OFF-OFF local motion correlators on the dynamic response in speed tuning ($n_c$ in Eq. \ref{Eq: medulla-ds} and \ref{Eq: lobula-ds}). 
The statistical results show that the dynamic response is stable and reflected by all the tested parameter and stimuli settings. 
The model is expected to respond more strongly to faster moving stimuli with more correlated detectors inside the ON and OFF channels. 
There is nevertheless little difference between $n_c = 4$ and $n_c = 8$, which indicates that $n_c = 4$ could be an optimal parameter set-up in our case. 
Such a structure can enhance the dynamic response by alleviating the impact by temporal frequency of visual stimuli though increasing the computational complexity.

\begin{figure}[t!]
	\begin{center}
		\subfloat[$V_t=27$ degrees/s, $V_b=-40$ degrees/s]{\includegraphics[width=0.45\textwidth]{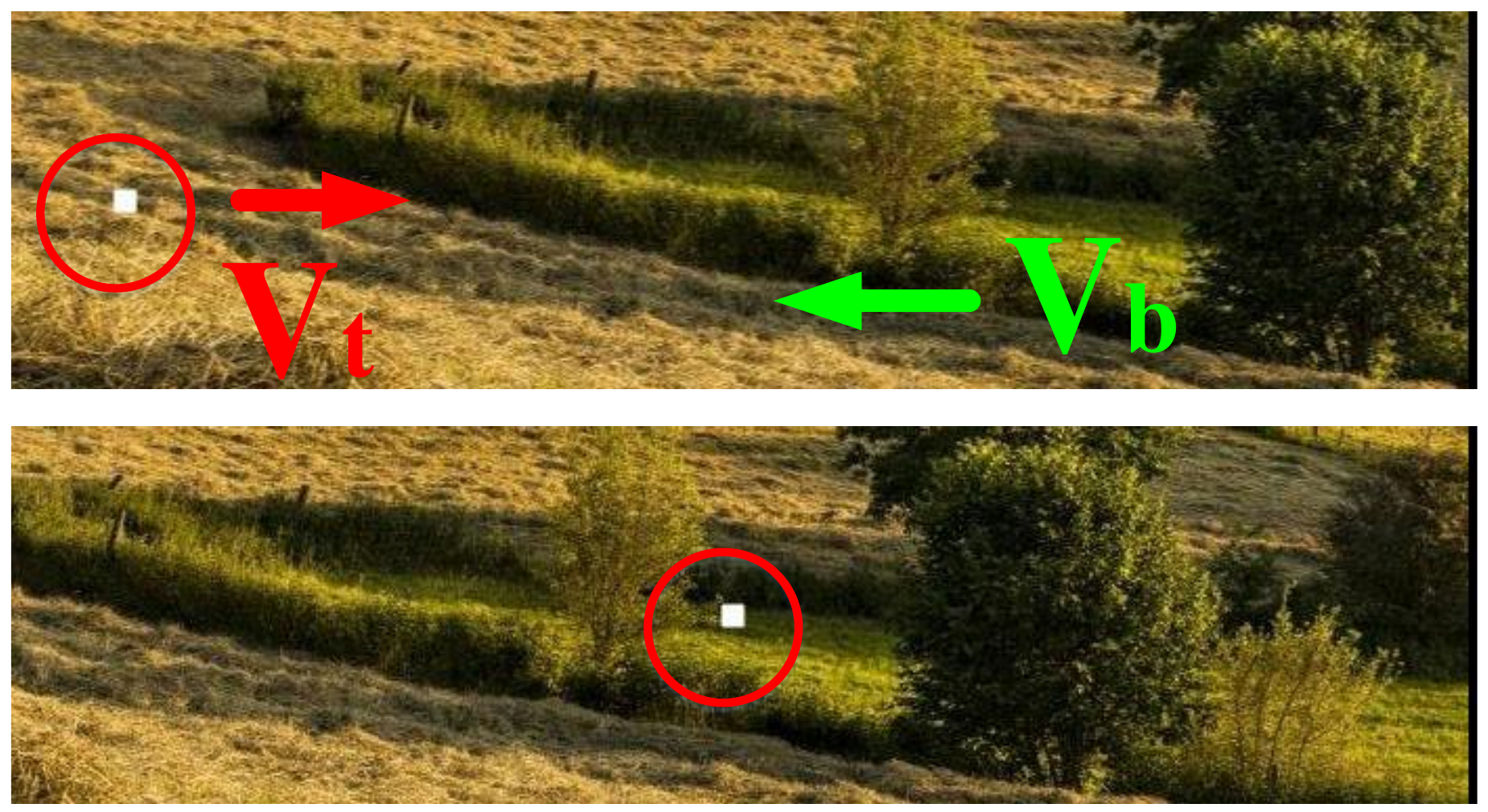}}
		\hfil
		\subfloat[$V_t=27$ degrees/s, $V_b=-40$ degrees/s]{\includegraphics[width=0.45\textwidth]{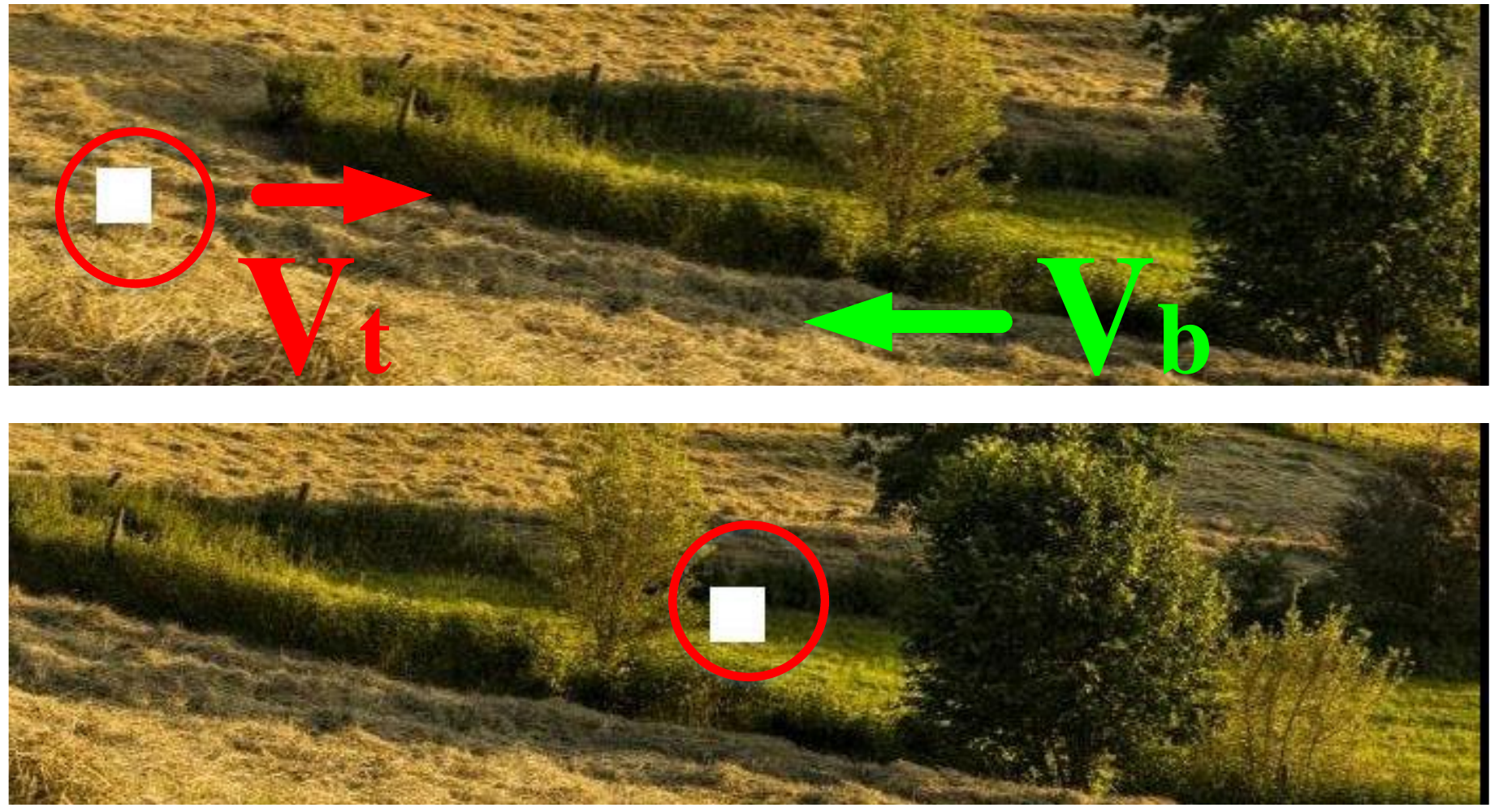}}
		\vfill
		%		\vspace{-10pt}
		\subfloat[model response (scene \#2)]{\includegraphics[width=0.49\textwidth]{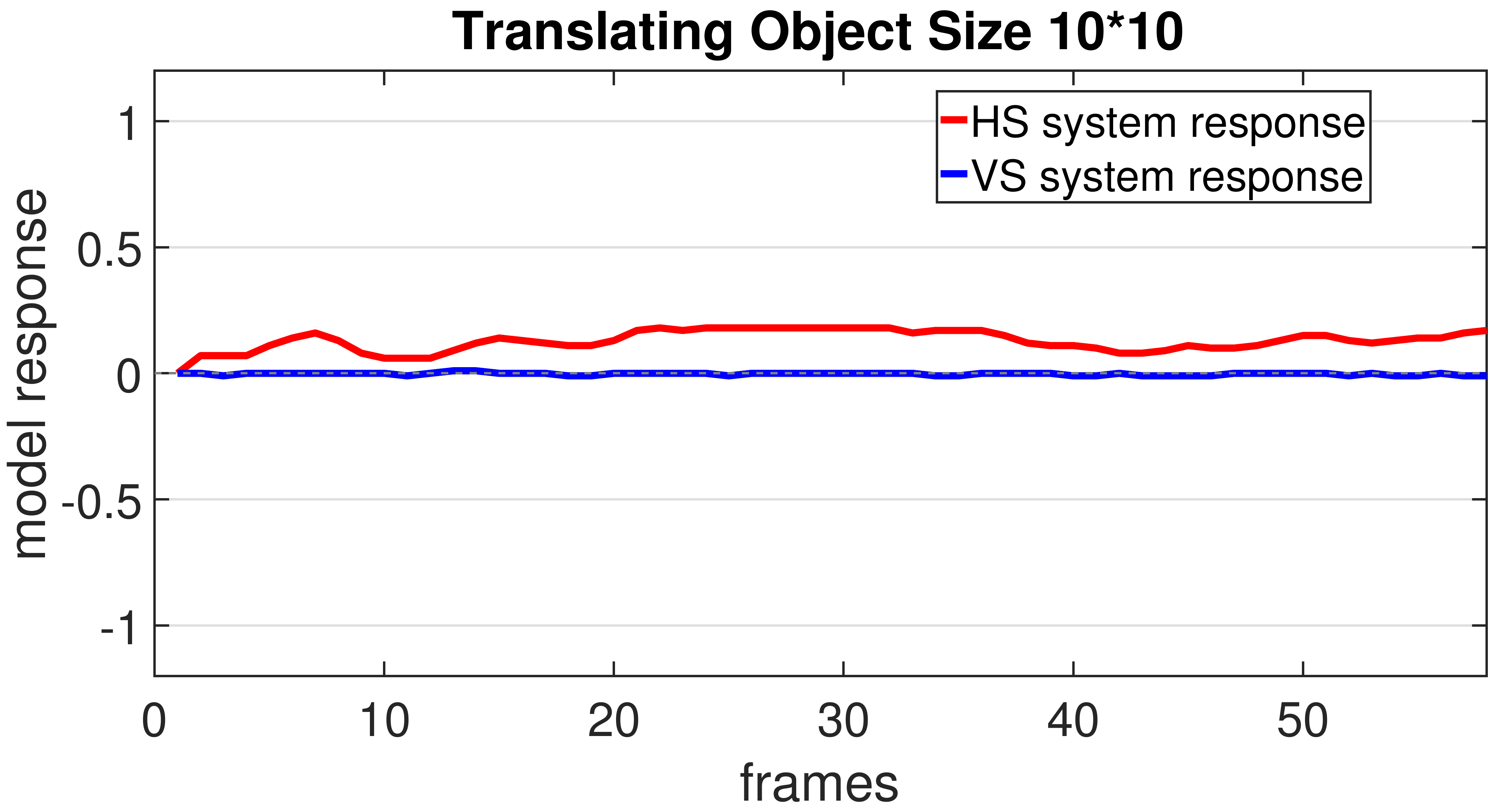}}
		\hfil
		\subfloat[model response (scene \#2)]{\includegraphics[width=0.49\textwidth]{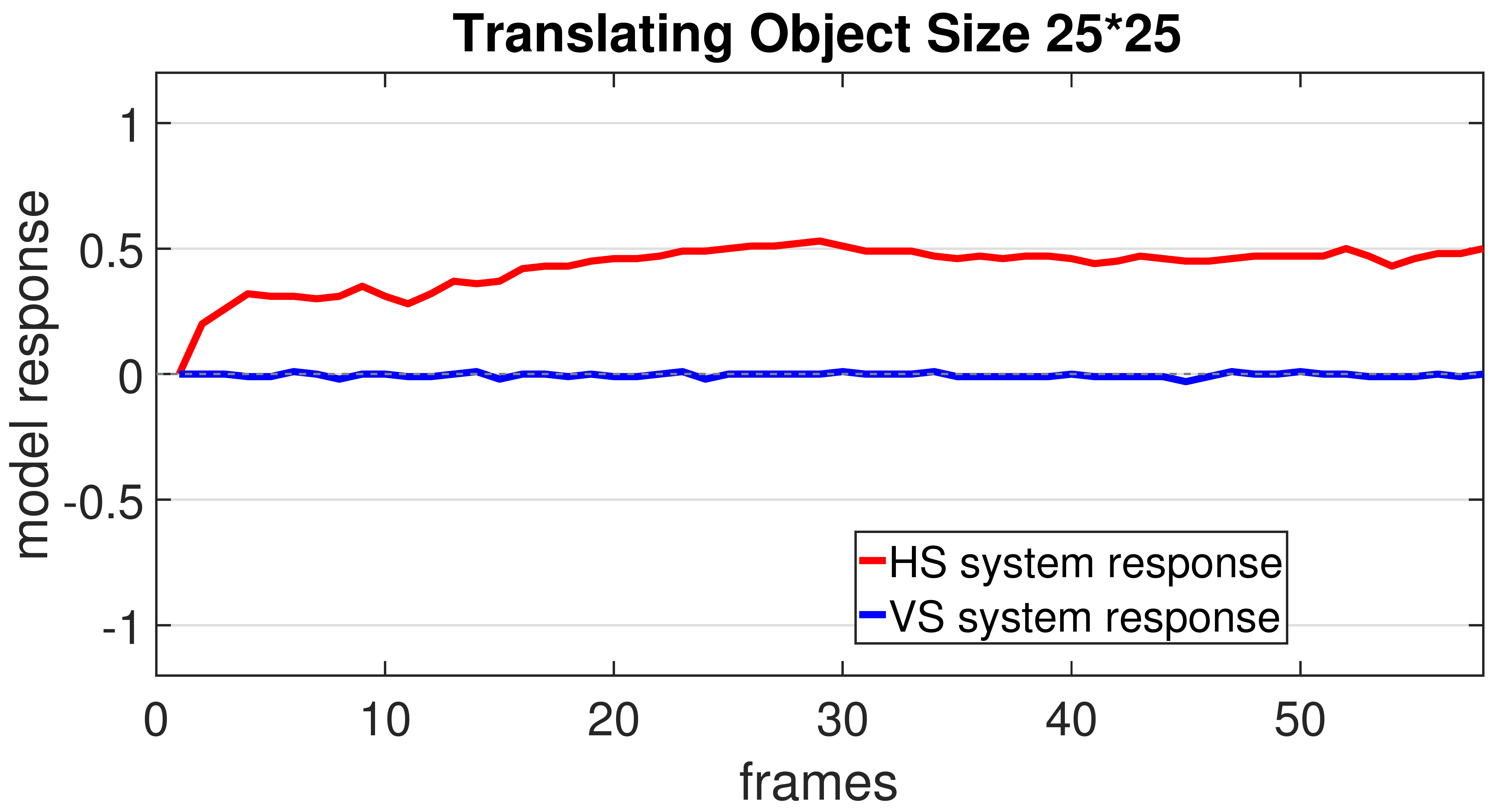}}
		\vfill
		\subfloat[$V_t=27$ degrees/s, $V_b=-40$ degrees/s]{\includegraphics[width=0.45\textwidth]{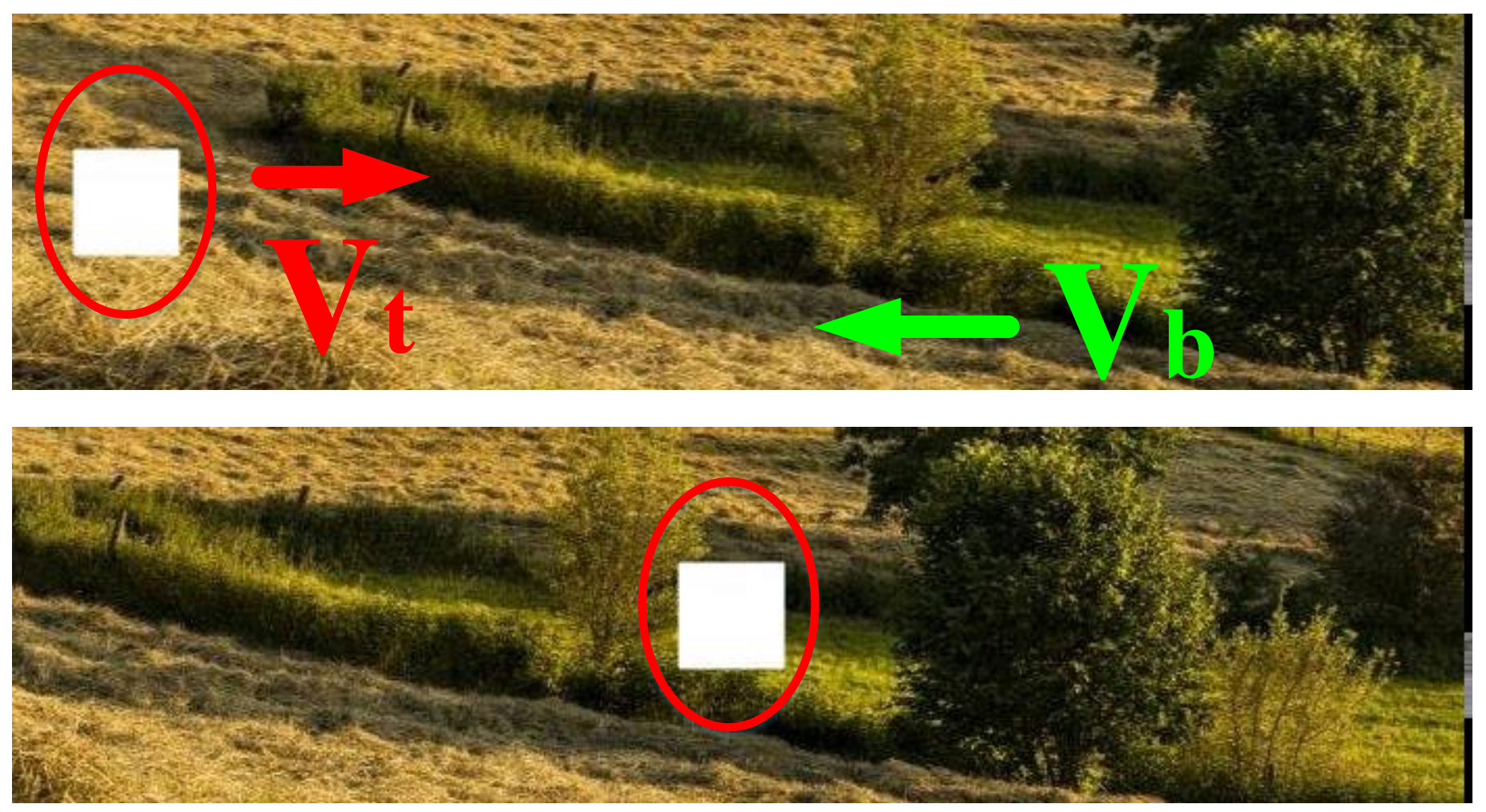}}
		\hfil
		\subfloat[$V_t=27$ degrees/s, $V_b=-40$ degrees/s]{\includegraphics[width=0.45\textwidth]{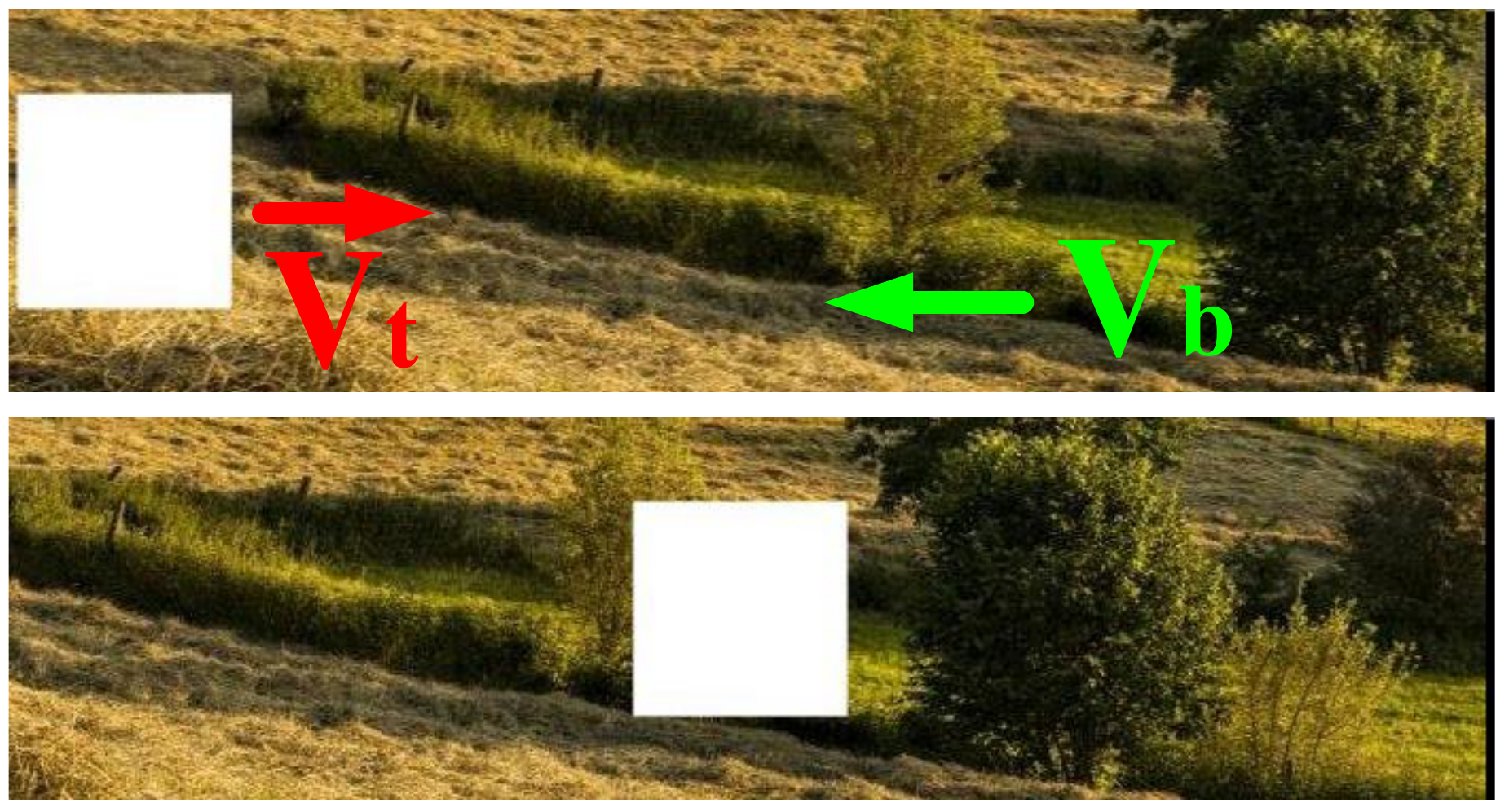}}
		\vfill
		%		\vspace{-10pt}
		\subfloat[model response (scene \#2)]{\includegraphics[width=0.49\textwidth]{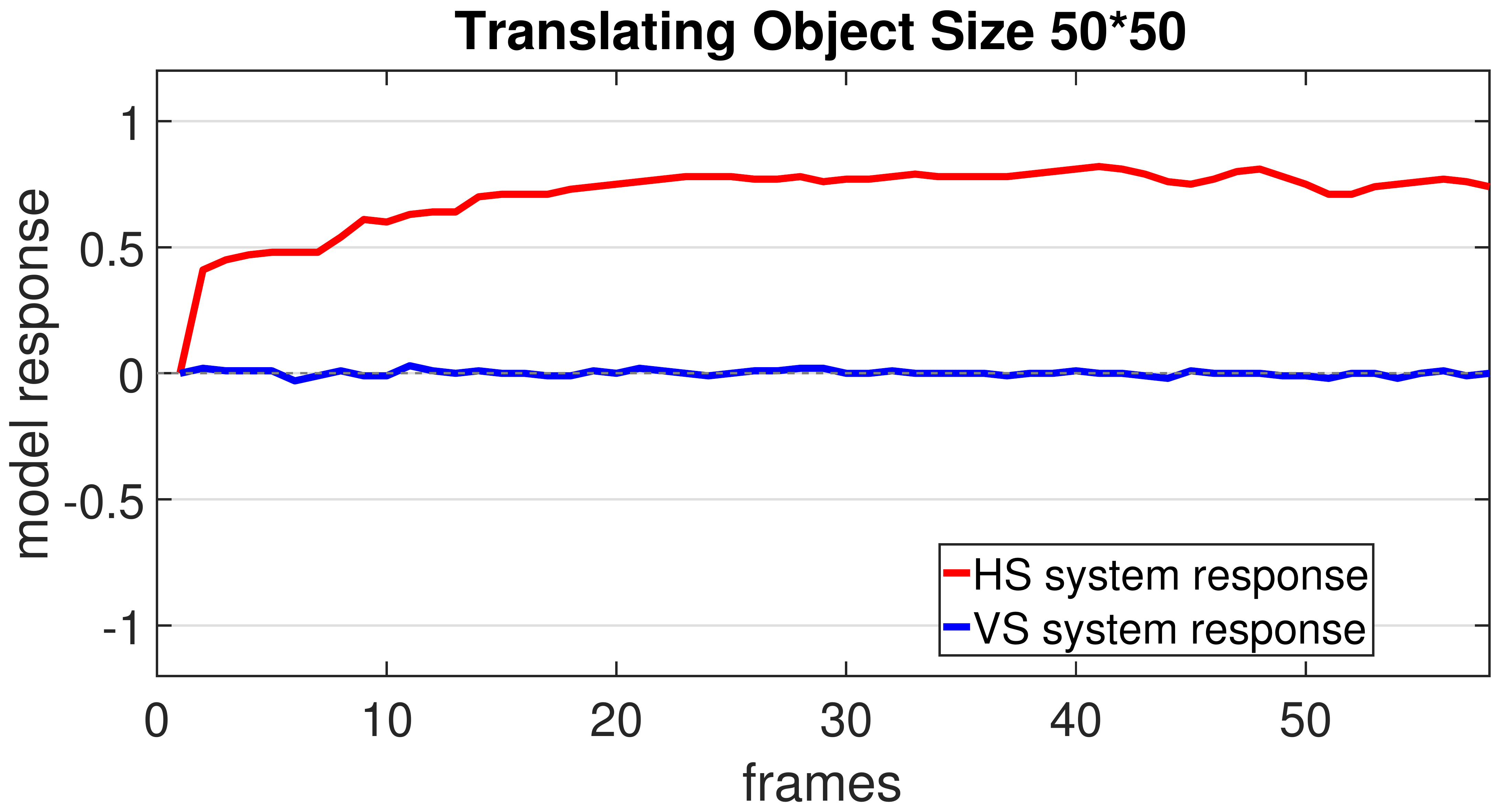}}
		\hfil
		\subfloat[model response (scene \#2)]{\includegraphics[width=0.49\textwidth]{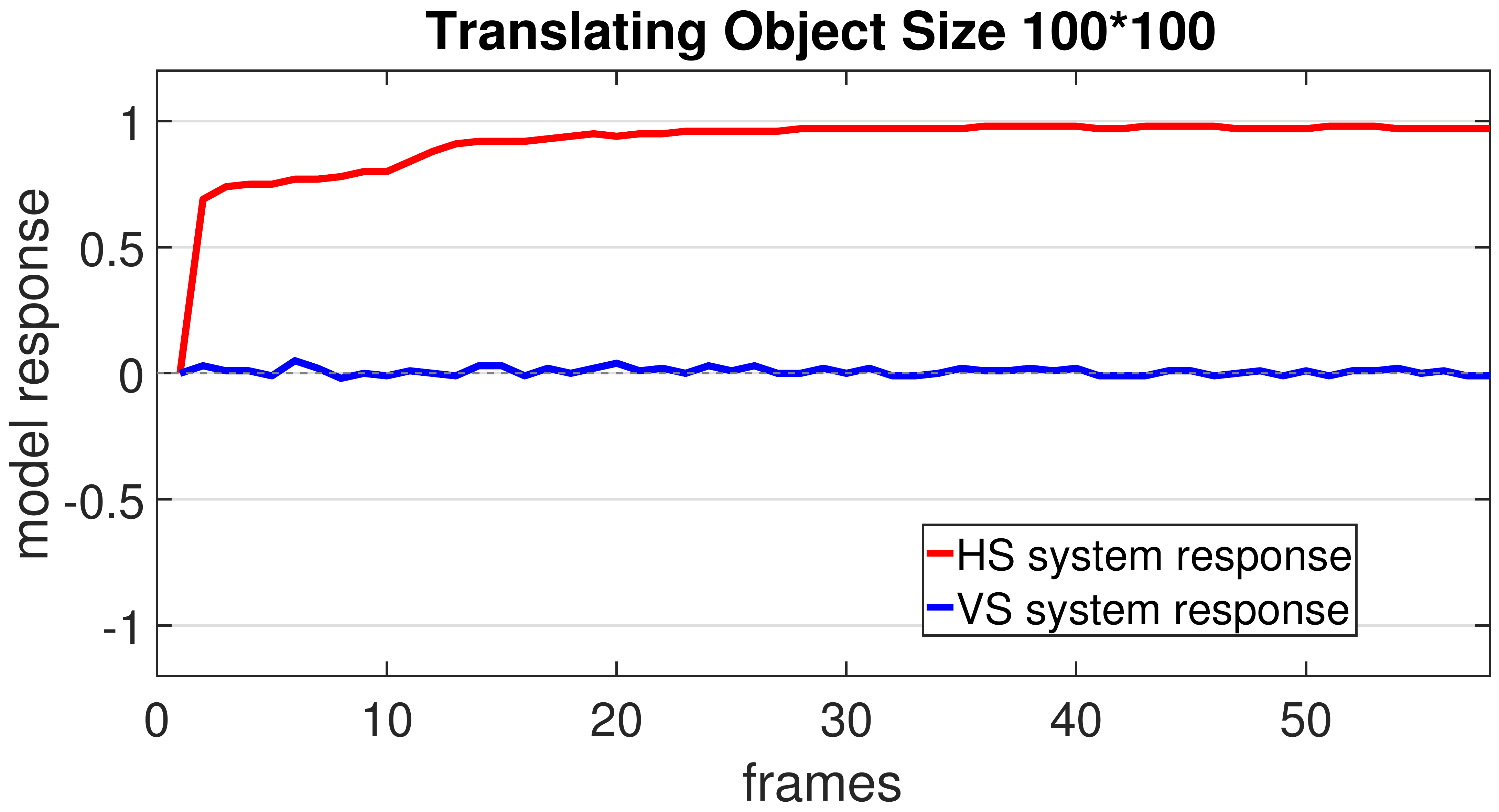}}
	\end{center}
	\caption{
		Proposed model responses by different sized white squares translating \textbf{rightward} at 27 degrees/s, in front of the cluttered background (scene \#2) moving \textbf{leftward} at -40 degrees/s.
	}
	\label{Fig: translating-object-size-investigation}
	%	\vspace{-10pt}
\end{figure}

Secondly, we compare the performance of model in the absence of proposed combination of spatial (vDoG) and temporal (FDSR) mechanisms refining ON and OFF contrast before generating the DS and DO responses. 
Fig. \ref{Fig: dynamics-investigation} illustrates the outputs in comparison with the Fig. \ref{Fig: Fig-offline-moving-clutter-translating-tests}. 
Obviously, without the coordination of proposed spatiotemporal mechanisms, the model is no longer capable of accurately decoding the direction of translating objects in front of the cluttered moving backgrounds. 
More concretely, the HS system represents negative responses to the ND translational OF caused by the backgrounds, and the VS system is also highly activated. 
The results further verify the importance of proposed spatiotemporal dynamics to fit with the desired robustness in cluttered moving backgrounds.

\begin{figure}[t!]
	\begin{center}
		\subfloat[HS system responses]{\includegraphics[width=0.49\textwidth]{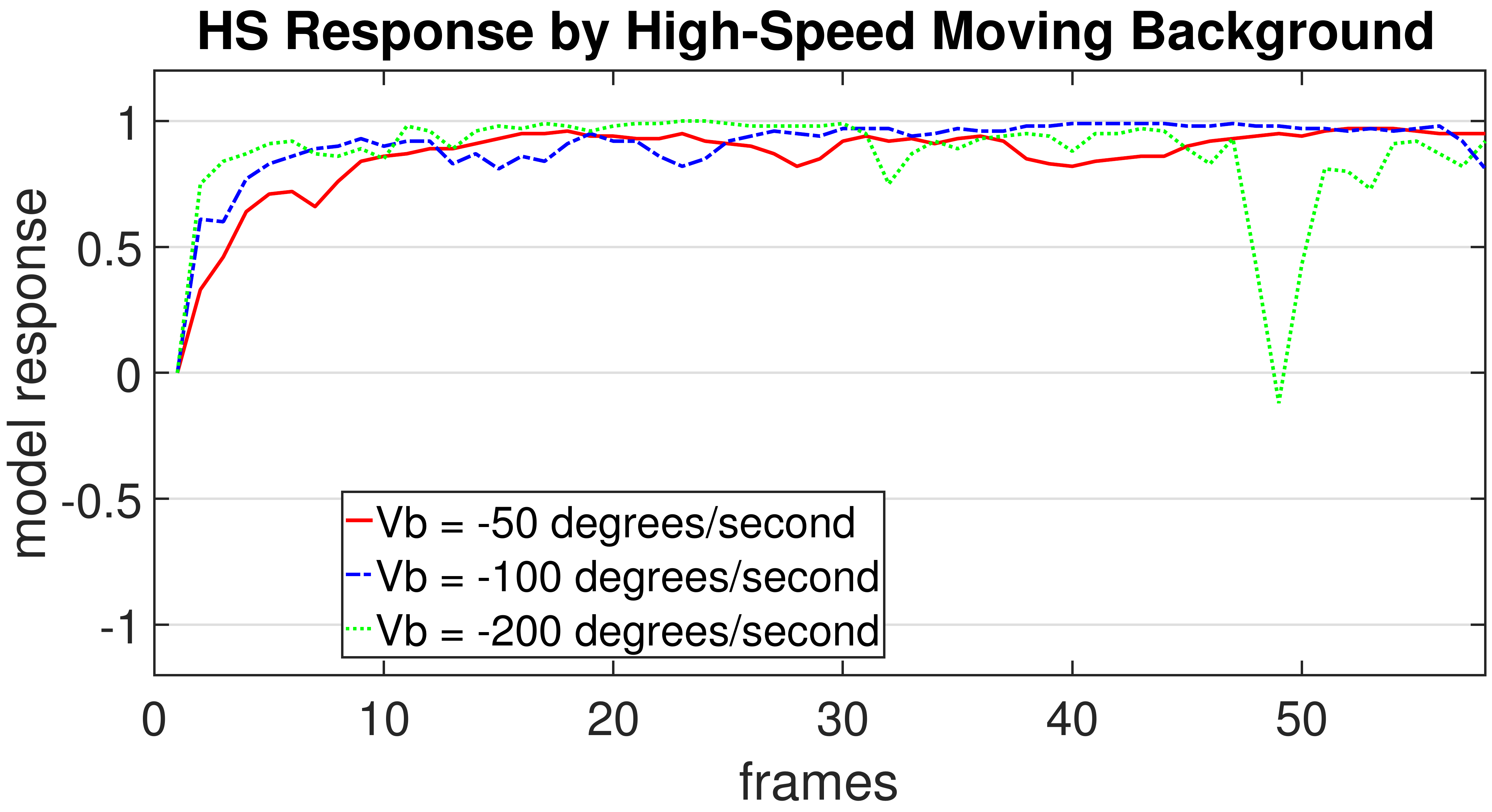}}
		\hfil
		\subfloat[VS system responses]{\includegraphics[width=0.49\textwidth]{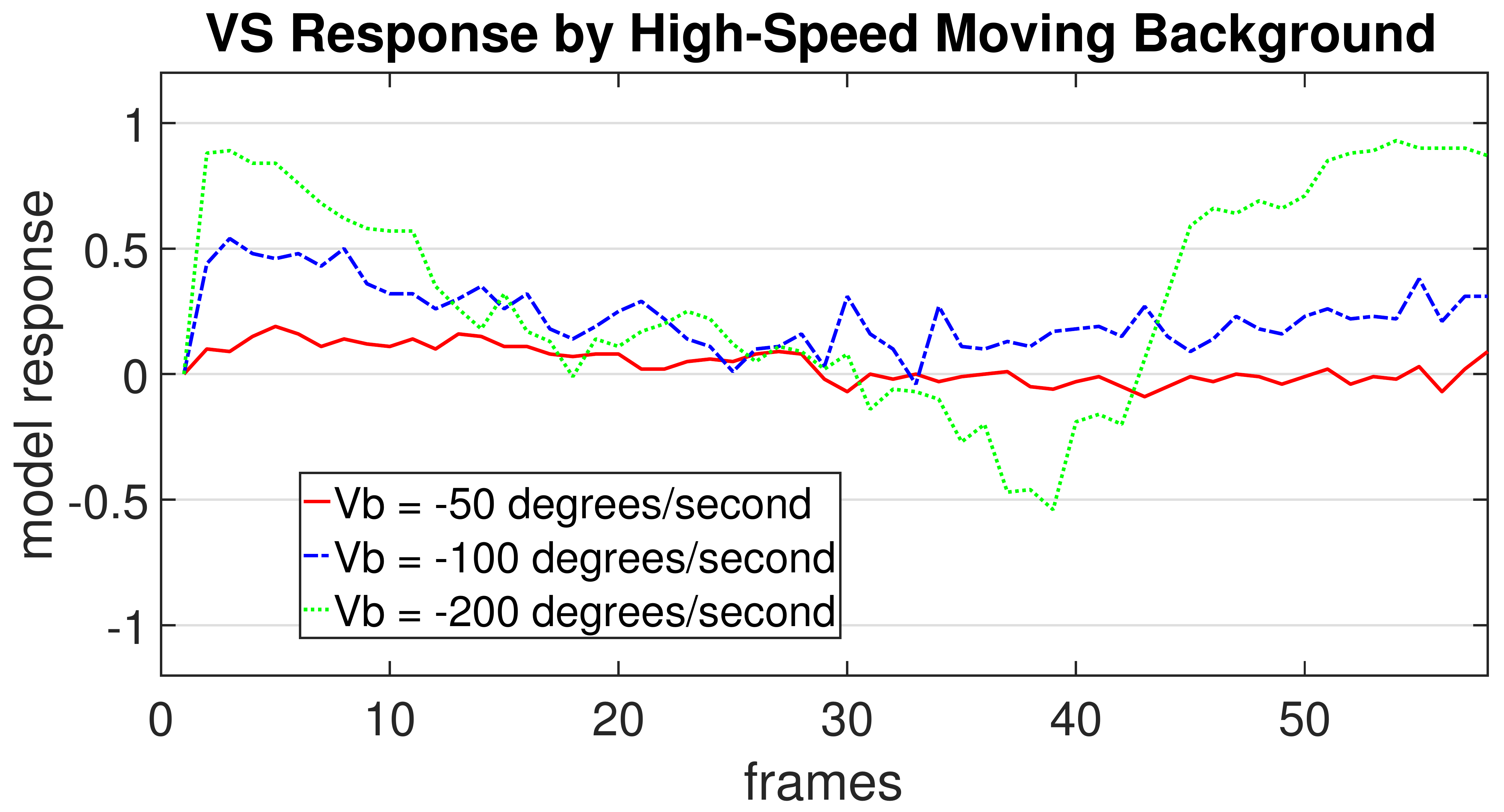}}
	\end{center}
	\caption{
		Proposed model responses by the white object ($25 \times 120$ pixels in size) translating rightward at 27 degrees/s, embedded in the high-speed and leftward moving background (scene \#2) at -50, -100, -200 degrees/s, respectively.
	}
	\label{Fig: high-speed-background-investigation}
%	\vspace{-10pt}
\end{figure}

Lastly, we also investigate the model performance on visual stimuli possessing different properties including varying sizes of translating objects and higher-speed moving backgrounds. 
Fig. \ref{Fig: translating-object-size-investigation} and \ref{Fig: high-speed-background-investigation} illustrate the results. 
Interestingly, the proposed model can detect the different sized targets moving in front of a cluttered, and fast-moving background. 
However, the model demonstrates responsive preference to larger over smaller sized targets representing stronger response of the HS system; whilst the VS system is rigorously suppressed by all the tested visual stimuli, as expected (see Fig. \ref{Fig: translating-object-size-investigation}).

When tested by the very high-speed moving cluttered background, some negative results are obtained: the HS system of the proposed model also responds correctly to the PD motion of foreground translation; the VS system nevertheless is activated more constantly than the afore-tested background angular velocities, especially at the highest velocity of -200 degrees/second  (see Fig. \ref{Fig: high-speed-background-investigation}). 
Therefore, the very high-speed cluttered moving background still poses a problem on decoding the direction of foreground moving object.

\section{Discussion}
\label{Sec: discussion}

\subsection{Characterisation of the Model}

We have demonstrated the effectiveness of the proposed \textit{Drosophila} motion vision pathways model for decoding the direction of translating objects in front of different visual backgrounds, from simple to more challenging cluttered moving ones. 
The visual system model articulates the signal processing behind the OF level, and satisfactorily reproduces the DS and DO responses revealed in the neural circuits. 
The direction of foreground translating objects is indicated by the global membrane potential of the wide-field HS and VS systems, with which the positive or negative response indicate the PD or ND motion (Fig. \ref{Fig: Fig-offline-clean-background-translating-tests}, \ref{Fig: Fig-offline-real-world-translating-tests} and \ref{Fig: Fig-offline-moving-clutter-translating-tests}). 
Importantly, the model shows robust direction selectivity and dynamic response to translating objects in front of cluttered moving backgrounds, at a range of tested angular velocities for both the foreground targets and the backgrounds (Fig. \ref{Fig: hs-vs-stats} and \ref{Fig: ensemble-parameter-investigation}). 
In addition, the model also shows responsive preference to faster-moving (Fig. \ref{Fig: hs-vs-stats} and \ref{Fig: ensemble-parameter-investigation}), larger-size (Fig. \ref{Fig: translating-object-size-investigation}), higher-contrast (Fig. \ref{Fig: Fig-offline-moving-clutter-translating-tests} and \ref{Fig: hs-vs-stats}) translating targets. 
Moreover, we have clarified the importance of proposed modelling of spatiotemporal dynamics to refine the ON and OFF contrasts prior to the generation of DS responses (Fig. \ref{Fig: dynamics-investigation}), and to improve dynamic response in speed tuning (Fig. \ref{Fig: ensemble-parameter-investigation}). 
Furthermore, we have also shown the existing limitation of the proposed visual system model for foreground translation perception, that is, the model could not suppress the very high-speed background translational OF, to an acceptable level (Fig. \ref{Fig: high-speed-background-investigation}). 
In fact, a single-type neural-pathway computation may be insufficient to handle this challenge, whereas the coordination of multiple neural pathways could be a potential solution.

Considering further improvements on this work, we also have the following observations: 
\begin{enumerate}
	\item A recent physiological study has suggested the visual interneurons in the medulla layer of fly motion vision pathways can rapidly adjust the contrast sensitivity \cite{Drews-dynamic-signal-compression}. 
	Therefore, the implementation of contrast normalisation in the proposed model could reduce the high contrast fluctuations in natural images.
	\item For the experimental setting, our model processes signals at only 30Hz that is 8 times lower than the fly's eye at about 250Hz, although our model works with 126000 pixels, 15 times over the fly with approximately 8400 pixels in total. 
	We will further investigate the proposed method by matching the settings with the fly visual systems and applying binocular vision.
\end{enumerate}

\subsection{Coordination of Multiple Neural Systems}

The experiments have also demonstrated the proposed model can perceive small translating object, and decode its direction under a same model setting. 
However, in contrast to the STMD models \cite{Wiederman2008(STMD-clutter),Wiederman2013(ON-OFF-correlation)}, the wide-field HS and VS systems have responsive preference to larger-size targets resulting in much stronger responses (Fig. \ref{Fig: translating-object-size-investigation}). 
A fascinating future work could be the integration of multiple neural pathways for size-varying target pursuit in natural environments.

Furthermore, the proposed visual system model perfectly complements the looming sensitive neural models like the LGMD \cite{Fu-2018(LGMD1-NN),Fu-LGMD2-TCYB}, on the aspect of direction selectivity (Fig. \ref{Fig: Fig-offline-clean-background-approaching-tests}). 
The coordination of them could facilitate the perception of different motion patterns in more challenging scenarios.

\section{Conclusion}
\label{Sec: conclusion}

This paper presents computational modelling of the \textit{Drosophila} motion vision pathways accounting for how the flies decode the direction of a moving target in front of highly variable backgrounds. 
The emphasis herein is laid behind the OF level: the proposed model mimics the visual processing from the photoreceptors, through the parallel ON and OFF pathways, to the LPTCs in four stratified sub-layers sensitive to motion in four cardinal directions. 
The wide-field HS and VS systems integrate the DS and DO responses from the LPTCs as the model outputs, with which the positive or negative response indicates the PD (rightward, downward) or ND (leftward, upward) translational motion. 
To extract merely the foreground translation and improve the dynamic response in a cluttered moving background, the proposed modelling of spatiotemporal dynamics including the coordination of motion pre-filtering mechanisms and the ensembles of local correlators inside the ON and OFF channels, works effectively. 
The experiments have verified the effectiveness of the proposed model with robust direction selectivity in various backgrounds, and also demonstrated its specific responsive preference. 
The proposed model processes signals in a feed-forward manner resembling the \textit{Drosophila} physiology; its computational efficiency and flexibility could fit with building neuromorphic sensors, either featuring compact size or achieving higher processing speed, for utility in mobile intelligent machines.

\bibliographystyle{spbasic}      % basic style, author-year citations
\bibliography{qinbingbib}   % name your BibTeX data base

\begin{thebibliography}{53}
\providecommand{\natexlab}[1]{#1}
\providecommand{\url}[1]{{#1}}
\providecommand{\urlprefix}{URL }
\expandafter\ifx\csname urlstyle\endcsname\relax
  \providecommand{\doi}[1]{DOI~\discretionary{}{}{}#1}\else
  \providecommand{\doi}{DOI~\discretionary{}{}{}\begingroup
  \urlstyle{rm}\Url}\fi
\providecommand{\eprint}[2][]{\url{#2}}

\bibitem[{Badwan et~al.(2019)Badwan, Creamer, Zavatone-Veth, and
  Clark}]{Badwan-Dynamic-nonlinearities-EMD}
Badwan BA, Creamer MS, Zavatone-Veth JA, Clark DA (2019) Dynamic nonlinearities
  enable direction opponency in \textit{Drosophila} elementary motion
  detectors. nature neuroscience 22:1318--1326

\bibitem[{Bagheri et~al.(2017)Bagheri, Cazzolato, Grainger, O’Carroll, and
  Wiederman}]{bagheri2017autonomous}
Bagheri ZM, Cazzolato BS, Grainger S, O’Carroll DC, Wiederman SD (2017) An
  autonomous robot inspired by insect neurophysiology pursues moving features
  in natural environments. Journal of Neural Engineering 14(4):046030

\bibitem[{Barlow and Levick(1965)}]{Barlow-1965(rabbit-DSNs)}
Barlow H, Levick W (1965) The mechanism of directionally selective units in
  rabbit's retina. Journal of Physiology 178:477--504

\bibitem[{Borst(2014)}]{Borst-2014(review-fly)}
Borst A (2014) Fly visual course control: Behaviour, algorithms and circuits.
  Nature Reviews Neuroscience 15:590--599

\bibitem[{Borst and Egelhaaf(1989)}]{EMD-1989(principles-review)}
Borst A, Egelhaaf M (1989) Principles of visual motion detection. Trends in
  Neurosciences 12:297--306

\bibitem[{Borst and Euler(2011)}]{Borst2011(review-motion)}
Borst A, Euler T (2011) Seeing things in motion: Models, circuits, and
  mechanisms. Neuron 71(6):974--994

\bibitem[{Borst and Helmstaedter(2015)}]{Borst2015(common-circuit-motion)}
Borst A, Helmstaedter M (2015) Common circuit design in fly and mammalian
  motion vision. Nature Neuroscience 18(8):1067--1076

\bibitem[{Borst et~al.(2010)Borst, Haag, and
  Reiff}]{Borst-2010(review-fly-vision)}
Borst A, Haag J, Reiff DF (2010) Fly motion vision. The Annual Review of
  Neuroscience 33:49--70

\bibitem[{Borst et~al.(2020)Borst, Haag, and Mauss}]{Borst-Review-2019-Fly}
Borst A, Haag J, Mauss AS (2020) How fly neurons compute the direction of
  visual motion. Journal of Comparative Physiology A 206:109--124

\bibitem[{Brinkworth and O'Carroll(2009)}]{OF-angular-velocity-PLoS-2009}
Brinkworth RSA, O'Carroll DC (2009) Robust models for optic flow coding in
  natural scenes inspired by insect biology. PLoS Computational Biology 5(11)

\bibitem[{Clark et~al.(2011)Clark, Bursztyn, Horowitz, Schnitzer, and
  Clandinin}]{Clark_2011(6Q-model-fly)}
Clark DA, Bursztyn L, Horowitz MA, Schnitzer MJ, Clandinin TR (2011) Defining
  the computational structure of the motion detector in drosophila. Neuron
  70(6):1165--1177

\bibitem[{Cope et~al.(2016)Cope, Sabo, Gurney, Vasilaki, and
  Marshall}]{Cope2016(model-angular-bee)}
Cope AJ, Sabo C, Gurney K, Vasilaki E, Marshall JA (2016) A model for an
  angular velocity-tuned motion detector accounting for deviations in the
  corridor-centering response of the bee. PLoS Computational Biology
  12(5):1--22

\bibitem[{Drews et~al.(2020)Drews, Leonhardt, Pirogova, Richter,
  Schuetzenberger, Braun, Serbe, and Borst}]{Drews-dynamic-signal-compression}
Drews MS, Leonhardt A, Pirogova N, Richter FG, Schuetzenberger A, Braun L,
  Serbe E, Borst A (2020) Dynamic signal compression for robust motion vision
  in flies. Current Biology 30:209--221

\bibitem[{Eichner et~al.(2011)Eichner, Joesch, Schnell, Reiff, and
  Borst}]{Eichner2011(2Q-motion)}
Eichner H, Joesch M, Schnell B, Reiff DF, Borst A (2011) Internal structure of
  the fly elementary motion detector. Neuron 70(6):1155--1164

\bibitem[{Escobar et~al.(2019)Escobar, Ohradzansky, Keshavan, Ranganathan, and
  Humbert}]{Alvarez-Small-Object-Avoidance-2019}
Escobar HD, Ohradzansky M, Keshavan J, Ranganathan BN, Humbert JS (2019)
  Bioinspired approaches for autonomous small-object detection and avoidance.
  IEEE Transactions on Robotics 35(5):1220--1232

\bibitem[{Fisher et~al.(2015)Fisher, Leong, Sporar, Ketkar, Gohl, Clandinin,
  and Silies}]{Fisher-L3(wide-field-local)}
Fisher YE, Leong JCS, Sporar K, Ketkar MD, Gohl DM, Clandinin TR, Silies M
  (2015) A class of visual neurons with wide-field properties is required for
  local motion detection. Current Biology 25(24):3178--3189

\bibitem[{Franceschini(2014)}]{Nicolas-2014(Review-Fly-Robot)}
Franceschini N (2014) Small brains, smart machines: From fly vision to robot
  vision and back again. Proceedings of the IEEE 102:751--781

\bibitem[{Franceschini et~al.(1989)Franceschini, Riehle, and
  Le~Nestour}]{Nicolas-1989(DSN-Insect-Neurons)}
Franceschini N, Riehle A, Le~Nestour A (1989) Directionally selective motion
  detection by insect neurons. In: Stavenga DG, Hardie RC (eds) Facets of
  Vision, Springer Berlin Heidelberg, pp 360--390

\bibitem[{Franceschini et~al.(1992)Franceschini, Pichon, and
  Blanes}]{Nicolas-insects-robot-vision}
Franceschini N, Pichon J, Blanes C (1992) From insect vision to robot vision.
  Philosophical Transactions of the Royal Society B 337(1281):283--294

\bibitem[{Frye(2015)}]{Frye2015(EMDs-basic)}
Frye M (2015) Elementary motion detectors. Current Biology 25(6):R215--R217

\bibitem[{Fu and Yue(2017{\natexlab{a}})}]{Fu-2017(ROBIO-fixation)}
Fu Q, Yue S (2017{\natexlab{a}}) Mimicking fly motion tracking and fixation
  behaviors with a hybrid visual neural network. In: Proceedings of the 2017
  IEEE international conference on robotics and biomimetics (ROBIO), IEEE, pp
  1636--1641

\bibitem[{Fu and Yue(2017{\natexlab{b}})}]{Fu2017(fly-DSNs-IJCNN)}
Fu Q, Yue S (2017{\natexlab{b}}) Modeling direction selective visual neural
  network with on and off pathways for extracting motion cues from cluttered
  background. In: Proceedings of the 2017 international joint conference on
  neural networks (IJCNN), IEEE, pp 831--838

\bibitem[{Fu et~al.(2016)Fu, Yue, and Hu}]{Fu-2016(LGMD2-BMVC)}
Fu Q, Yue S, Hu C (2016) Bio-inspired collision detector with enhanced
  selectivity for ground robotic vision system. In: British machine vision
  conference, BMVA Press, pp 1--13

\bibitem[{Fu et~al.(2017)Fu, Hu, Liu, and Yue}]{Fu2017a(LGMDs-IROS)}
Fu Q, Hu C, Liu T, Yue S (2017) Collision selective {LGMDs} neuron models
  research benefits from a vision-based autonomous micro robot. In: Proceedings
  of the 2017 IEEE/RSJ international conference on intelligent robots and
  systems (IROS), IEEE, pp 3996--4002

\bibitem[{{Fu} et~al.(2018){Fu}, {Bellotto}, {Hu}, and {Yue}}]{Fu-ROBIO-2018}
{Fu} Q, {Bellotto} N, {Hu} C, {Yue} S (2018) Performance of a visual fixation
  model in an autonomous micro robot inspired by drosophila physiology. In:
  Proceedings of the 2018 IEEE international conference on robotics and
  biomimetics (ROBIO), IEEE, pp 1802--1808

\bibitem[{Fu et~al.(2018{\natexlab{a}})Fu, Hu, Liu, and Yue}]{Fu-TAROS(review)}
Fu Q, Hu C, Liu P, Yue S (2018{\natexlab{a}}) Towards computational models of
  insect motion detectors for robot vision. In: Giuliani M, Assaf T,
  Giannaccini ME (eds) Towards autonomous robotic systems conference, Springer
  International Publishing, pp 465--467

\bibitem[{Fu et~al.(2018{\natexlab{b}})Fu, Hu, Peng, and
  Yue}]{Fu-2018(LGMD1-NN)}
Fu Q, Hu C, Peng J, Yue S (2018{\natexlab{b}}) Shaping the collision
  selectivity in a looming sensitive neuron model with parallel {ON} and {OFF}
  pathways and spike frequency adaptation. Neural Networks 106:127--143,
  \doi{https://doi.org/10.1016/j.neunet.2018.04.001}

\bibitem[{Fu et~al.(2019{\natexlab{a}})Fu, Hu, Peng, Rind, and
  Yue}]{Fu-LGMD2-TCYB}
Fu Q, Hu C, Peng J, Rind FC, Yue S (2019{\natexlab{a}}) A robust collision
  perception visual neural network with specific selectivity to darker objects.
  IEEE Transactions on Cybernetics pp 1--15, \doi{10.1109/TCYB.2019.2946090}

\bibitem[{Fu et~al.(2019{\natexlab{b}})Fu, Wang, Hu, and Yue}]{Fu-ALife-review}
Fu Q, Wang H, Hu C, Yue S (2019{\natexlab{b}}) Towards computational models and
  applications of insect visual systems for motion perception: A review.
  Artificial Life 25(3):263--311

\bibitem[{Gabbiani and Jones(2011)}]{Circuit-genetic(genetic-push-motion)}
Gabbiani F, Jones PW (2011) A genetic push to understand motion detection.
  Neuron 70(6):1023--1025

\bibitem[{Haag et~al.(2016)Haag, Arenz, Serbe, Gabbiani, and
  Borst}]{Fly2016(direction-selectivity-fly)}
Haag J, Arenz A, Serbe E, Gabbiani F, Borst A (2016) Complementary mechanisms
  create direction selectivity in the fly. eLife 5:1--15

\bibitem[{Hassenstein and Reichardt(1956)}]{HR-1956(EMD-1956)}
Hassenstein B, Reichardt W (1956) Systemtheoretische analyse der zeit-,
  reihenfolgen- und vorzeichenauswertung bei der bewegungsperzeption des
  riisselkiifers chlorophanus. Zeitschrift fur Naturforschung pp 513--524

\bibitem[{Iida and Lambrinos(2000)}]{Iida_2000(fly-visual-odometer)}
Iida F, Lambrinos D (2000) Navigation in an autonomous flying robot by using a
  biologically inspired visual odometer. Sensor Fusion and Decentralized
  Control in RoboticSystem III Photonics East 4196:86--97

\bibitem[{Joesch et~al.(2010)Joesch, Schnell, Raghu, Reiff, and
  Borst}]{Joesch-2010(ON-OFF-fly)}
Joesch M, Schnell B, Raghu SV, Reiff DF, Borst A (2010) {ON} and {OFF} pathways
  in drosophila motion vision. Nature 468(7321):300--304

\bibitem[{Joesch et~al.(2013)Joesch, Weber, Eichner, and
  Borst}]{Joesch_2013(functional-ON-OFF)}
Joesch M, Weber F, Eichner H, Borst A (2013) Functional specialization of
  parallel motion detection circuits in the fly. Journal of Neuroscience
  33(3):902--905

\bibitem[{Maisak et~al.(2013)Maisak, Haag, Ammer, Serbe, Meier, Leonhardt,
  Schilling, Bahl, Rubin, Nern, Dickson, Reiff, Hopp, and
  Borst}]{Maisak_2013(T4-T5-fly)}
Maisak MS, Haag J, Ammer G, Serbe E, Meier M, Leonhardt A, Schilling T, Bahl A,
  Rubin GM, Nern A, Dickson BJ, Reiff DF, Hopp E, Borst A (2013) A directional
  tuning map of drosophila elementary motion detectors. Nature
  500(7461):212--216

\bibitem[{Mauss et~al.(2015)Mauss, Pankova, Arenz, Nern, Rubin, and
  Borst}]{Mauss-Opposing-Motions-Fly}
Mauss AS, Pankova K, Arenz A, Nern A, Rubin GM, Borst A (2015) Neural circuit
  to integrate opposing motions in the visual field. Cell 162:351--362

\bibitem[{Moeckel and Liu(2007)}]{Moeckel-Liu-time-to-travel-model}
Moeckel R, Liu SC (2007) Motion detection circuits for a time-to-travel
  algorithm. In: Proceedings of the 2007 IEEE international symposium on
  circuits and systems, IEEE, pp 3079--3082

\bibitem[{Riehle and Franceschini(1984)}]{Riehle-fly-ON-OFF-1984}
Riehle A, Franceschini NH (1984) Motion detection in flies: Parametric control
  over on-off pathways. Experimental Brain Research 54(2):390--394

\bibitem[{Rind(1990)}]{DSNs-1990(Rind-locust-DSNs)}
Rind F (1990) Identification of directionally selective motion-detecting
  neurones in the locust lobula and their synaptic connections with an
  identified descending neurone. Journal of Experimental Biology 149:21--43

\bibitem[{Rister et~al.(2007)Rister, Pauls, Schnell, Ting, Lee, Sinakevitch,
  Morante, Strausfeld, Ito, and
  Heisenberg}]{Rister-2007(dissection-motion-channel)}
Rister J, Pauls D, Schnell B, Ting CY, Lee CH, Sinakevitch I, Morante J,
  Strausfeld NJ, Ito K, Heisenberg M (2007) Dissection of the peripheral motion
  channel in the visual system of drosophila melanogaster. Neuron
  56(1):155--170

\bibitem[{Serres and Ruffier(2017)}]{Serres2017(review-optic-flow)}
Serres JR, Ruffier F (2017) Optic flow-based collision-free strategies: From
  insects to robots. Arthropod Structure and Development 46(5):703--717

\bibitem[{Straw(2008)}]{Straw-VisionEgg}
Straw AD (2008) Vision egg: an open-source library for realtime visual stimulus
  generation. Frontiers in Neuroinformatics 2(4)

\bibitem[{Strother et~al.(2014)Strother, Nern, and
  Reiser}]{Strother2014(direct-observation-on-off)}
Strother JA, Nern A, Reiser MB (2014) Direct observation of on and off pathways
  in the drosophila visual system. Current Biology 24(9):976--983

\bibitem[{Strother et~al.(2017)Strother, Wu, Wong, Nern, Rogers, Le, Rubin, and
  Reiser}]{Fly-DS-2017(emergence-direction-selectivity)}
Strother JA, Wu ST, Wong AM, Nern A, Rogers EM, Le JQ, Rubin GM, Reiser MB
  (2017) The emergence of directional selectivity in the visual motion pathway
  of drosophila. Neuron 94(1):168--182.e10

\bibitem[{Vanhoutte et~al.(2017)Vanhoutte, Mafrica, Ruffier, Bootsma, and
  Serres}]{Vanhoutte-time-of-travel-OF-micro-flying-robot}
Vanhoutte E, Mafrica S, Ruffier F, Bootsma RJ, Serres J (2017) Time-of-travel
  methods for measuring optical flow on board a micro flying robot. Sensors
  17(3):571

\bibitem[{Wang et~al.(2019{\natexlab{a}})Wang, Fu, Wang, Peng, Baxter, Hu, and
  Yue}]{huatian-IJCNN-2019}
Wang H, Fu Q, Wang H, Peng J, Baxter P, Hu C, Yue S (2019{\natexlab{a}})
  Angular velocity estimation of image motion mimicking the honeybee tunnel
  centring behaviour. In: Proceedings of the 2019 IEEE International Joint
  Conference on Neural Networks, IEEE

\bibitem[{Wang et~al.(2019{\natexlab{b}})Wang, Fu, Wang, Peng, and
  Yue}]{huatian-AIAI-2019}
Wang H, Fu Q, Wang H, Peng J, Yue S (2019{\natexlab{b}}) Constant angular
  velocity regulation for visually guided terrain following. In: Artificial
  Intelligence Applications and Innovations, Springer International Publishing,
  pp 597--608

\bibitem[{Wiederman and O’Carroll(2013)}]{wiederman2013biologically}
Wiederman SD, O’Carroll DC (2013) Biologically inspired feature detection
  using cascaded correlations of {OFF} and {ON} channels. Journal of Artificial
  Intelligence and Soft Computing Research 3(1):5--14

\bibitem[{Wiederman et~al.(2008)Wiederman, Shoemaker, and
  O'Carroll}]{Wiederman2008(STMD-clutter)}
Wiederman SD, Shoemaker PA, O'Carroll DC (2008) A model for the detection of
  moving targets in visual clutter inspired by insect physiology. PLoS One
  3(7):1--11

\bibitem[{Wiederman et~al.(2013)Wiederman, Shoemaker, and
  O'Carroll}]{Wiederman2013(ON-OFF-correlation)}
Wiederman SD, Shoemaker PA, O'Carroll DC (2013) Correlation between {OFF} and
  {ON} channels underlies dark target selectivity in an insect visual system.
  Journal of Neuroscience 33(32):13225--13232

\bibitem[{Zanker and Zeil(2005)}]{Zanker_2005(motion-signal-outdoor)}
Zanker JM, Zeil J (2005) Movement-induced motion signal distributions in
  outdoor scenes. Network: Computation in Neural Systems 16(4):357--376

\bibitem[{Zanker et~al.(1999)Zanker, Srinivasan, and
  Egelhaaf}]{Zanker-1999(EMD-speed-tuning)}
Zanker JM, Srinivasan MV, Egelhaaf M (1999) Speed tuning in elementary motion
  detectors of the correlation type. Biological Cybernetics 80(2):109--116

\end{thebibliography}

\end{document}